\documentclass[]{aa} 
\usepackage{graphicx, textcomp}
\usepackage{multirow}
\usepackage{natbib}
\usepackage[switch]{lineno}
\usepackage[]{threeparttable}
\usepackage{tabularx}
\usepackage{txfonts}
 \usepackage{hyperref}
 \hypersetup{
      colorlinks   = true,
      citecolor    = blue
 }
\authorrunning{Boccardi B. at al.}

\begin{document}

   \title{Jet formation studies in AGN: a search for new targets}

\author{B. Boccardi\inst{1}, L. Ricci\inst{2, 1}, E. Madika\inst{1}, V. Bartolini\inst{1}, U. Bach\inst{1}, P. Grandi\inst{3}, E. Torresi\inst{3}, T.P. Krichbaum\inst{1}, J. A. Zensus\inst{1}}
\institute{Max-Planck-Institut f\"{u}r Radioastronomie, Auf dem H\"{u}gel 69, D-53121 Bonn, Germany \and Julius-Maximilians-Universit{\"a}t W{\"u}rzburg, Fakult{\"a}t für Physik und 
Astronomie, Institut für Theoretische Physik und Astrophysik, Lehrstuhl für Astronomie, Emil-Fischer-Str. 31, D-97074 W{\"u}rzburg \and INAF -- Osservatorio di Astrofisica e Scienza dello Spazio di Bologna, Via Gobetti 101, I-40129 Bologna, Italy}

\date{Received November 23, 2024; 
Accepted December 23, 2024}

 
  \abstract
  {In recent years, the jet formation region in active galaxies has been imaged through mm-VLBI in few ideal targets, first and foremost M\,87. An important leap forward for understanding jet launching could be made by identifying a larger number of suitable objects, characterized by different accretion modes and jet powers. While numerous, potentially excellent targets exist in the nearby Universe, their VLBI properties in the mm-band are mostly unknown due to general faintness of this population.}
 {In this article, we present 1\,cm and 7\,mm VLBI data of a sample of sixteen poorly explored radio galaxies, comprising both High-Excitation (HEG) and Low-Excitation Galaxies (LEG) and spanning a large range in radio power. Several among them are $\gamma$-ray emitters. The combination of the sources vicinity ($z<0.1$) with a large black hole mass ($\log{M_{\rm BH}}\gtrsim 8.5$) results in a high spatial resolution in units of Schwarzschild radii ($<10^3-10^4$\,$R_{\rm S}$), necessary for probing the region where the jet is initially accelerated and collimated. We aim at identifying the best candidates for follow-up observations with current and future VLBI facilities.}
{The observations were performed with the High Sensitivity Array, including Effelsberg and the phased-VLA, which has allowed us to characterize the sub-parsec properties of these faint jets and to estimate their core brightness temperature and orientation.}
{The number of sources imaged on scales $\lesssim 10^3$\,$R_{\rm S}$ is more than doubled by our study. All targets were detected at both frequencies, and several present two-sided jet structures. Several LEG jets show hints of limb-brightening. The core brightness temperatures are generally below the equipartition value, indicating that equipartition has not yet been reached and/or that the emission is de-boosted. Among LEG, we identify  3C\,31, 3C\,66B, and 3C\,465 as the most promising, combining a relatively high flux density ($>50$\,$\rm mJy$) with superb spatial resolution ($<500$\,$R_{\rm S}$) at 7\,mm. The powerful HEG 3C\,452 is interesting as well due to its highly symmetric, two-sided jet base. Most sources are expected to become prime targets for future experiments with the ngEHT and ngVLA.}
{}


   \maketitle

\section{Introduction}
Extra-galactic jets in active galactic nuclei (AGN),  observed for the first time a century ago in the optical \citep{Curtis1918}, 
have been subject of a renewed interest from the scientific community in recent years \citep[see][for a review]{Blandford2019}. In the mm-radio band, the development of sensitive interferometric arrays providing angular resolutions down to tens of $\rm{micro-arcseconds}$ has enabled the exploration of the immediate surroundings of the supermassive black hole, where jets are formed \citep[][and references therein]{Boccardi2017}. The most exemplary case is the imaging of M\,87 on event horizon scales, performed with the Event Horizon Telescope (EHT) at 230 GHz \citep{EHT2019}. 

M\,87 is the only extra-galactic source where such an exceptional goal can be achieved using current arrays. However, several other objects have been investigated on slightly larger spatial scales, where the jet is undergoing acceleration up to relativistic velocities and collimation down to narrow opening angles. Studies of this region, which extends up to $10^3-10^7$ Schwarzschild radii ($R_{\rm S}$) from the black hole \citep[e.g.,][]{Kovalev2020, Boccardi2021} are important for constraining theoretical models of jet launching \citep{BZ1977, BP1982} and serve as a crucial input for numerical simulations \citep[e.g.,][]{Chatterjee2019, Lalakos2022}. As it is for M\,87, the ideal targets for such studies are nearby radio galaxies hosting very massive black holes, with jets oriented at relatively large angles with respect to the observer. Indeed, these properties enable the jet base to be imaged with high spatial resolution in units of $R_{\rm S}$. Most of the radio galaxies examined in detail so far have jets with relatively low power and host low-luminosity nuclei, i.e., they are optically classified as low-excitation galaxies \citep[LEG, see][]{HB2014}. Some notable examples, besides M\,87 itself, are Cen\,A \citep{Janssen2021}, NGC\,1052 \citep{Baczko2022}, and NGC\,4261 \citep{Yan2023}. The classification as LEG signals the presence of a radiatively inefficient, hot accretion flow at the center of these sources.
\begin{table*}[!ht]
\centering
\footnotesize
\caption{Sources in the sample and their main properties.}
\begin{tabular}{cccccccc}
\hline
IAU name & Other & $z$ &$\rm pc/mas$& Log ($M_\mathrm{BH}$) & Accr. mode  & Radio morph. & Log ($P_{\rm t}$)\\  \hline 
0055+30    & \object{NGC\,315}    & 0.0165  &0.33& 9.32${^1}$   &  LEG${^{12}}$ & FRI & 24.24     \\ 
0104+32    & \object{3C\,31}      & 0.0170  &0.34& 9.62${^2}$   &   LEG${^{13}}$ &FRI & 24.80   \\   
0106+13    & \object{3C\,33}     & 0.0597  &1.14& 8.83${^3}$   &  HEG${^{13}}$ &FRII &  26.42    \\ 
0220+43    & \object{3C\,66B}     & 0.0213  &0.43& 9.32${^4}$   &  LEG${^{13}}$ & FRI &  25.30   \\ 
0222+36    & \object{B2\,0222+36} & 0.0333  &0.66& 8.75${^5}$   &   LEG${^{14}}$& C & 23.93    \\ 
0313+41    & \object{IC\,310}     & 0.0188   &0.38& 8.48${^6}$  &   LEG${^{15}}$ & FRI &  23.73  \\
0331+39    & \object{4C\,39.12}   & 0.0206  &0.41& 8.86${^7}$   &   LEG${^{16}}$  & FRI & 24.20    \\ 
0844+31    & \object{IC\,2402}    & 0.0673  &1.27& 8.99${^8}$   &   LEG${^{14}}$ & FRII & 25.53\\         
1142+20    & \object{3C\,264}     & 0.0216  &0.43& 8.67${^4}$   &   LEG${^{13}}$ & FRI  & 25.17  \\ 
1217+29    & \object{NGC\,4278}   & 0.0021  &0.04& 8.53${^9}$   &   LEG${^{14}}$ & C &  21.75   \\ 
1626+39    & \object{3C\,338}     & 0.0304  & 0.60& 8.89${^4}$   &   LEG${^{13}}$ & FRI  & 25.57   \\ 
1658+30    & \object{4C\,30.31}   & 0.0344  &0.68& 8.93${^{10}}$  &   LEG${^{12}}$&FRI  &  24.63  \\ 
1833+32    & \object{3C\,382}    & 0.0579  &1.11& 8.98${^{11}}$&   HEG${^{13}}$ & FRII & 26.04    \\ 
1842+45    & \object{3C\,388}     & 0.0917  &1.69& 9.18${^5}$   &   LEG${^{13}}$& FRII  & 26.47   \\ 
2243+39    & \object{3C\,452}     & 0.0811  &1.51& 8.89${^3}$   &   HEG${^{13}}$ & FRII & 26.64   \\ 
2335+26    & \object{3C\,465}     & 0.0302  &0.60 & 9.14${^4}$   &   LEG${^{13}}$ & FRI  &  25.61  \\ \hline
\end{tabular}
\tablefoot{Col. 1: IAU name. Col. 2: Other name. Col. 3: Redshift. Col. 4: Parsecs per milli-arcsecond. Col. 5: Black hole mass in solar masses. Col. 6: Accretion mode. Col. 7: Radio morphology. 'C' indicates a compact structure on kpc scales. Col. 8: 408 MHz radio power in $\rm W/Hz$.}
\tablebib{
Col.5: 
[1] \cite{Boizelle2021}, [2] \cite{North2019}, [3] \cite{Koss2022}, [4] \cite{Balmaverde2008}, [5] \cite{Woo2002}, [6] \cite{Alecsic2014}, [7]  derived from the relation by \cite{Tremaine2002} based on the stellar velocity dispersion from \cite{vdBosch2015}, [8] \cite{Sikora2013}, [9] \cite{Ly2004}, [10]  derived from the relation by \cite{Tremaine2002} based on the stellar velocity dispersion from \cite{Bettoni2009}, [11] \cite{Fausnaugh2017}. Col. 7: [12] \cite{2004A&A...413...97G}, [13] \cite{Buttiglione2010}, [14] \cite{Liuzzo2013}, [15] \cite{1996AJ....111...53O}.  [16] based on the Eddington ratio estimated in X-rays by \cite{2020ApJ...900..124B}. Col. 8: see \cite{Liuzzo2009}. Originally classified as compact, 4C\,39.12 was re-classified as FRI by \cite{Liuzzo2013}. Col. 9: see \cite{Liuzzo2009}. For IC\,310 only, we estimated the power from the 365 MHz flux density \citep{Texas1996}, assuming $\alpha=-0.7$.}
 \end{table*}

Significant progress in our understanding of jet formation could come by extending such studies to objects characterized by different accretion modes and jet powers. A first attempt in this direction was made by \cite{Boccardi2021}, who have examined the properties of the jet acceleration and collimation region in a small sample comprising not only LEG but also HEG. High-excitation galaxies are thought to be powered by cold thin disks, and generally produce high-power jets developing Fanaroff-Riley (FR) II radio morphologies \citep{Fanaroff1974}. Some interesting trends emerged in this study: jets in HEG were suggested to collimate for larger distances, and to present of a wider sheath surrounding the spine with respect to LEG. However, the numbers are still too low to draw solid conclusions. In particular, jets in HEG are under-represented due to their higher redshifts and lower flux densities and, except for Cygnus\,A \citep{Boccardi2016b, Boccardi2016a}, they have not been imaged on scales smaller than ${\sim}10^4-10^5$\,$R_{\rm S}$ in the mm-band.

The nearby Universe host numerous, potentially excellent new targets \citep[e.g.,][]{Rama2023}. These include radio galaxies often well-studied on kilo-pc scales, but poorly explored on pc and sub-pc scales due to the jet faintness, which in the past precluded an imaging through very-long-baseline interferometry (VLBI) at wavelengths shorter than a few cm. The most sensitive VLBI arrays are nowadays capable of detecting some of these faint targets also in the mm-band, thanks to the use of wide bandwidth receiving systems and to the inclusion of elements with large collecting area, such as the phased-VLA (Very Large Array).
In this article, we investigate the high-frequency VLBI properties of a sample of poorly explored radio galaxies spanning different luminosity classes.  The goal of this work is to identify the most interesting targets, so that follow-up studies can be performed through dedicated observations. In the longer term, this and future experiments will contribute to the compilation of larger samples for performing first population studies of jet formation in AGN. This is especially relevant in view of the planned construction of next-generation (ng) VLBI facilities, such as the ngEHT \citep{Johnson2023} and the ngVLA \citep{Murphy2018}, which will push much further the resolution and sensitivity limits.
\section{Sample description}
 \begin{table*}[!ht]
\centering
\footnotesize
\caption{Log of observations and characteristics of the VLBI clean maps for each source.}
\begin{tabular}{ccc|ccc|ccc}
\hline
&&&&1\,cm&&&7\,mm&\\
\hline
P.C.  & Date &   Source  & Beam & $S_{\mathrm {p}}$ & $S_{\mathrm {t}}$& Beam  & $S_{\mathrm {p}}$ & $S_{\mathrm {t}}$\\ 
 &  & &$\mathrm {[mas\times mas, deg]}$ & $\mathrm {[mJy/beam]}$&  $\mathrm {[mJy]}$& $\mathrm {[mas\times mas, deg]}$ & $\mathrm {[mJy/beam]}$&  $\mathrm {[mJy]}$\\ 
\hline
BB393A&2018/10/23&3C338  &$0.56\times 0.22, -17$&26&56 &$0.30\times 0.13, -6$&19&28\\ 
&&4C30.31 &$0.57\times 0.22, -16$&43&67&$0.29\times 0.12, -11$&38&54\\  
&&3C382 &$0.56\times 0.22, -16$&63&124 &$0.28\times 0.13, -7$&56&81 \\   
&&3C388 &$0.44\times 0.21, -19$&33&45&$0.28\times 0.14, -4$&25&25\\
\hline
BB393B&2018/11/24&IC2402 &$0.66\times 0.26, -22$&35&41 &$0.35\times 0.14, -15$&26&34 \\      
&&3C264&$0.77\times 0.25, -10$&93&151 &$0.39\times 0.14, -9$ &60&99  \\
&&NGC4278  &$0.61\times 0.23, -8$&18&61&$0.35\times 0.12, -11$&14&23 \\ 
\hline
BB393C&2018/11/24&NGC315&$0.59\times0.27, -11$&184&447&$0.30\times0.16, -13$&116&268\\
&&3C31 &$0.60\times 0.23, -15$&63&76&$0.33\times 0.13, -15$&52&64\\ 
&&3C452 &$0.56\times 0.22, -19$&21&68&$0.29\times0.16, -10$&29&53 \\
&&3C465  &$0.66\times0.23, -17$&71&129&$0.33\times0.13, -17$&46&89  \\      
\hline
BB393D&2018/12/02&3C33 &$0.72\times 0.29, -10$&12&17 &$0.42\times 0.24, 16$&19&19 \\
&&3C66B &$0.47\times 0.25, -20$&85&147 &$0.26\times 0.15, -16$&64&111\\ 
&&0222+36 &$0.52\times 0.32, -14$&23&45 &$0.39\times 0.23, -40$&17&27\\
&&IC310 &$0.47\times 0.33, -21$&45&70&$0.32\times 0.24, 14$&34&46 \\
&&4C39.12 &$0.49\times 0.22, -21$&52&75&$0.27\times 0.12, -22$&48&70 \\
\hline
\end{tabular}
\tablefoot{Col. 1: Project code. Col. 2: Date of observation. Col. 3: Source name. Col. 4: Beam FWHM and position angle at 1\,cm.  Col. 5:  Peak intensity  at 1\,cm. Col. 6: Total flux density at 1\,cm. 
Col. 7, Col. 8, Col. 9 : Same as in Col. 4, 5, and 6, respectively, but at 7\,mm. All values are for untapered data with uniform weighting.}
\label{table:vlbi_obs}
\end{table*}

The sample comprises sixteen objects (Table 1). With one exception (IC\,310),  these belong to the Bologna sample \citep{Giovannini2001, Giovannini2005, Liuzzo2009}, which implies that they have been imaged with VLBI at least once at 6\,cm. This ensured the existence of a compact core and enabled us to make detectability predictions for this experiment. One basic criterion was indeed for the sources to be bright enough for self-calibration in the fringe-fitting. Objects with sufficiently high declination have been preferred to facilitate good mutual visibility on intercontinental baselines. The sample originally included the giant radio galaxy 3C\,236, which was however not observed due to a mis-identification in the source catalog used in the observation scheduling.  In turn, IC\,310 was not originally included, but was observed by taking advantage of the pointing gaps scheduled for the largest telescopes. 

The targets are found at $z<0.1$ and host large black holes ($\log{M_{\rm BH}}\gtrsim 8.5$), which results in a spatial resolution $<10^3-10^4$\,$R_{\rm S}$ (Fig. 2) \footnote{A $\Lambda$CDM cosmology with H$_\mathrm{0}$= 71 \rm{km} \rm{s}$^{-1}$ Mpc $^{-1}$, $\Omega_\mathrm{M}=0.27$, $\Omega_{\mathrm{\Lambda}}=0.73$ \citep{2009ApJS..180..330K} is assumed.}. 
Due to the low-redshift cut, the sample is dominated by LEG. However, three HEG are also present (3C\,33, 3C\,382, and 3C\,452), and this is still a significant number when considering that Cygnus\,A was so far the only HEG imaged on similar scales.
The total radio power spans a broad interval, $21.75<\log(P_{\rm t})<26.64$ at 408\,$\rm MHz$, and the sources also present a variety of large scale radio morphologies, with nine FRI\,s, five FR\,IIs, and two compact (C), i.e., not developing extended structures with clear morphology on kilo-pc scales. The sample includes five $\gamma$-ray sources \citep[NGC\,315, 4C\,39.12, 3C\,264, IC\,310, and NGC\,4278,][]{4LAC, Cao2024}, with the latter three emitting up to TeV-energies \footnote{http://tevcat.uchicago.edu/}.
\section{Data set and analysis}
The VLBI observations were performed with High Sensitivity Array (HSA) at 1\,cm and 7\,mm (experiment code BB393). In addition to the VLBA (Very Long Baseline Array), the phased-VLA and the Effelsberg radio telescope participated in the experiment. The observations were organized in four blocks (A, B, C, D), each of 12 hours. The on-source time for each target was typically in the range of 30-45 min at 1\,cm, and 45-60 min at 7\,mm. In addition to the calibrators, compact and bright sources in the targets vicinity were observed for the VLA-phasing, as the targets themselves were in most cases inadequate for this purpose (too faint and/or resolved on VLA scales).   Data were recorded in dual polarization mode with a bandwidth of 256 MHz per polarization (two sub-bands per polarization, each with 64 channels), resulting in an aggregate bit rate of 2 Gbit/s.

The data calibration was performed in AIPS \citep[Astronomical Image Processing System,][]{1990apaa.conf..125G} through the standard procedure, including an opacity correction at both frequencies. The amplitude calibration was challenged by problems in the system temperature values recorded at the VLA, which were non-sense in parts of the experiments. These were replaced by estimates made considering the sources elevation. 
In the fringe fitting at 1\,cm, a solution interval of 1 minute was adopted for all sources except 3C\,338, for which 2 minutes allowed to recover significantly more solutions. At 7\,mm, we used a solution interval of 1 minute for all sources. 

The imaging was carried out using the software package DIFMAP \citep[DIFference MAPping,][]{DIFMAP}. The main properties of the clean maps are reported in Table 2, and the entire set of images is presented in Appendix A. 

Through the MODELFIT subroutine we have then modelled the main emission features in each source through circular Gaussian components,  which has allowed us to derive their integrated flux density $S$, radial distance $r$, position angle $pa$, and size $d$, the latter assumed to correspond to the full width at half maximum (FWHM) of the Gaussian  (Appendix B, Tables B.1-B.15). For each of these quantities, we have calculated the associated uncertainties following the approach of \cite{Lee2008}, which is based on the determination of the signal-to-noise ratio (S/N) of each emission feature. The S/N also determines the resolution limit for each component, which we have considered when estimating the brightness temperature of the VLBI cores at the two frequencies (Table B.16). A detailed description of this analysis can be found in Appendix B. We notice that this error analysis only considers the statistical image errors. Systematic errors, for instance affecting the amplitude calibration, are harder to quantify.  High frequency imaging of such faint objects could be affected by large uncertainties in the amplitudes, up to ${\sim}20-30 \%$, and in the image fidelity. However, the comparison between results obtained in the two bands indicates a good agreement. This applies both to the observed structures and to the flux density distribution, with the targets generally showing flat spectrum cores and steep spectrum jets, as expected.

\section{Results}
All sources were detected and imaged at both bands.  In Fig. 1 we present the maps obtained for three representative sources: the FR\,II-HEG 3C\,452 (left panel), showing a twin jet both at 7\,mm and 1\,cm, the FRI-LEG 3C\,31 (central panel), revealing jet limb-brightening on scales $<10^3$\,$R_{\rm S}$, and the compact-LEG NGC\,4278 (right panel), the closest and weakest of the targets, which we find to be characterized by a complex core structure and poorly collimated two-sided emission (Fig. A.9). Notes on each source are reported, together with the full set of images, in Appendix A. Below we discuss the main properties of the observed sub-parsec scale structures, highlighting the most relevant findings. Results obtained from this experiment for the brightest target, NGC\,315, were already published by  \cite{Boccardi2021, Ricci2022}. In the following, we include this source in the discussion of the sample properties, but we refer to these papers for the images and detailed analysis of the source.

\begin{figure*}[!ht]
    \centering
\includegraphics[height=4.6cm]{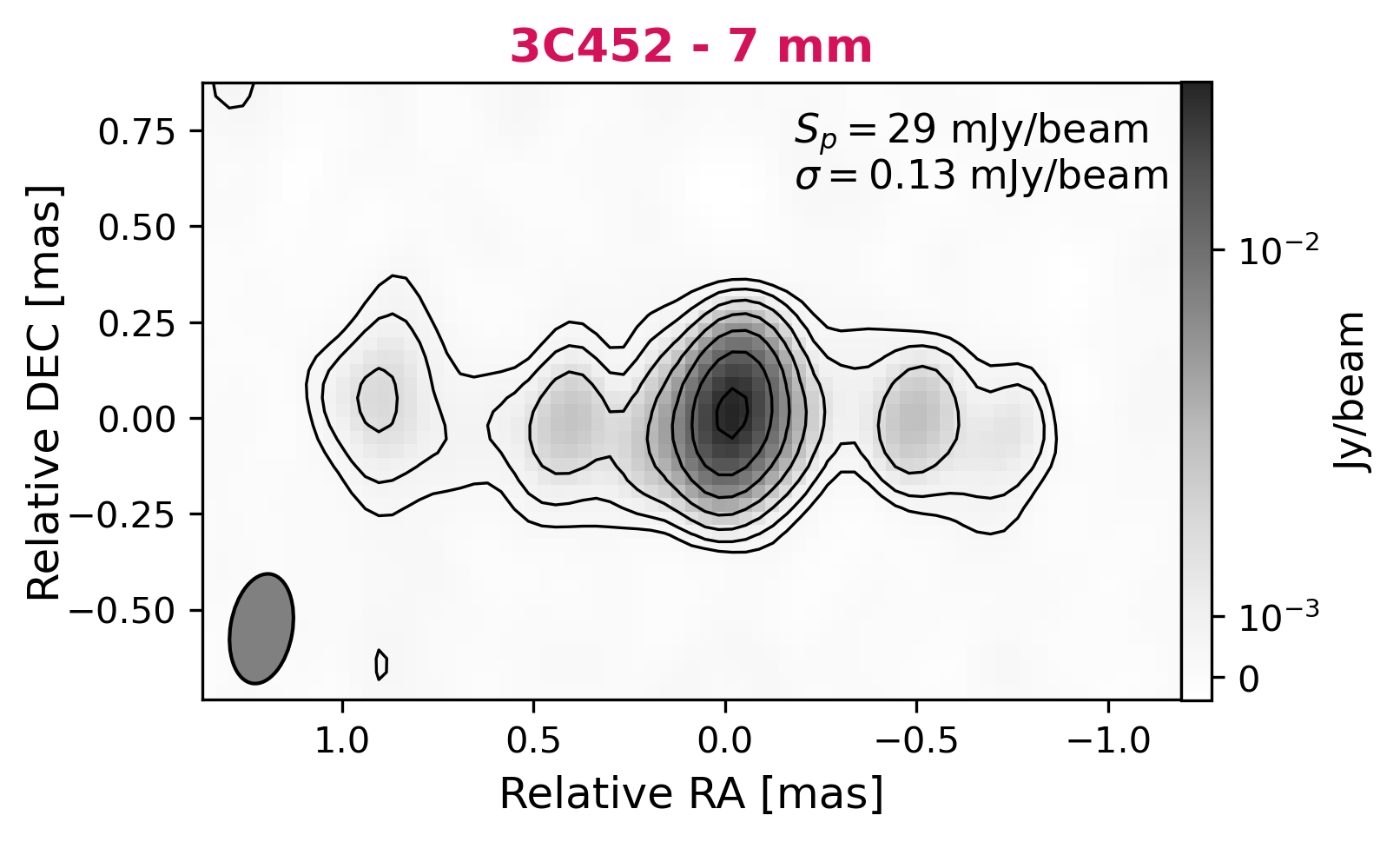}
\hspace{-0.15cm}  \includegraphics[height=6.4cm]{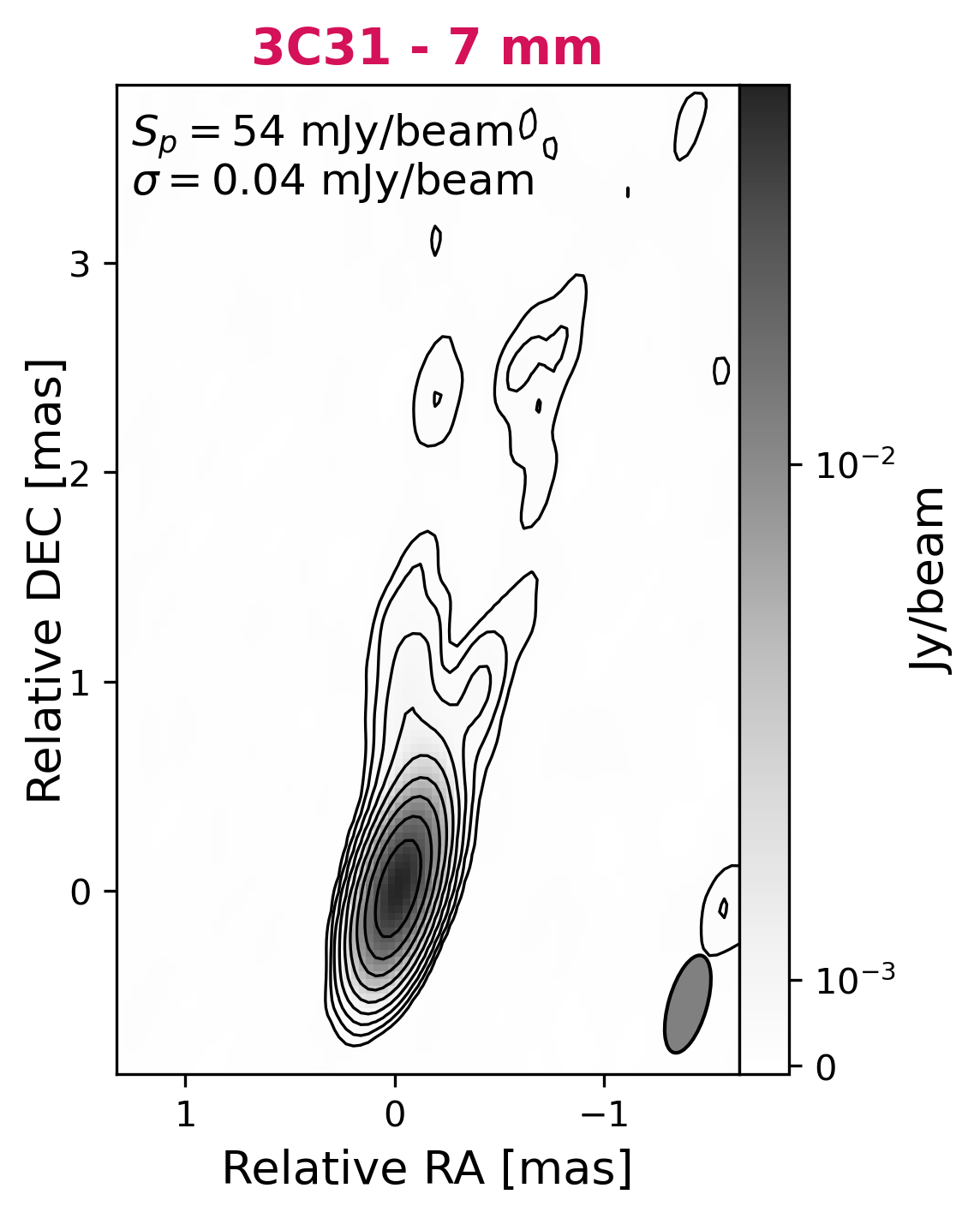}
\hspace{-0.15cm}\includegraphics[height=6.4cm]{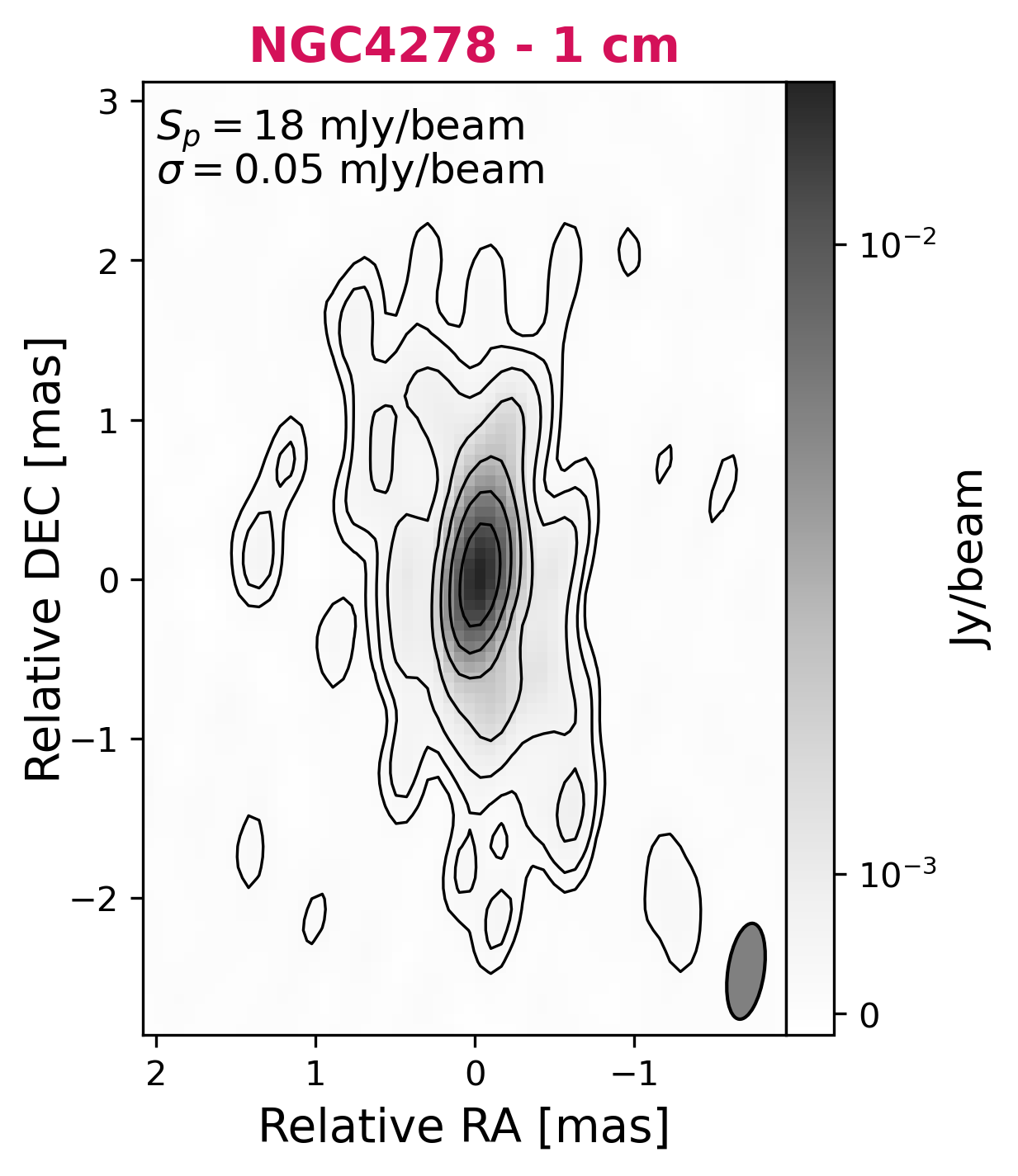}
  \caption{Left: 7\,mm HSA image of 3C\,452 (FRII-HEG). The beam size and position angle are $0.29\times0.16$\,$\rm mas$, $-10^{\circ}$. Center: 7\,mm HSA image of 3C\,31 (FRI-LEG). The beam size and position angle are $0.48\times0.18$\,$\rm mas$, $-16^{\circ}$. Right: 1\,cm HSA image of NGC\,4278 (compact-LEG). The beam size and position angle are $0.61\times0.23$\,$\rm mas$, $-8^{\circ}$.  Contours start at -3$\sigma$ and increase by a factor of two until allowed by the  peak intensity. These maps 
were produced with uniform weighting in the case of 3C\.452 and NGC\,4278, and with natural weighting in the case of 3C\,31, which better highlights the faint limb-brightened structure of this jet. The full set of images is shown in Appendix A.}
    \label{}
\end{figure*}

\subsection{Flux density and spatial resolution}
The total flux densities span an interval 17\,mJy$\,<S_{\rm tot}<$\,447\,mJy at 1\,cm and 19\,mJy\,$<S_{\rm tot}<$\,268\,mJy at 7\.mm (Table 2). Assuming the BH masses in Table 1, the maximum angular resolution achieved at 7\.mm, ${\sim}0.1$\,$\rm mas$, translates into a spatial resolution in the range  85-2034\,$R_{\rm S}$. 

In Fig. 2, we display the ranges of flux densities and resolutions in the sample, as well as those of better known, previously studied objects. This shows that, while this is clearly a fainter population, the number of sources imaged on scales $\lesssim 10^3$\,$R_{\rm S}$ is more than doubled by the present study.  Given the very large mass estimated by \cite{North2019}, the highest spatial resolution is achieved for 3C\,31, followed by NGC\,4278 and NGC\,315, which all classified as LEG. The lowest resolution is instead obtained, as expected, for the powerful HEG 3C\,33 and 3C\,452, which are found at higher redshift. These are, nevertheless, still good candidates for probing the jet acceleration and collimation region and, due to the high level of symmetry between jet and counter-jet (Sect. 4.3), they are excellent targets for testing the AGN unified schemes.  

 We note that the spatial resolution values in Fig. 2 scale linearly with the BH mass, and mass estimates can often vary significantly, by a factor of two or more \citep[e.g.,][]{Barth2016, Gravity2018}, depending on the observational method. The black hole masses in Table 1 were selected based on a careful literature search. In the presence of multiple studies for a given source, we have preferred estimates from methods which are considered more accurate, such as reverberation mapping in the case of 3C\,382 \citep{Fausnaugh2017}, or resolved kinematics of cold gas in the case of 3C\,31 \citep{North2019}. Nevertheless, by considering a ``worst case scenario'' in which the masses are smaller by one order of magnitude, the spatial resolution range for this sample would increase by the same factor.  A resolution ${\lesssim}10^4$\,$R_{\rm S}$ would still be sufficient, in most cases, to probe the jet acceleration and collimation region.

\begin{figure*}[!ht]
    \centering
    \includegraphics[width=0.65\textwidth]{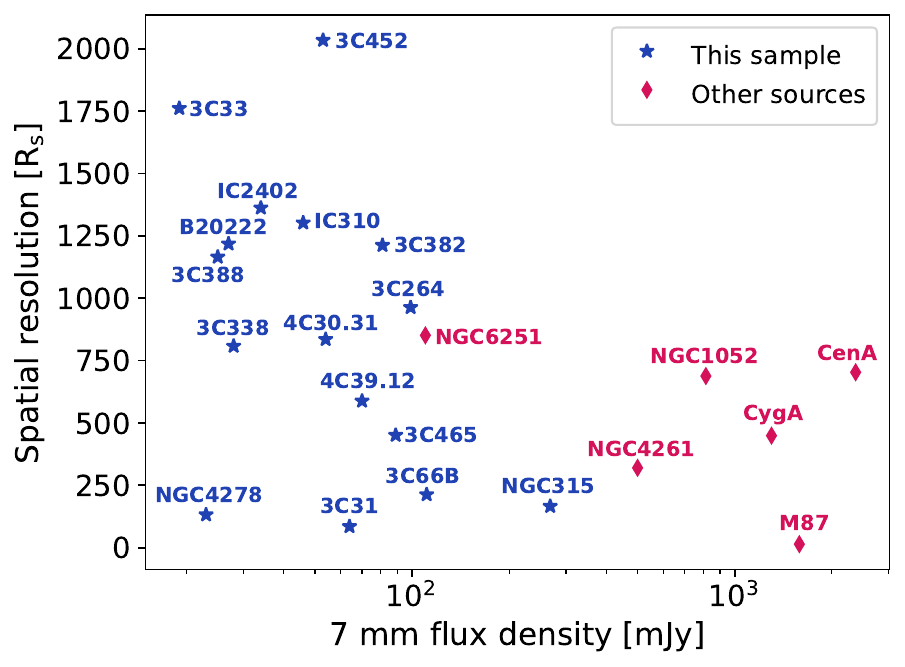}
    \caption{Spatial resolution in $R_{\rm S}$ for an angular resolution of $0.1$\,$\rm mas$ versus flux density at 7\,mm. In blue we plot the sources in our sample, in red well known radio galaxies previously studied on these scales \citep{Boccardi2021}. For Cyg\,A, NGC\,1052, and M\,87 we assume the average 7\,mm flux densities from \cite{Boccardi2016a}, \cite{Baczko2019}, \cite{Walker2018}, respectively. For NGC\,6251, we refer to \cite{Cheng2020}. For Cen\,A, no 7\,mm VLBI flux density is reported in the literature, and we estimate it from the 1\,cm one \citep{Mueller2011} assuming $\alpha=-0.5$.}
    \label{}
\end{figure*}

\subsection{Core brightness temperature}

We estimate the core brightness temperature $T_{\rm B}^{\rm c} [\rm K]$ of each source at the two frequencies by taking into account the parameters of the core MODELFIT components. The observed brightness temperature $T_{\rm B}^{\rm obs}$, which differs from the intrinsic one by the Doppler factor $\delta$ ($T_{\rm B}^{\rm obs}=\delta T_{\rm B}^{\rm intr}$), is expressed as a function of the 
component flux density $S$ [\rm Jansky], size $d$ [\rm mas], redshift $z$, and frequency $\nu_{\rm obs}$ [\rm GHz] as:

\begin{equation}
T_{\rm B}^{\rm obs}=\frac{1.22\times 10^{12}\cdot S(1+z)}{d^2\nu_{\rm obs}^2}
\end{equation}

 The obtained values are reported in Table B.16 and displayed in the histogram in Fig. 3, where we overlay the distributions observed at 1\,cm and 7\,mm. The 7\,mm histogram does not include the value obtained for 3C\,33, due to the fact the core size was found to be smaller than the resolution limit. In this single case we estimate a lower limit for the brightness temperature,  $T_{\rm B}^{\rm c}\geq1.5\times10^{10}$\,$\rm K$.

 With respect to results typically observed in blazars, which have core temperatures often exceeding the Inverse Compton limit \citep[$T_{\rm B}\sim 5\times 10^{11}$\,$\rm K$,][]{1969ApJ...155L..71K}, our sources are characterized by lower values, with few exceptions well below this limit and even below the equipartition value $T_{\rm B}\sim 5\times 10^{10}$\,$\rm K$. This is, approximately, the temperature expected to signal equipartition conditions between the particles and the magnetic energy densities \citep{Readhead1994}, and the one typically estimated in blazars in the quiet state after correcting for Doppler boosting \citep{Homan2021}. There are several, possibly concurrent effects, which can explain our low values. 1) Due to the fact that we are probing the jet base at short distance from the BH, equipartition conditions might not have been reached yet, i.e., the jet is still strongly magnetized on these scales. This may be supported by the fact that, in most cases (in  13/16 sources), we measure lower $T_{\rm B}$ values at 7\,mm with respect to 1\,cm, i.e., at 7\,mm we are probing regions even closer to the central engine, and therefore more strongly magnetized, than at 1\,cm. 2) Due to the larger viewing angle of these jets, the Doppler factors can be smaller than 1 also in the approaching side for sufficiently high intrinsic speeds (see Sect. 4.4), which will result in $T_{\rm B}^{\rm obs}<T_{\rm B}^{\rm intr}$. 3) Finally, it is possible that some of these cores have become fully optically thin at the observed frequencies. This could be especially the case for the lowest power sources, where less extreme physical conditions at the jet base could induce milder opacity effects.

\subsection{Jets morphology and orientation}
Except for 3C\,388 and 3C\,33, which appear point-like at 7\,mm, extended jet emission is detected in all cases, with  half the objects (8/16) showing two-sided structures. 
For them, in Table 3 we report the jet-to-counter-jet intensity ratio $R_{\rm J/CJ}$, which we use to estimate some jet parameters. 

 We perform this analysis at 1\,cm, since a larger number of counter-jets is detected in this band. We define a source as ``two-sided" if at least two Gaussian features are fitted in the region upstream of the core. 
The approaching jet flux density is then calculated as the sum of the integrated flux density of all components in the brightest side, including the core, and the receding jet flux density as the sum of the features in the opposite side.
Objects like 3C\,66B and 3C\,264, which present a single upstream feature in the proximity of the core, are of uncertain classification. Indeed, such features could still be part of the approaching jet, marking optically thick regions between the black hole and the VLBI core. Core shift and spectral studies will be necessary to clarify their nature. The upstream emission in NGC\,315 was also modeled with a single component in this specific epoch, thus we do not classify the source as 'two-sided' based on these images. However, for this target 
we have performed detailed alignment and spectral analysis considering a broad multi-wavelength data set, which clearly confirmed the presence of a counter-jet \citep{Boccardi2021,Ricci2022}.

By considering that $R_{\rm J/CJ}$ depends on the viewing angle $\theta$, the intrinsic speed $\beta$ and the spectral index $\alpha$ as

\begin{equation}
 R_{\rm J/CJ}=\frac{(1+\beta\cos\theta)} {(1-\beta\cos\theta)}^{2-\alpha}, 
 \end{equation}

we can use this observable for constraining $\beta$ and $\theta$  (we assume $\alpha=-0.7$, with $S_\nu\propto\nu^{\alpha}$) under the hypothesis of intrinsic symmetric  and constant speed. In particular, we can derive a lower limit for the speed $\beta_{\rm min}$ by setting $\theta=0$, and an upper limit for the angle $\theta_{\rm max}$ by setting $\beta=1$. We can also derive exact values for $\theta$ by making an assumption on the Lorentz factor $\Gamma$. Relatively low jet speeds are inferred in kinematic studies of radio galaxies, with $\Gamma$ typically in the range ${\sim}1-2$ \cite[e.g.,][]{Lister2019}. By assuming $\Gamma\sim1.4$, corresponding to $\beta=0.7$\,$\rm c$, as an average value for this sample we obtain  reasonable estimates for the jet viewing angle of the two-sided sources, which is in the range $25^{\circ}-81^{\circ}$  (Table 3). However, the true $\theta$ might deviate significantly from these estimates in few cases. The low $R_{\rm J/CJ}$ observed in the lowest power  object NGC\,4278 is also compatible with a much lower jet speed  ($\beta_{\rm min}{\sim}0.1$\,$\rm c$) and viewing angle. For this source, a smaller jet viewing angle would be consistent with its recent detection at TeV energies by the Large High Altitude Air Shower Observatory \citep{Cao2024}, as well as with previous cm-VLBI results \citep{Giroletti2005}. Moreover, it would explain the rather complex morphology of the diffuse two-sided emission we have imaged in this source  (Fig. 1 and Appendix A.9), which could be ascribed to projection effects. 

 As mentioned, Eq. 2 is strictly valid only in the hypothesis of intrinsically symmetric emission from two flows propagating at constant speed. Since we may be imaging the acceleration region of these jets, $R_{\rm J/CJ}$ is likely to increase with distance from the BH, due to the increasing Lorentz factor. This effect was indeed observed in NGC\,315 based on these data \citep{Ricci2022}. Asymmetries between the approaching and receding sides, either intrinsic \citep[e.g.,][]{Baczko2019} or induced by external absorbers \citep[e.g.,][]{Haga2015}, are also possible. A detailed analysis of the variation of $R_{\rm J/CJ}$ with distance can therefore provide a unique insights on these effect, particularly if combined with spectral information. We are currently following this approach for the sources in Table 3, in a work which will be subject of a future publication.

\begin{figure}[!ht]
    \centering
    \includegraphics[width=0.47\textwidth]{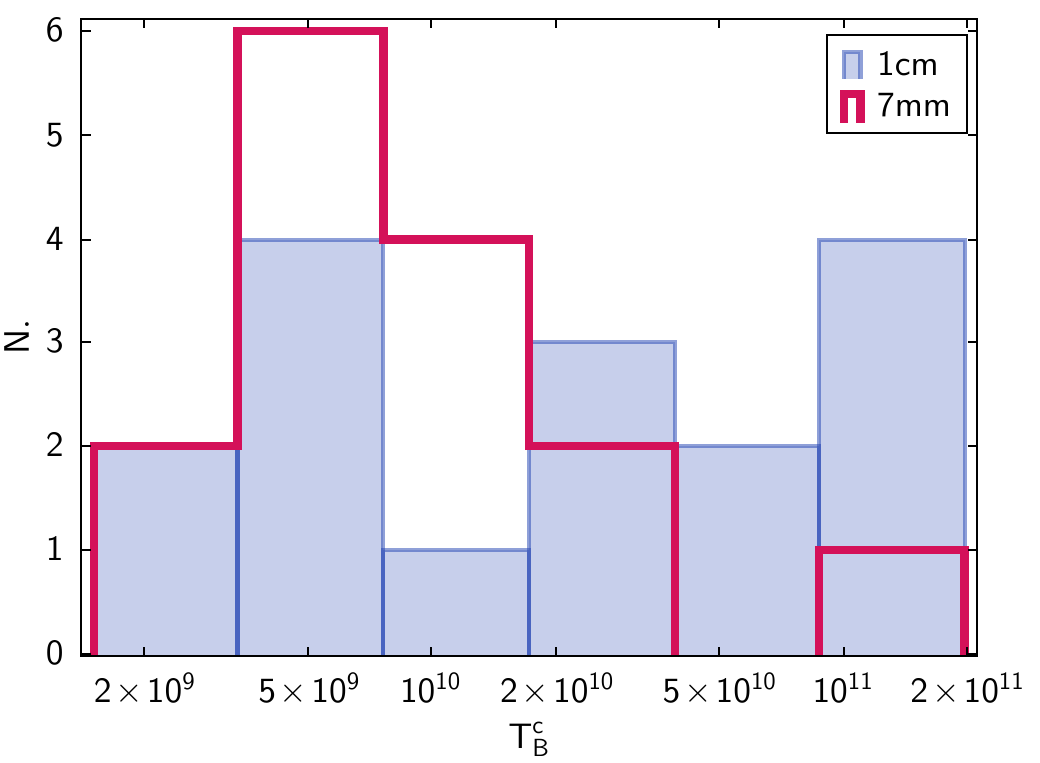}
    \caption{Distributions of the observed core brightness temperature at 1\,cm and 7\,mm. The values on the x-axis are displayed in logarithmic scale. In linear scale, the six bins span the ranges 0.150-0.339, 0.339-0.763, 0.763-1.720, 1.720-3.878, 3.878-8.741, 8.741-19.703 in units of $10^{10}\,\rm K$.}
    \label{}
\end{figure}

\subsection{Jets internal structure}
 As discussed in Sect. 4.3,  a moderate Lorentz factor seem adequate to explain the properties of the observed radio emission in our galaxies (with the possible exception of NGC\,4278, which could require even lower speeds). However, the true maximum $\Gamma$ in most of these jets can be expected to be much higher. Indeed, the measurement of low speeds in misaligned sources is likely due to the presence of transverse velocity gradients in the jet, with the fast central spine becoming faint for larger angles due to Doppler de-boosting.  The presence of this gradient may manifest in the limb-brightening of the jet emission, previously observed in several LEG \citep[][and references therein]{Boccardi2017}. Based on our images, sources showing hints of this feature,  most often in the innermost 7\,mm jet, are 3C\,31 (Fig. 1), 3C\,66B,  4C\,39.12, 3C\,264, and 3C\,465 (Appendix A.1, A.3, A.6, A.8, A.15, respectively), which all belong to the LEG class.  NGC\,315, which is also a LEG, did not clearly showed limb-brightening in our images, but this was observed in more recent data by \cite{Park2024}. Before concluding that limb-brightening only characterises jets in LEG, we should however notice that these are also the best resolved in the transverse direction due to their proximity with respect to HEG. Indeed, the only HEG jet which is well resolved transversely, the jet in Cygnus A, is also limb-brightened \citep{Boccardi2016a}. Future, higher resolution VLBI studies of jets in HEG will help clarifying this aspect.

\section{Prospects for future observations}
Through this work we have identified new interesting targets for jet formation studies, several of which are bright enough to be examined in detail by performing dedicated experiments using current VLBI arrays.
3\,mm-imaging with the upgraded global mm-VLBI array \citep[GMVA, e.g.,][]{Kim2023} including the phased-ALMA (Atacama Large Millimeter Array) is also expected to be feasible for mm-flux densities of few tens of mJy.
Besides NGC\,315, which has already been subject of several studies \citep[see also][]{2021ApJ...909...76P, Park2024}, the targets which best combine a superb spatial resolution ($<500$\,$R_{\rm S}$) with a relatively high flux density ($>50$\,$\rm mJy$) at 7\,mm are 3C\,31, 3C\,66B, 3C\,465, which all belong to the LEG class. While fainter, the LEG NGC4278 is also extremely well resolved, providing a unique view into the base of a low-power, poorly collimated jet on milli-parsec scales. Among HEG, 3C\,33 and 3C\,452 show highly symmetric two-sided jets, and 3C\,452 is also rather bright ($53$\,$\rm mJy$ at 7\,mm). This makes it a highly interesting target to investigate the nuclear regions of a powerful source.  In the longer term, we expect most of these objects to become prime targets of next-generation arrays.  In the mm-band, the ngVLA \citep{Murphy2018} will be similar to the GMVA in terms of angular resolution, but its sensitivity will be improved by orders-of-magnitude.  It will be sensitive to thermal emission as well, which will allow us to probe for the first time the full outflow stratification properties in these objects. Several of our targets are also of interest for future observations with the ngEHT \citep{Johnson2023, Rama2023}, which will aim at probing the black hole shadow in other extragalactic objects besides M\,87.

\begin{table}[!ht]
\centering
\caption{Two-sided sources and estimated jet parameters.}
\begin{tabular}{c|cccccc}
\hline
Name        &   $R_{\rm J/CJ}$ &  $\beta_{\rm min}$ & $\theta_{\rm max}$& $\theta_{\rm \beta=0.7}$  \\
             &&  [\rm{c}] &[$\rm deg]$ &[$\rm deg]$  \\ \hline 
3C\,31      &$55.8\pm16.1$&$0.63\pm0.03$&$50.8\pm2.4$&  25\\   
3C\,33      &$6.5\pm1.5$&$0.33\pm0.04$&$70.6\pm2.3$& 62\\ 
B2\,0222+36 &$4.2\pm0.5$&$0.26\pm0.02$&$75.0\pm1.2$ & 68 \\ 
IC\,2402    &$32.2\pm10.2$&$0.57\pm0.04$&$55.5\pm2.8$& 36\\         
NGC\,4278  &$1.9\pm0.2$&$0.11\pm0.02$&$83.5\pm1.4$ & 81  \\ 
3C\,338     &$6.4\pm1.6$&$0.33\pm0.04$&$70.7\pm2.5$ & 62  \\ 
3C\,452     &$2.9\pm0.3$&$0.20\pm0.02$&$78.7\pm1.0$ &  74  \\ 
3C\,465    &$6.3\pm0.7$&$0.33\pm0.02$&$70.8\pm1.0$ & 62\\ \hline
\end{tabular}
\tablefoot{Col. 1: Source name. Col. 2: Jet-to-counter-jet intensity ratio at 1\,cm. Col. 3: Lower limit for the jet intrinsic speed. Col. 4: Upper limit for the jet viewing angle. Column 5: Jet viewing angle assuming an intrinsic speed of $0.7$\,$\rm c$ (or $\Gamma{\sim}1.4)$.}
 \end{table}

 \begin{acknowledgements} 
The authors thank the anonymous referee for providing valuable feedback.
The authors thank DaeWon Kim for reading the article. BB acknowledges the financial support of a Otto Hahn research group from the Max Planck Society. This research has been partially funded by the Deutsche Forschungsgemeinschaft (DFG, German Research Foundation) – project number 443220636 (DFG research unit FOR 5195: "Relativistic Jets in Active Galaxies").
The VLBA is a facility of the National Science Foundation under cooperative agreement by Associated Universities, Inc.  
 \end{acknowledgements}

\bibliographystyle{aa}
\bibliography{reference.bib}
\clearpage
\begin{appendix}
\onecolumn
    \section{Images and notes on individual sources} 
    In the following, we present the VLBI images at 1\,cm (22\,GHz) and 7\,mm (43\,GHz) for the sources in Table 1. The maps obtained for NGC\,315 were already presented and analyzed by \cite{Boccardi2021, Ricci2022}, so they are not shown here. We report all the maps produced with uniform weighting (with { \it uvweight 2, -1}), whose properties are also summarized in Table 2, as well as those produced with natural weighting (with {\it uvweight 0, -1} or {\it uvweight 0, -2}), in case of the presence of extended jet emission which becomes better represented in this weighting scheme. Negative contours mark isophotes at -3 times the noise $\sigma$ in each map, while positive contours start at $3\sigma$ and increase by a factor of two until allowed by the peak intensity.

   \subsection{3C\,31}
  Considering the very large black hole mass of $4.2\times10^9$\,$M_{\odot}$, estimated by \citep{North2019}, the FRI-LEG 3C\,31 is the best resolved source in our sample, with an angular resolution of $0.1$\,$\rm mas$ corresponding to only  $85$\,$R_{\rm S}$. At both frequencies we detect a narrow approaching jet in the north-west direction, showing some position angle changes along its propagation and signs of limb-brightening (Fig. A.1). The emission from a weak counter-jet is hinted in all maps, but we were able to describe it in the MODELFIT  only at 1\,cm. 
           \begin{figure*}[!h]
        \centering
       \includegraphics[height=5.8cm]{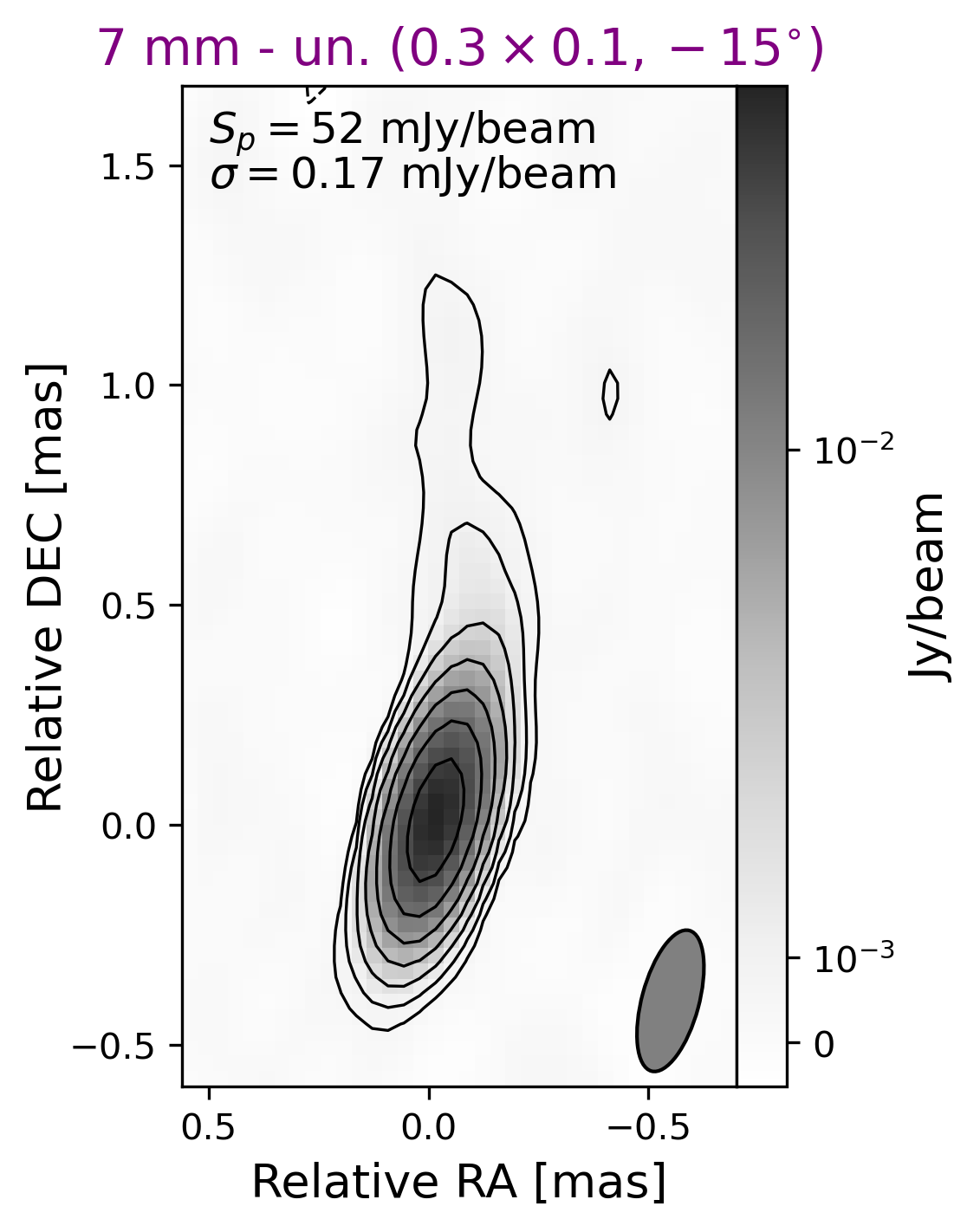}
        \includegraphics[height=5.8cm]{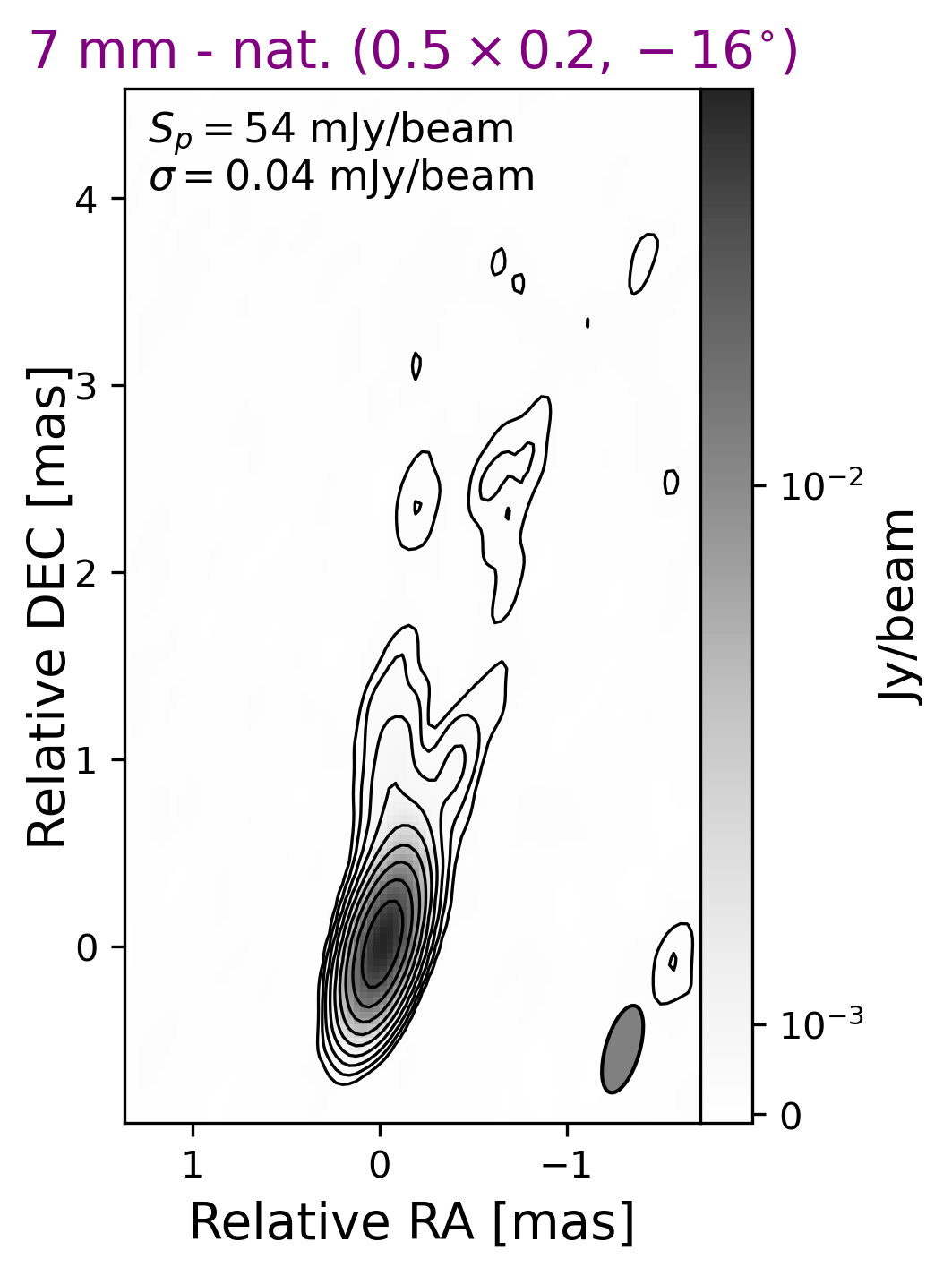}
        \includegraphics[height=5.8cm]{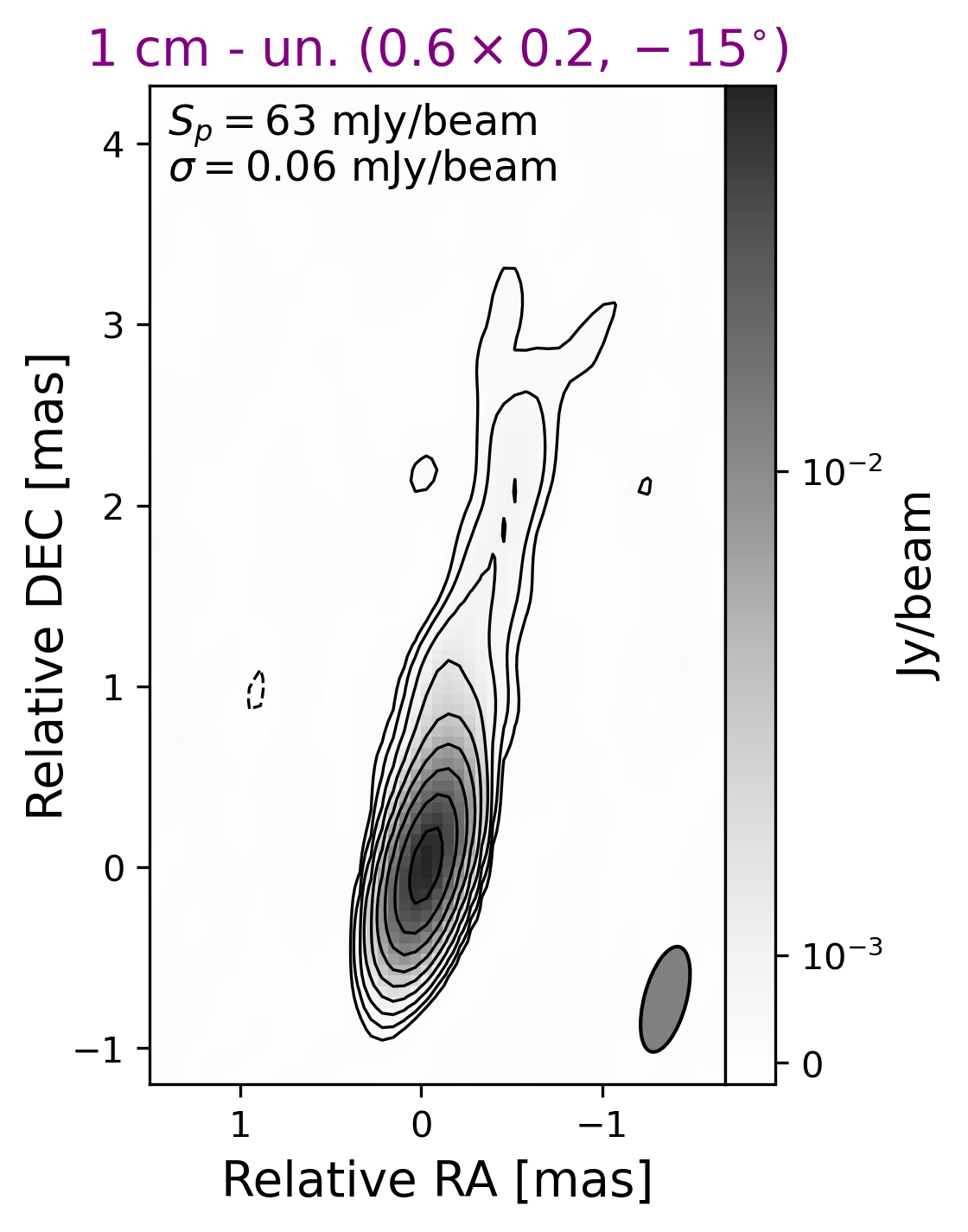}
        \includegraphics[height=5.8cm]{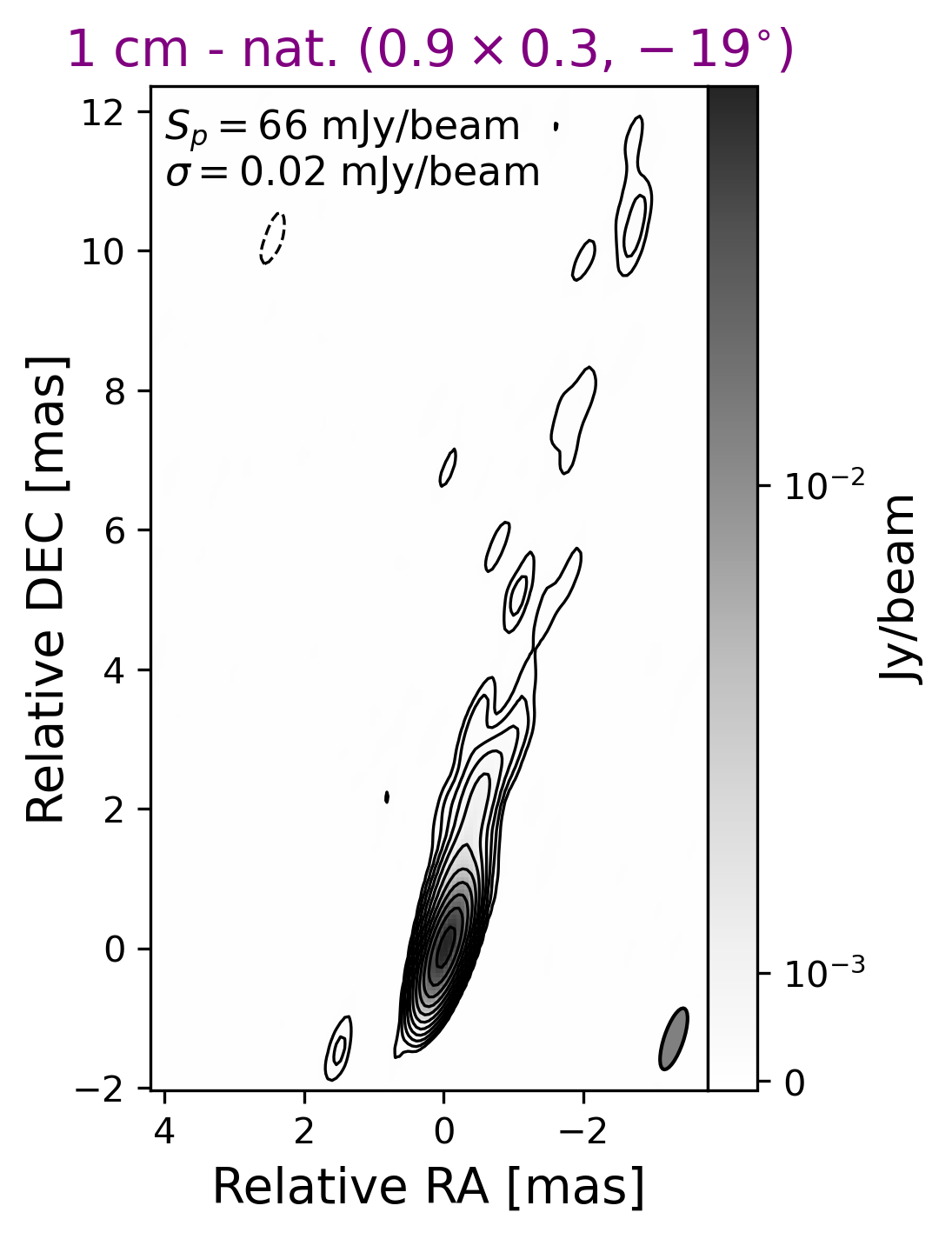}
  \caption{VLBI images of 3C\,31. From left to right: 7\,mm images with uniform and natural weighting, 1\,cm images with uniform and natural weighting.}
        \label{fig:enter-label}
    \end{figure*}  
    
            \subsection{3C\,33}
  3C33 is one the three HEGs in our sample, and it develops a classic FRII morphology on kilo-parsec scales. It is also the faintest target, with a flux density $\lesssim20$\,$\rm mJy$ in both bands. At 7\,mm the source appears point-like, while at 1\,cm we detect a highly symmetric two-sided jet propagating in the north-east/south-west direction (Fig, A.2), in agreement with the 6\,cm image presented by \cite{Giovannini2001}. The strong symmetry indicates an orientation close to the plane of the sky, consistent with the classification of the source as a narrow-line radio galaxy, i.e., having an obscured nucleus \citep[e.g.,][]{Buttiglione2010}.

  \begin{figure*}[!h]
        \centering
        \includegraphics[height=6.5cm]{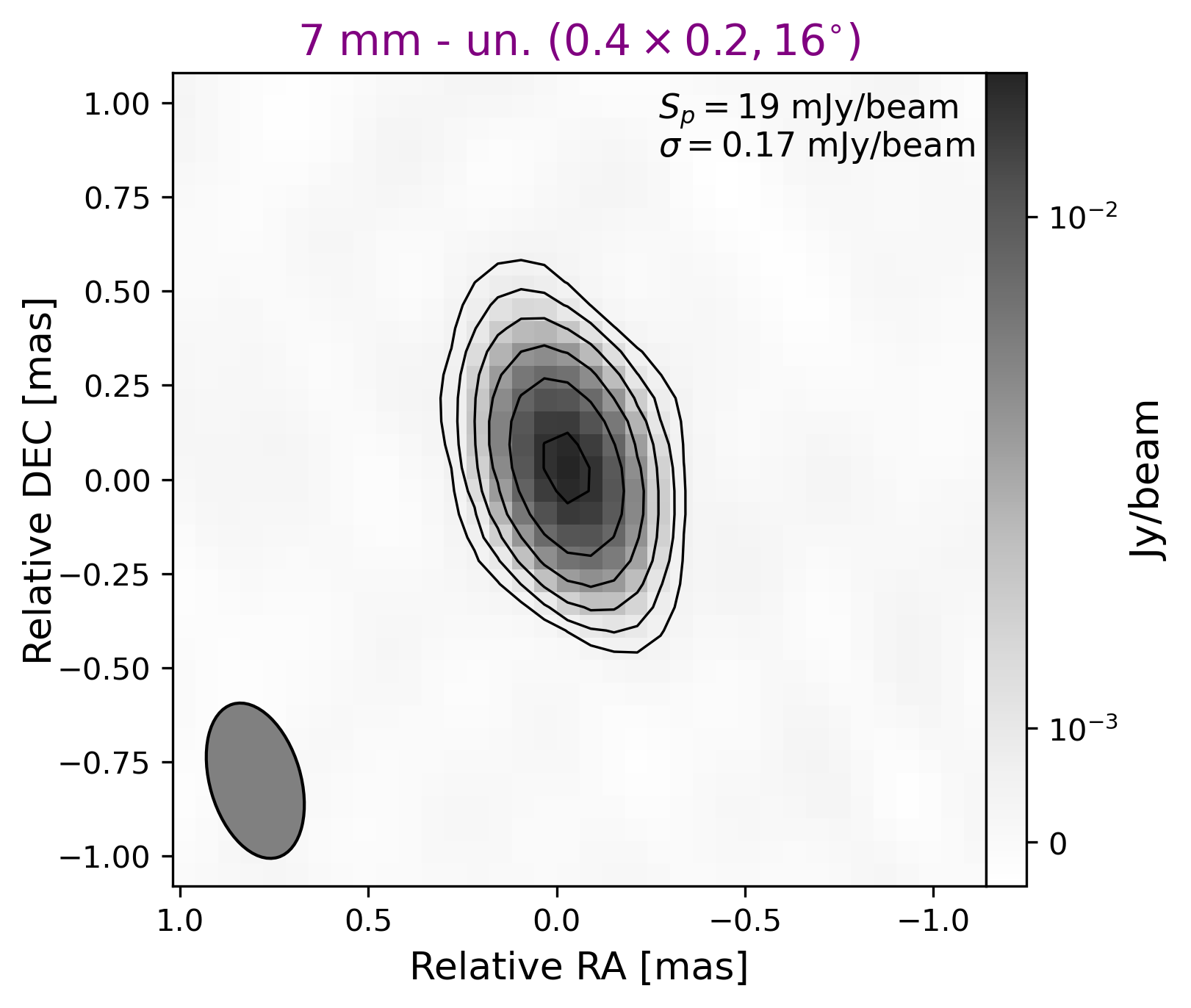}
        \includegraphics[height=6.5cm]{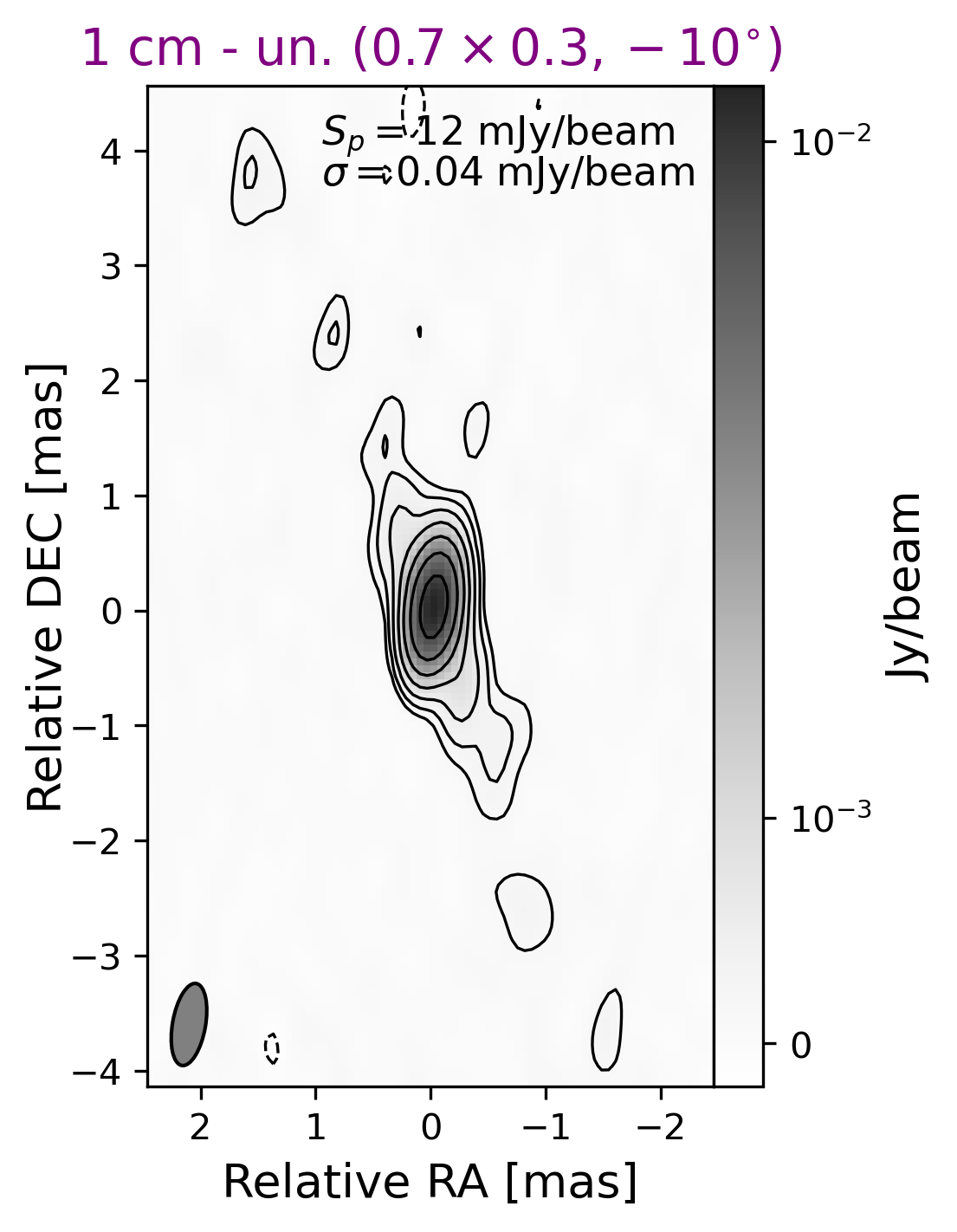}
        \includegraphics[height=6.5cm]{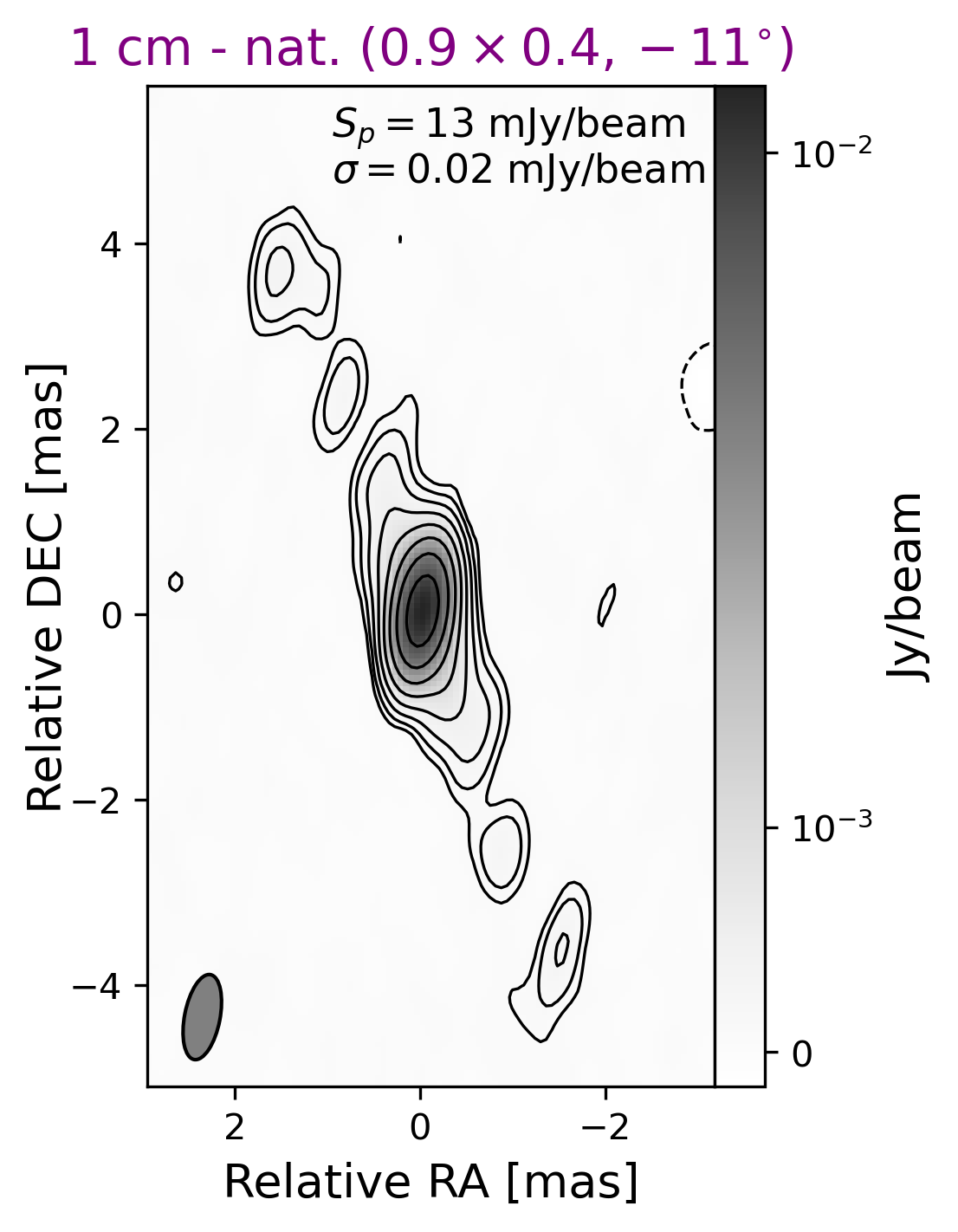}
        \caption{VLBI images of 3C\,33.  Left: 7\,mm image with uniform weighting. Center: 1\,cm image with uniform weighting. Right: 1\,cm image with natural weighting.}
        \label{fig:enter-label}
    \end{figure*}    
    
\clearpage

     \subsection{3C\,66B}
     3C\,66B is another FRI-LEG where the jet base can be imaged with very high spatial resolution, since an angular resolution of $0.1$\,$\rm mas$ corresponds to only $213$\,$R_{\rm S}$. It is an interesting target also because of the putative existence of a binary supermassive black hole at its center, proposed by \cite{2003Sci...300.1263S}. We detect an approaching jet propagating towards north-east at both bands, and some emission upstream of the core at 1\,cm only (Fig. A.3). Further high resolution observations will be needed to confirm the presence of a counter-jet in this source. Hints of limb-brightening are observed in the innermost jet at 7\,mm. 
     \begin{figure*}[!h]
        \centering
       \includegraphics[height=4.7cm]{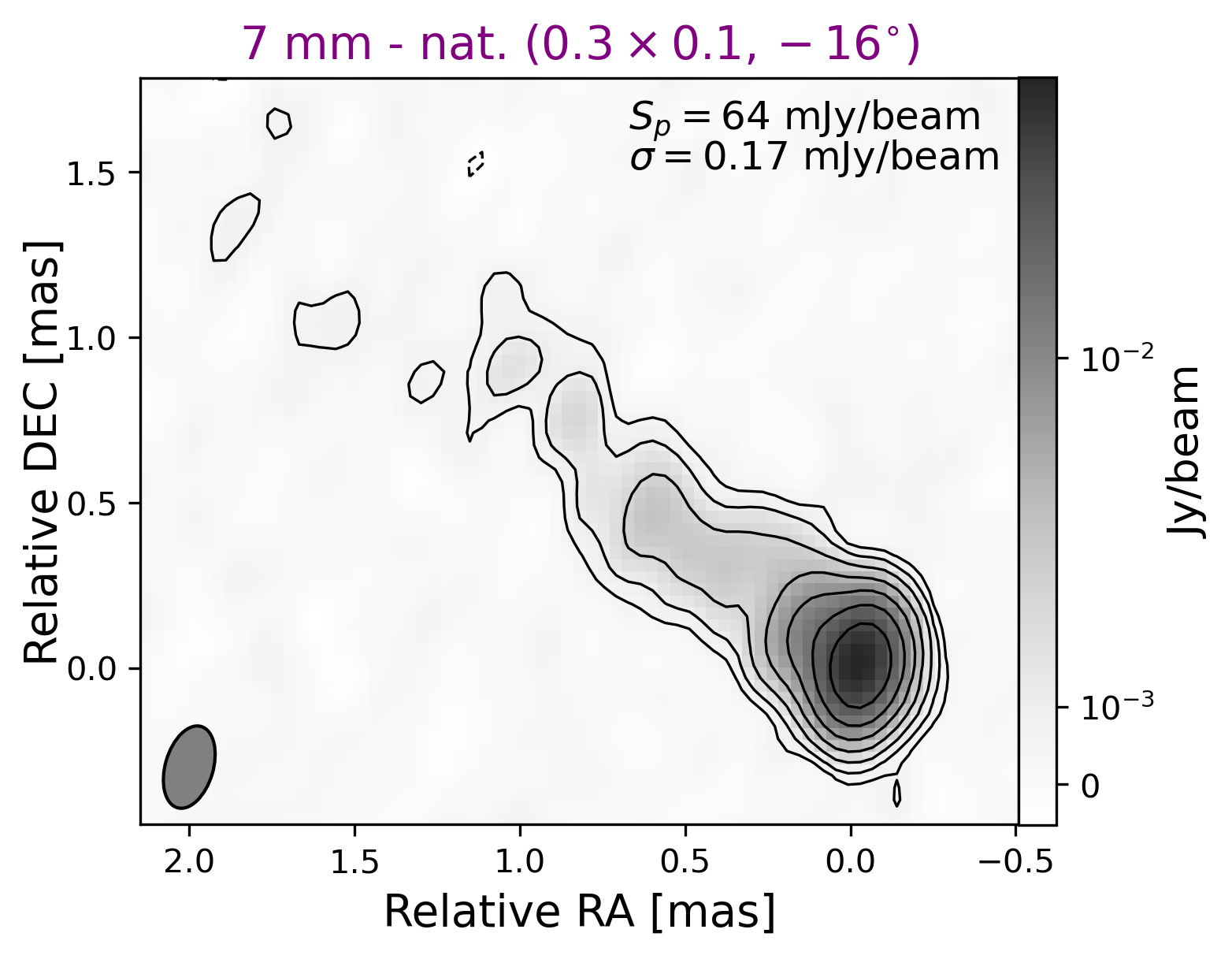}
        \includegraphics[height=4.7cm]{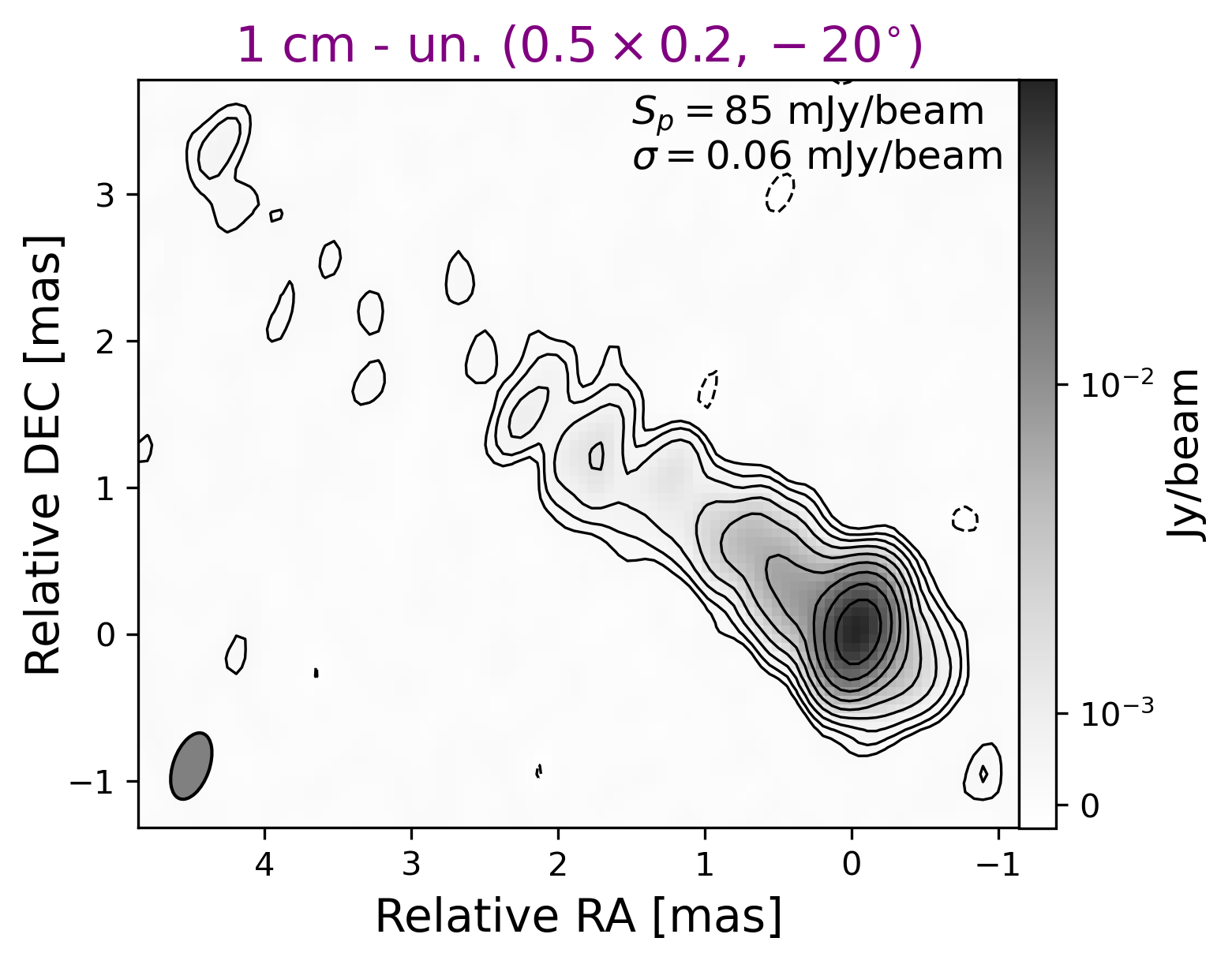}
        \includegraphics[height=4.7cm]{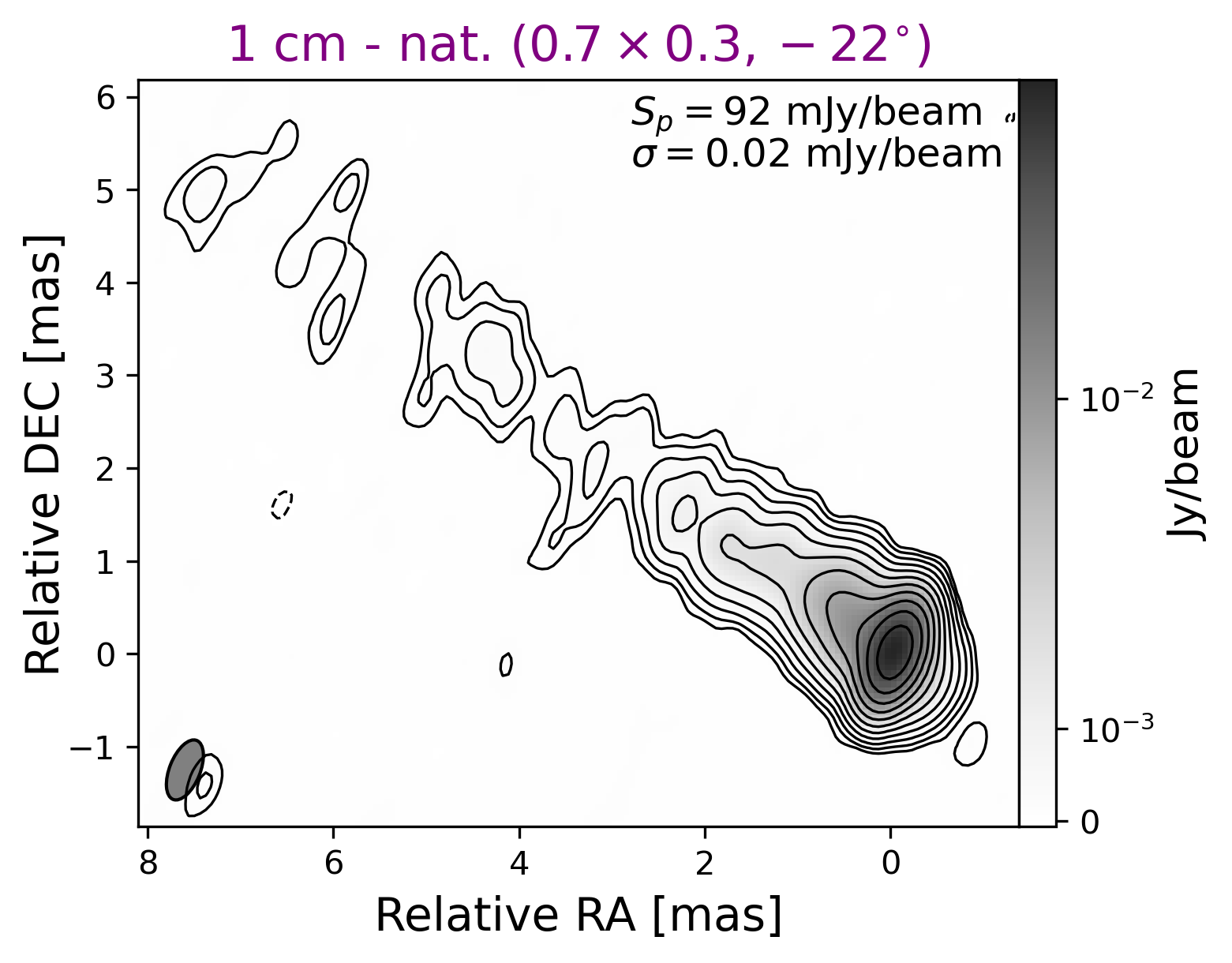}
        \caption{VLBI images of 3C\,66B.  Left: 7\,mm image with uniform weighting. Center: 1\,cm image with uniform weighting. Right: 1\,cm image with natural weighting.}
        \label{fig:enter-label}
    \end{figure*}

        \subsection{B2\,0222+36}
          B2\,0222+36 is a LEG showing a compact structure on kilo-parsec scales. High frequency VLA observations reveal faint two-sided emission, immersed in a ${\sim}10$\,$\rm kpc$ halo observed at lower frequencies \citep{Giroletti2005b}. On sub-parsec scales, we observe at both bands an interesting two-sided structure  (Fig. A.4) with a low jet to counter-jet ratio ($R_{\rm J/CJ}{\sim}4$), in agreement with previous 1.6 GHz VLBI results from \citep{Giroletti2005b}. The strong symmetry may not necessarily indicate a large viewing angle, but also an intrinsically low speed. 
          \begin{figure*}[!h]
        \centering
        \includegraphics[height=6.9cm]{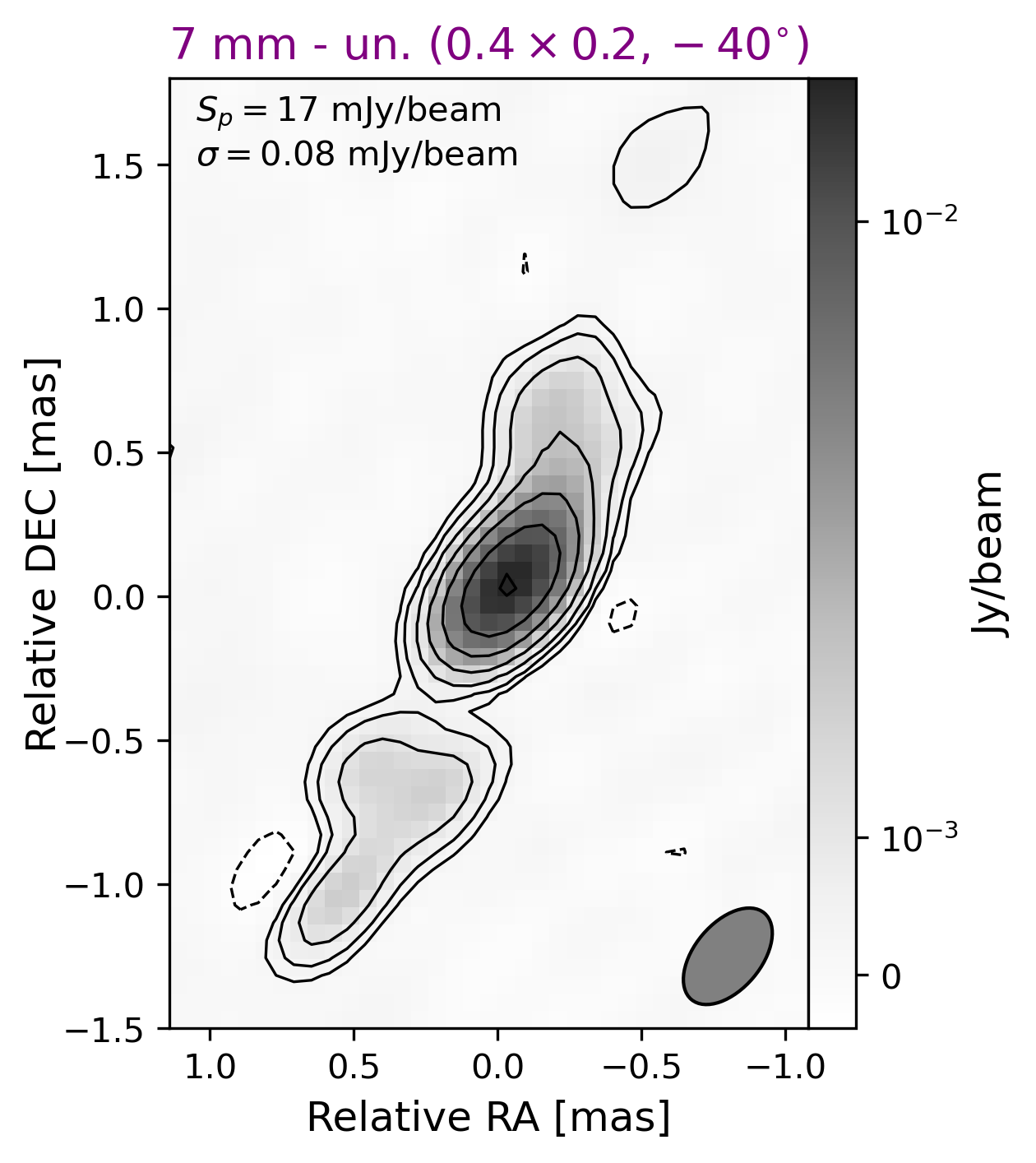}
        \includegraphics[height=6.9cm]{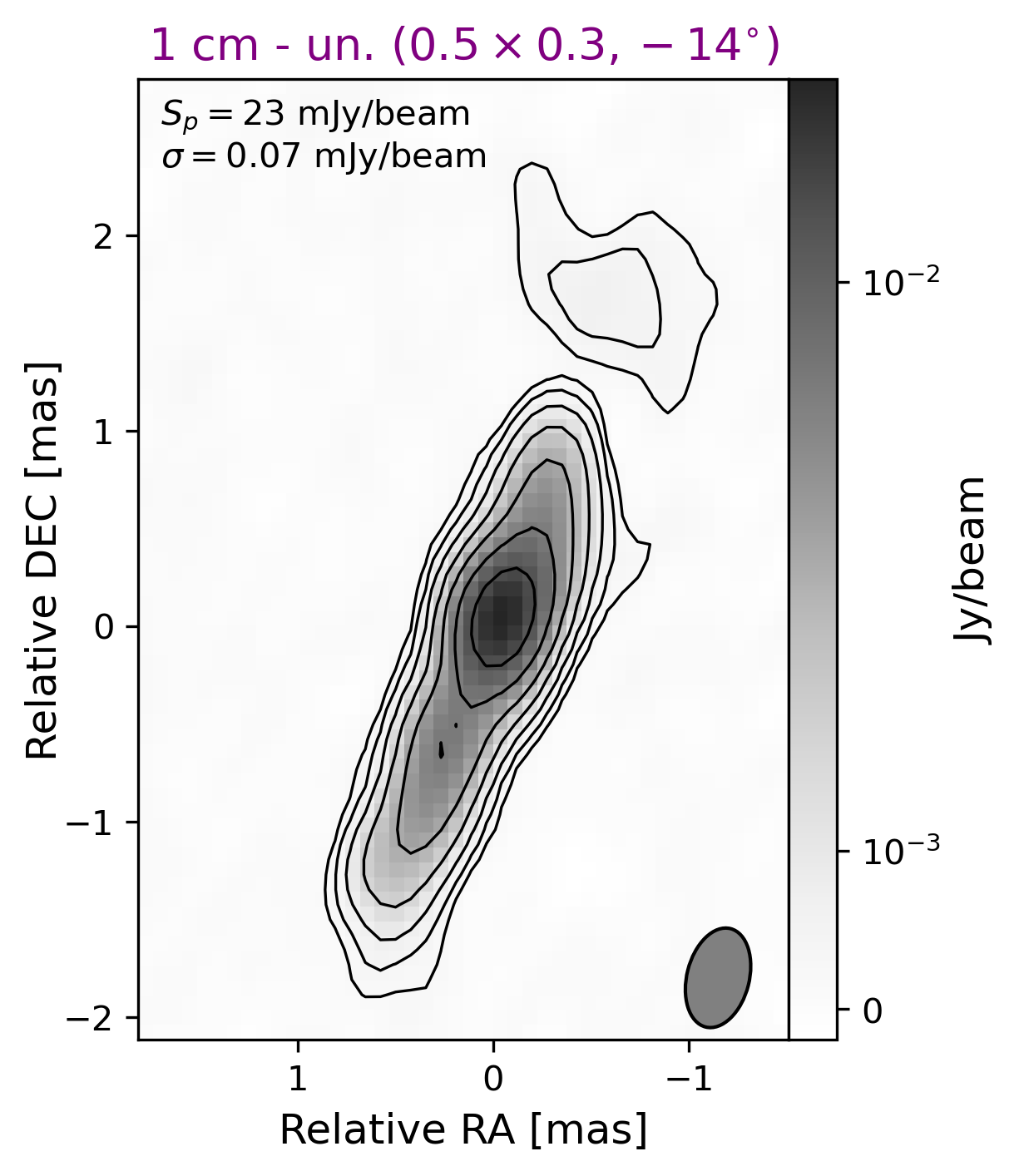}
        \includegraphics[height=6.9cm]{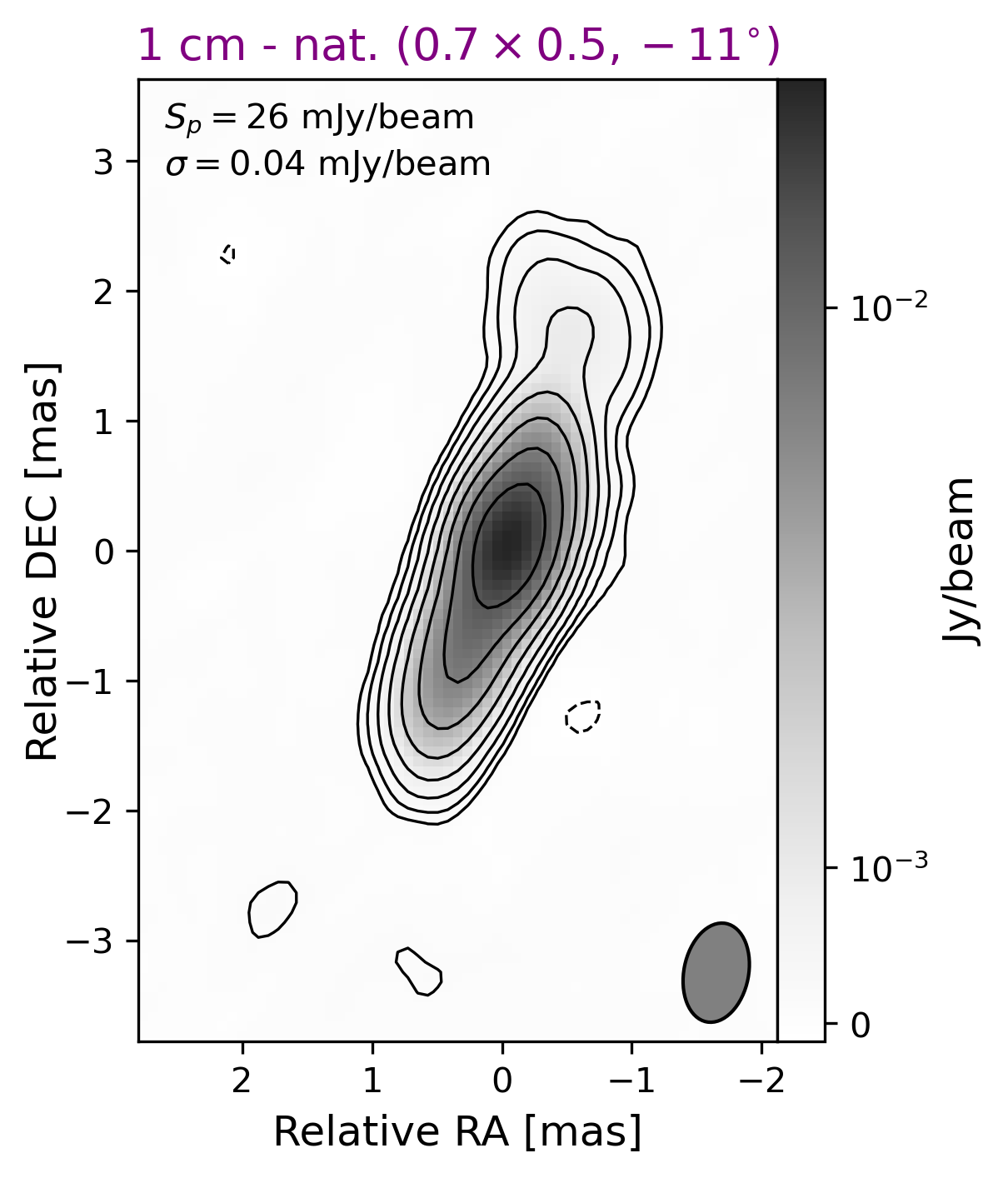}
        \caption{VLBI images of B2\,0222+36.  Left: 7\,mm image with uniform weighting. Center: 1\,cm image with uniform weighting. Right: 1\,cm image with natural weighting.}
        \label{fig:enter-label}
    \end{figure*}  
        
\clearpage
         
         \subsection{IC\,310}
        The FRI-LEG IC\,310 is a well known $\gamma$-ray source \citep{4LAC} characterized by minute-scale time variability at TeV energies \citep{2014Sci...346.1080A}. Our 1\,cm and 7\,mm VLBI images, featuring a higher rms noise due to the lack of the VLA and Effelsberg during its observation, show a one-sided jet oriented towards south-west (Fig. A.5), in agreement with cm-VLBI results \citep[e.g.,][]{Lister2019}.
     \begin{figure*}[!h]
        \centering
       \includegraphics[height=4.8cm]{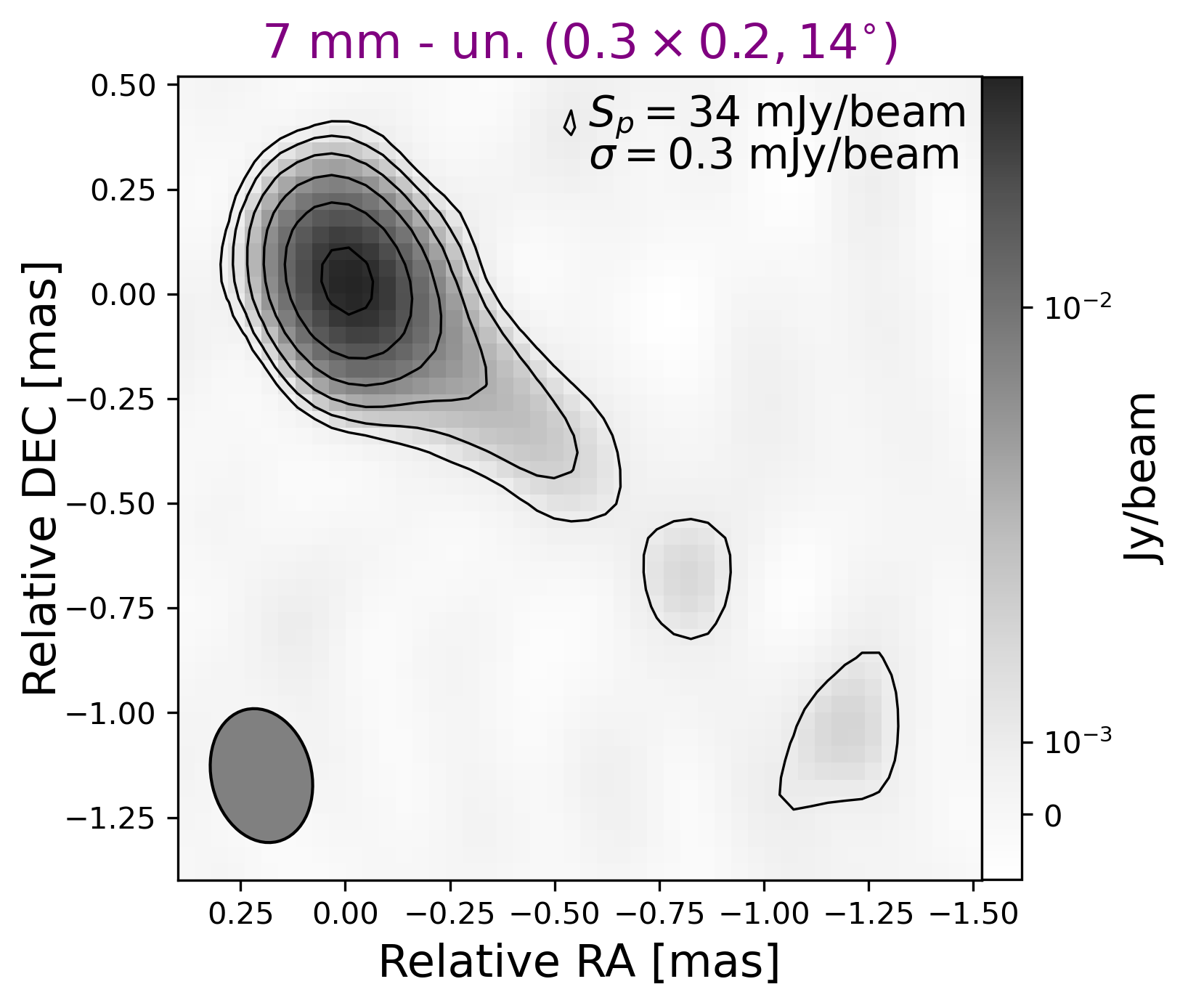}
        \includegraphics[height=4.8cm]{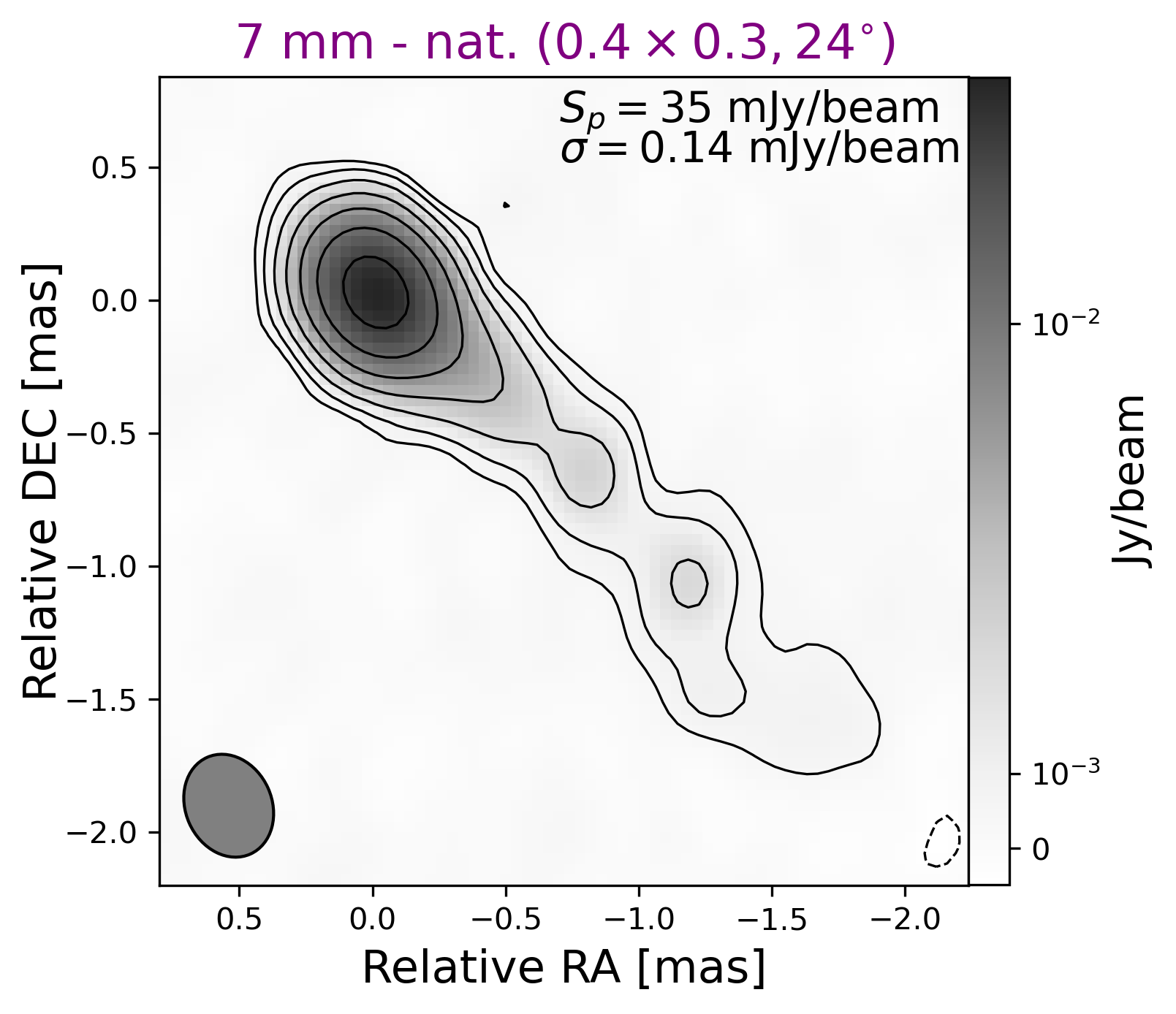}\\
        \includegraphics[width=5.5cm]{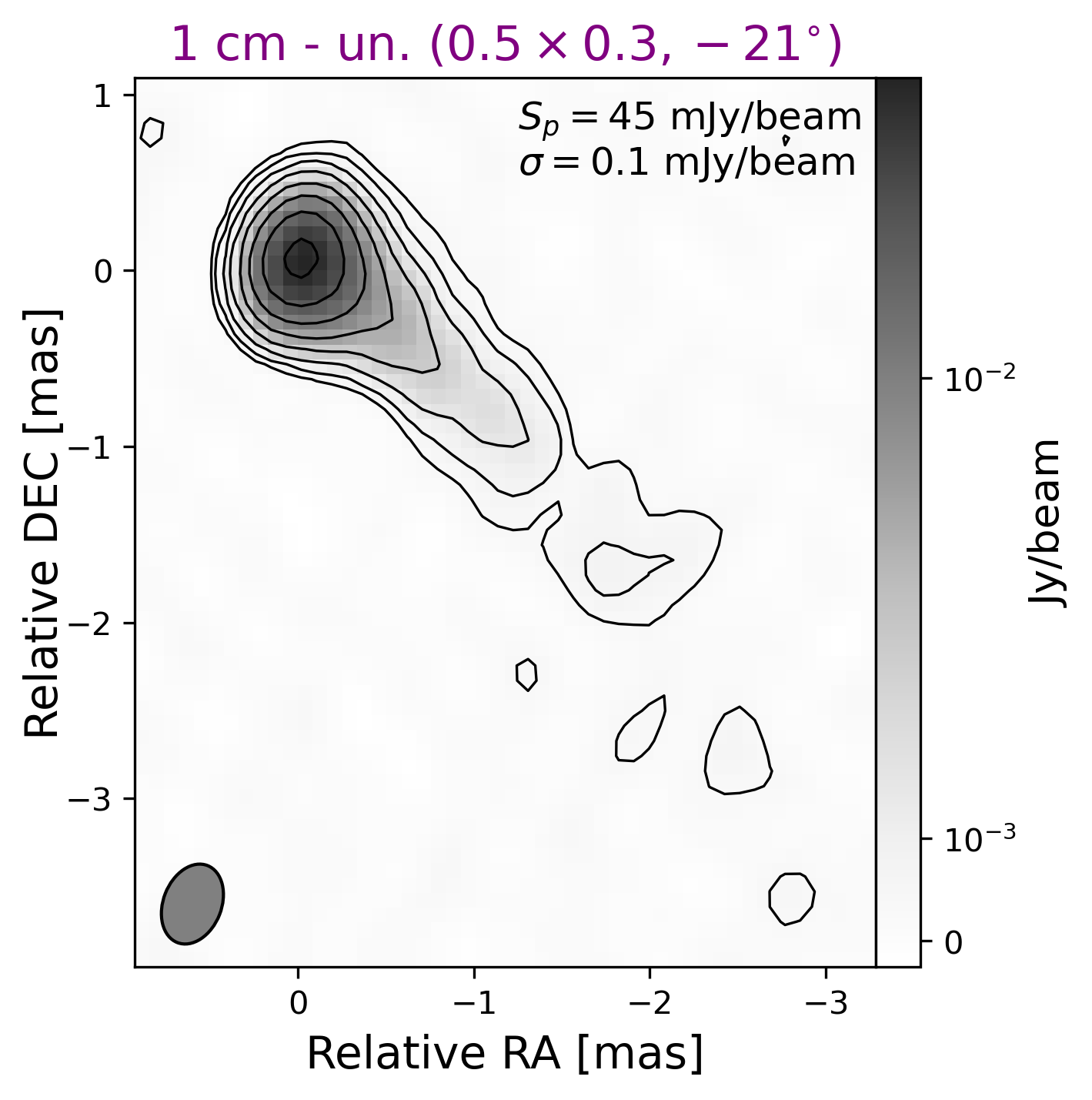}
        \includegraphics[width=5.5cm]{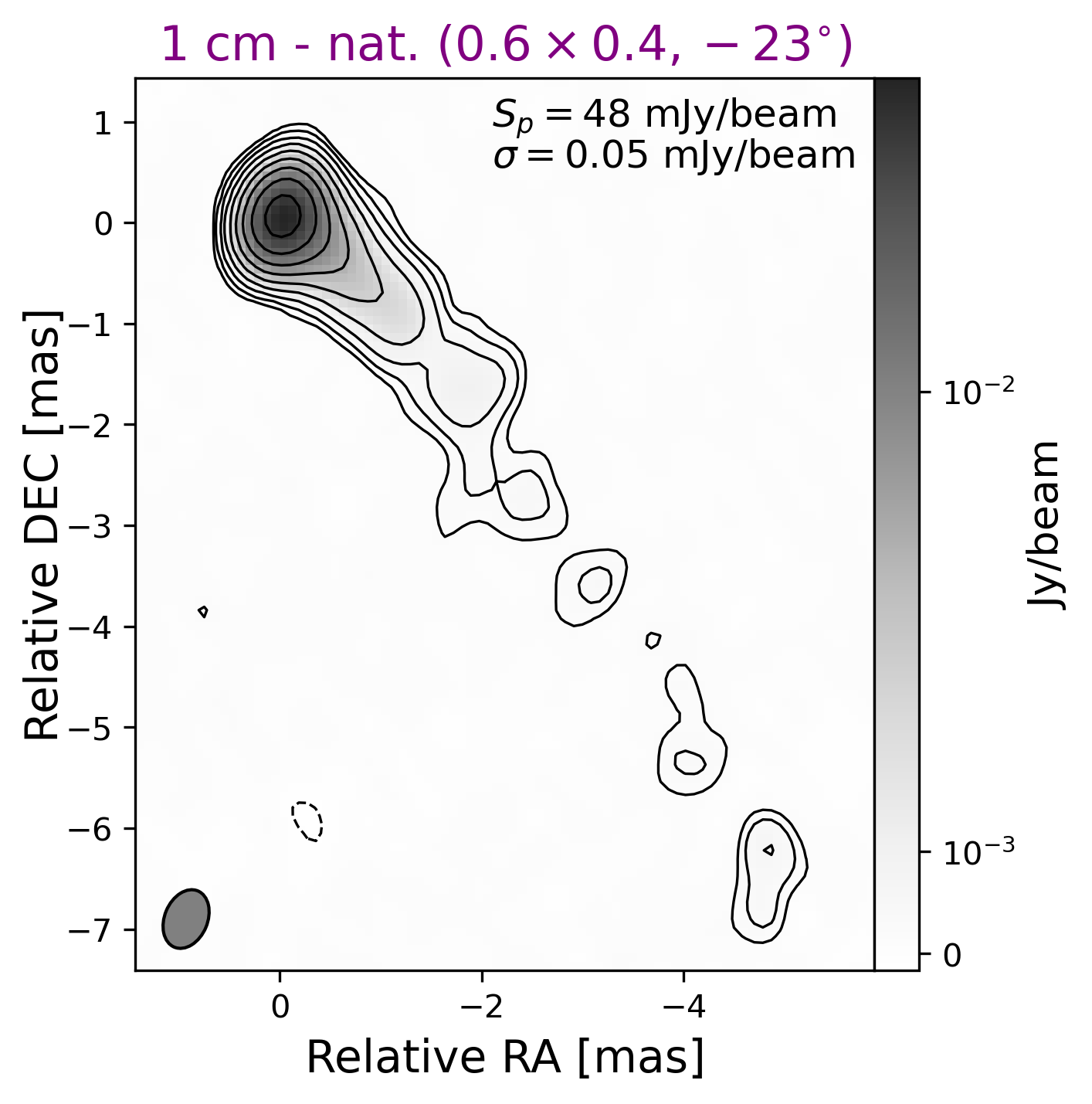}
        \caption{VLBI images of IC\,310. Top: 7\,mm image with uniform (left) and natural weighting (right). Bottom: 1\,cm image with uniform (left) and natural weighting (right).}
        \label{fig:enter-label}
    \end{figure*}

       \subsection{4C\,39.12}
   The FRI-LEG 4C\,39.12 is one of the radio galaxies detected by the { Fermi} satellite at $\gamma$-ray energies \citep{4LAC}. Its one-sided jet oriented towards south-east, which we detect at both bands (Fig. A.6),  is indeed indicative of a relatively small jet viewing angle. At 1\,cm, the source shows rather extended jet emission, which becomes well resolved transversely and appears to trace a helical pattern.  This could indicate that the jet in 4C\,39.12 is also limb-brightened, with the two jet rails intersecting at a distance of ${\sim}7$\,$\rm mas$ from the core. Deeper observations will be needed to confirm this feature.
        \begin{figure*}[!h]
        \centering
       \includegraphics[height=5.85cm]{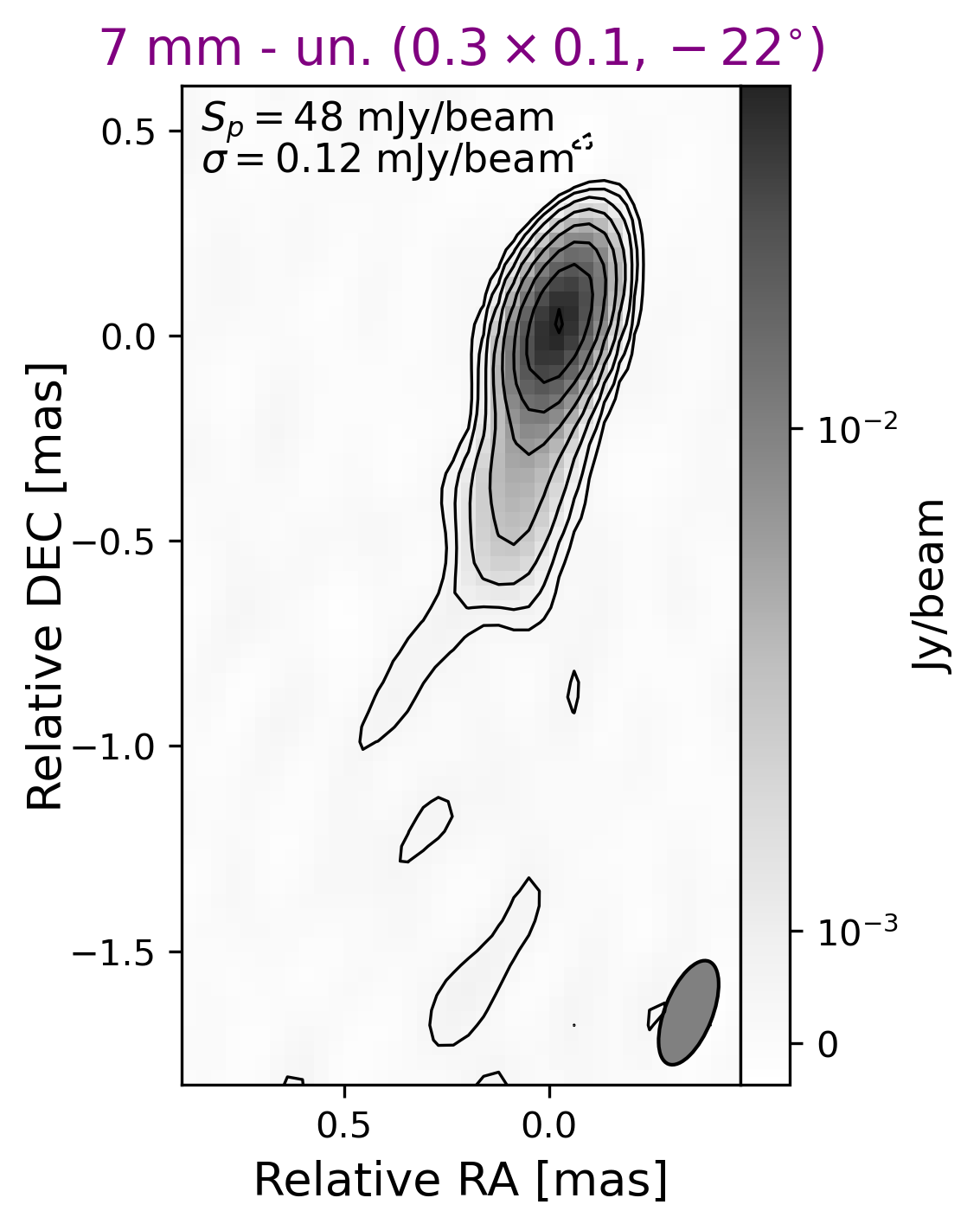}
        \includegraphics[height=5.75cm]{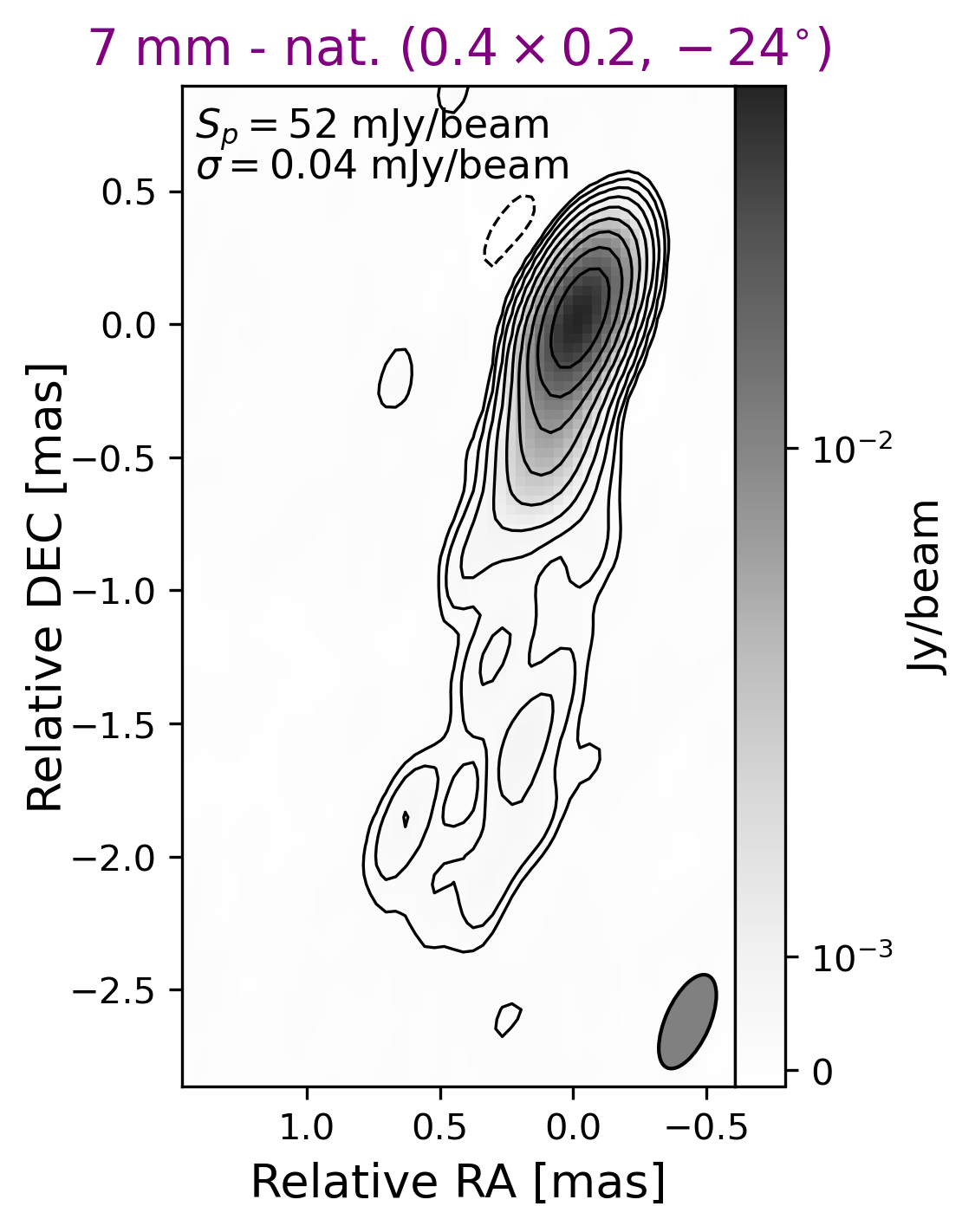}
        \includegraphics[height=5.75cm]{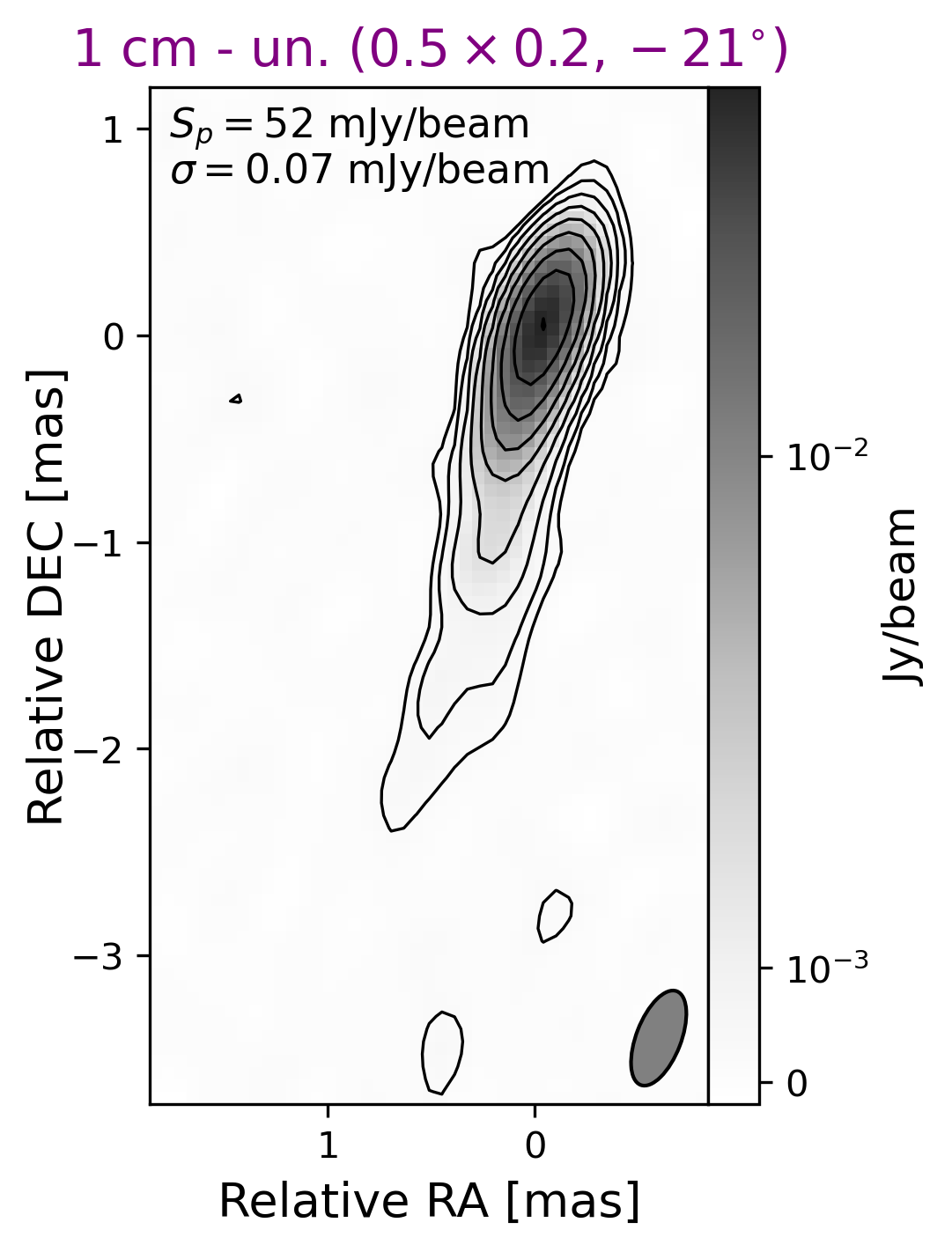}
        \includegraphics[height=5.75cm]{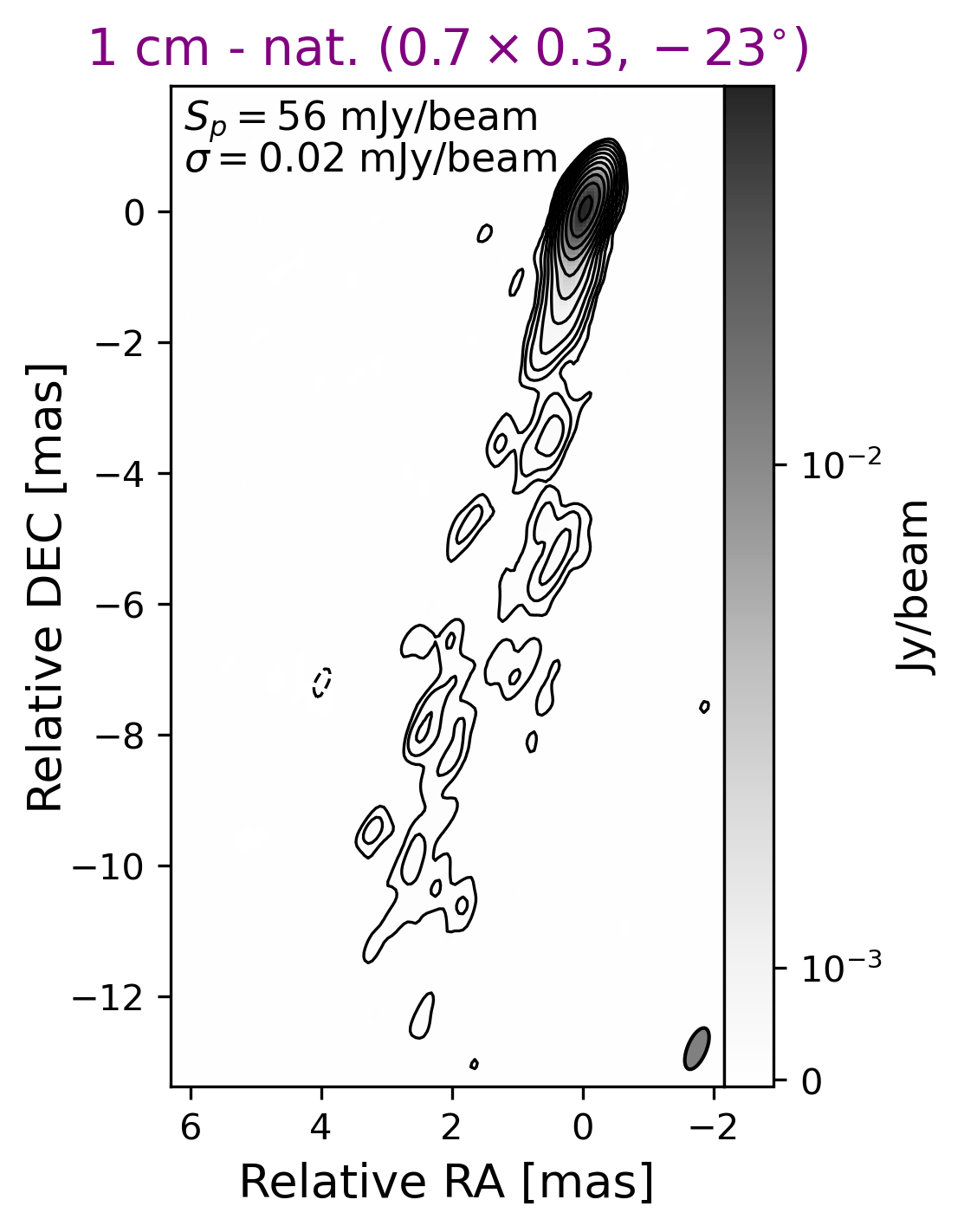}
        \caption{VLBI images of 4C\,39.12. From left to right: 7\,mm images with uniform and natural weighting, 1\,cm images with uniform and natural weighting.}
        \label{fig:enter-label}
    \end{figure*}  
   
  \clearpage
  
      \subsection{IC\,2402}
    IC2402 is one of the faintest sources in our sample. Both at 1\,cm and at 7\,mm, it shows a two-sided jet structure in the north-west/south-east direction, with the north-west side being the most extended and, likely, the approaching one (Fig. A.7). We report the classification of the source as a FRII-LEG (Table 1), although extended FRI lobes are also detected beyond the FRII hotspots, indicating restarted activity \citep{Giovannini2005}. 
        \begin{figure*}[!h]
        \centering
       \includegraphics[height=6.8cm]{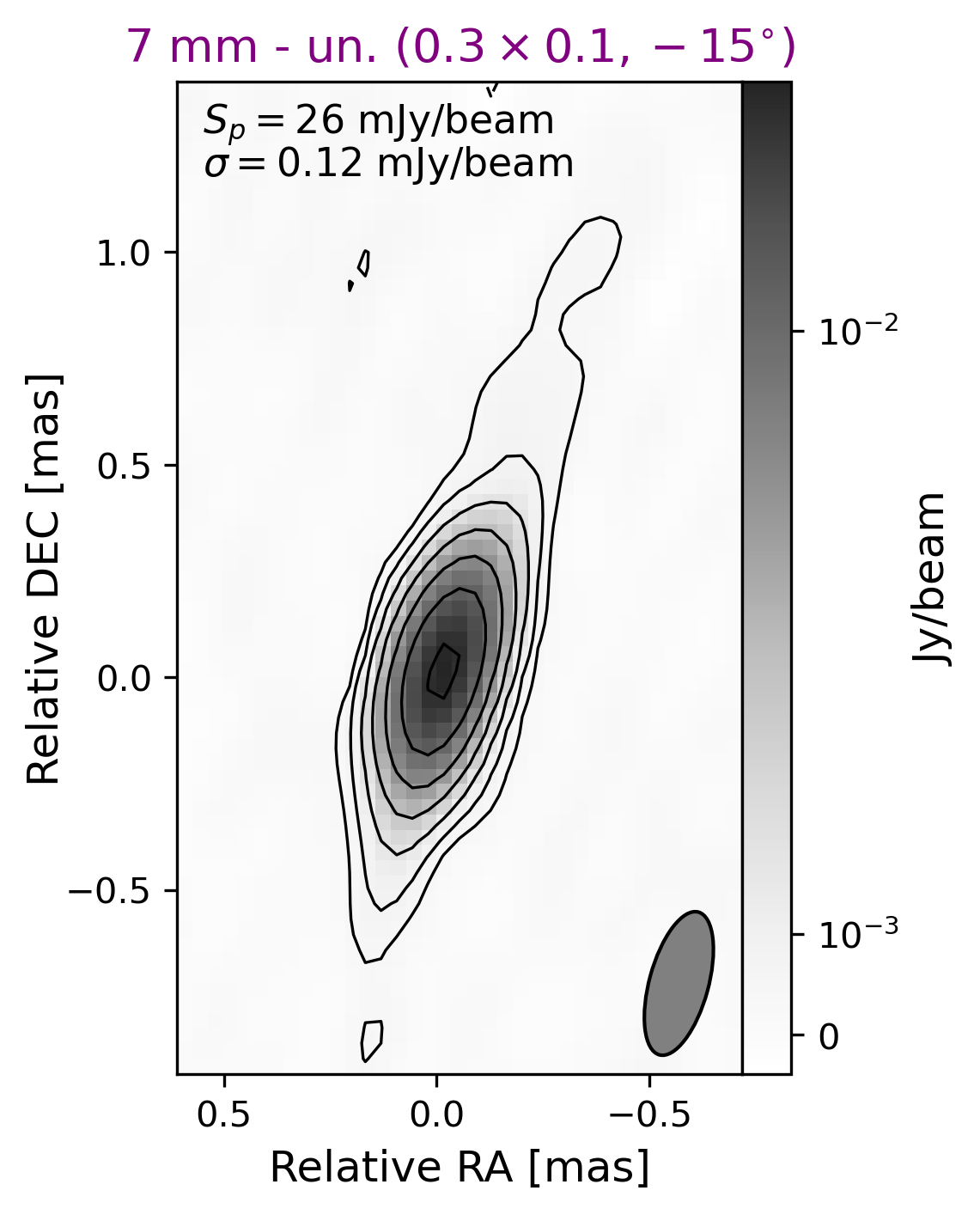}
        \includegraphics[height=6.8cm]{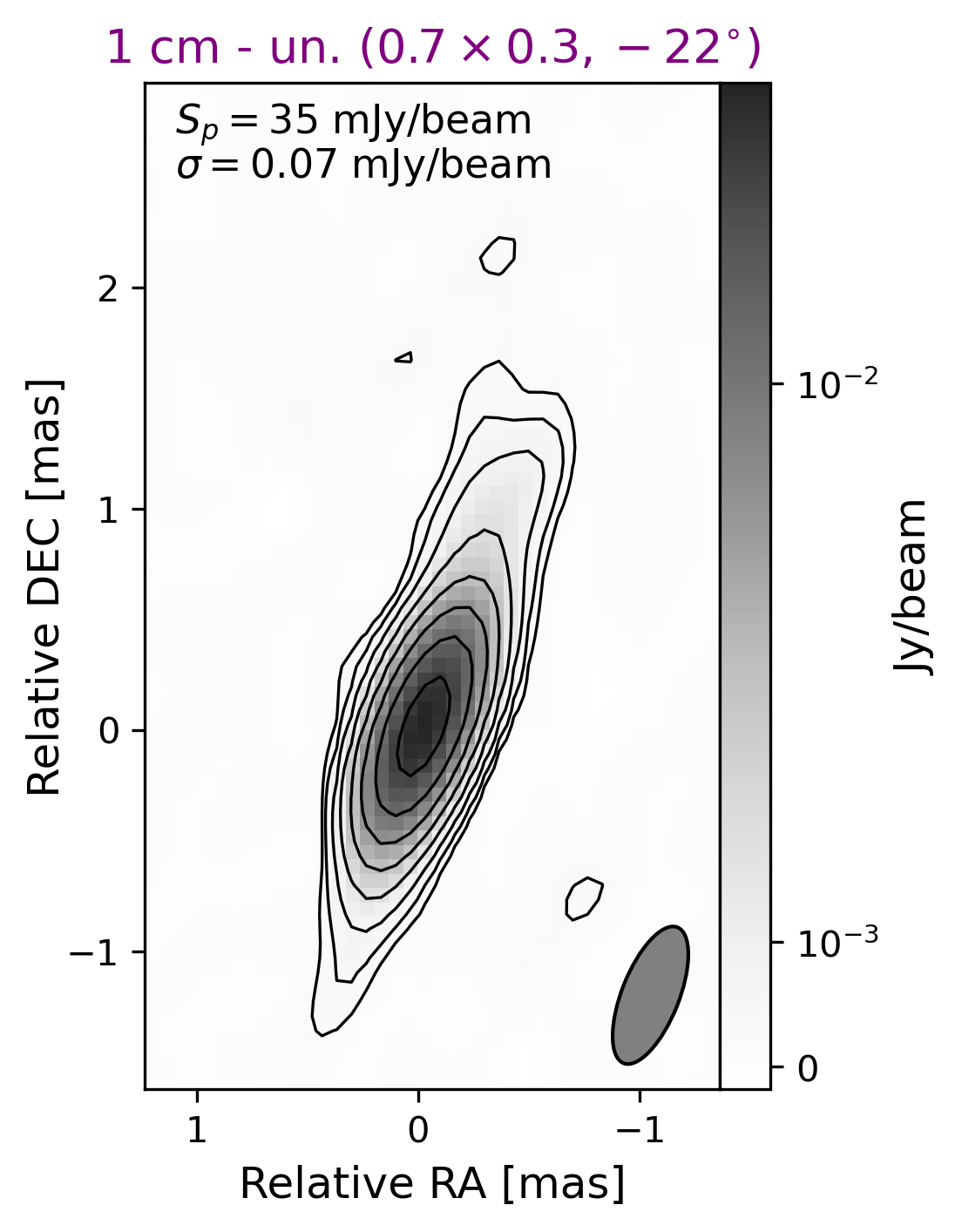}
        \includegraphics[height=6.8cm]{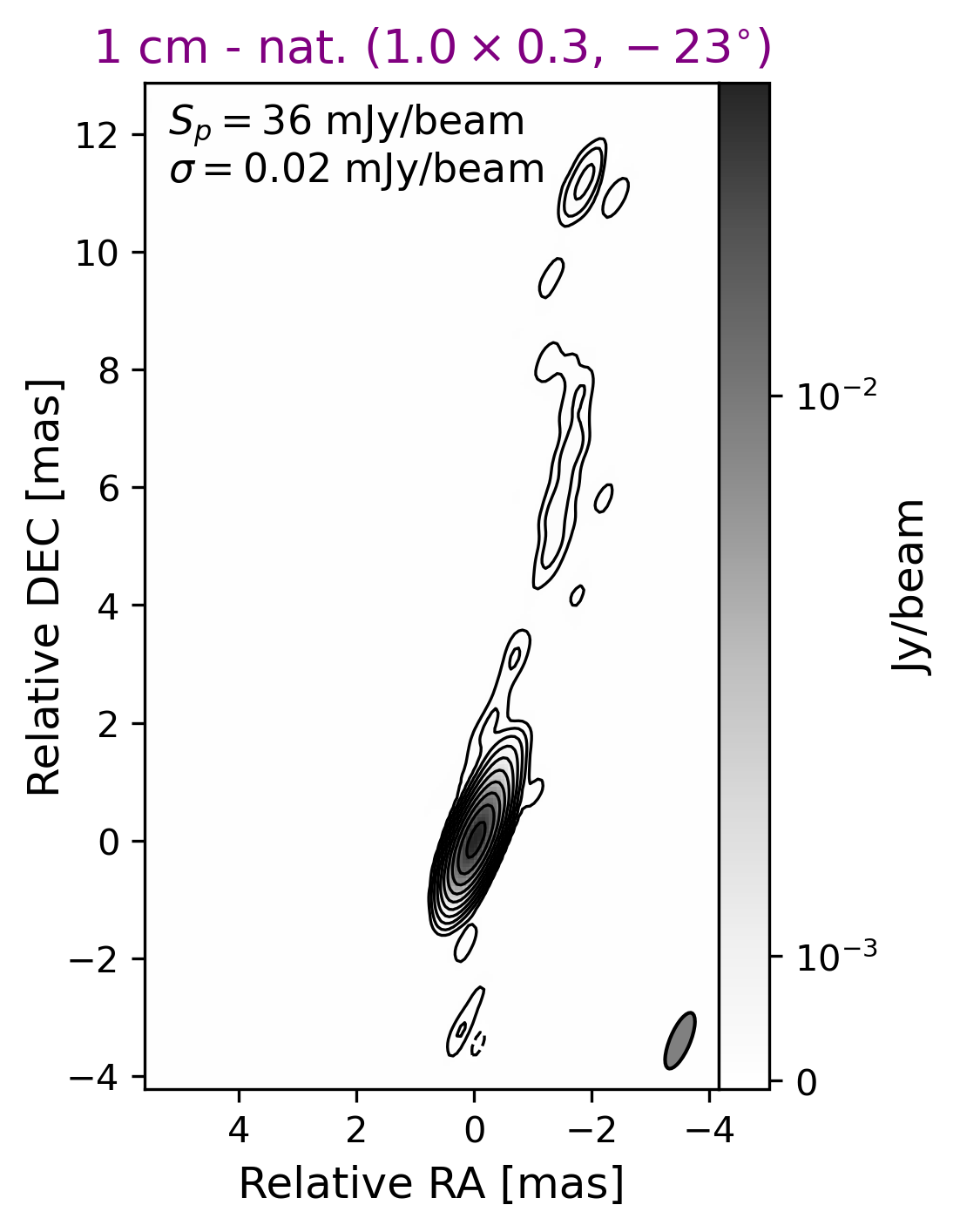}
        \caption{VLBI images of IC\,2402.  Left: 7\,mm image with uniform weighting. Center: 1\,cm image with uniform weighting. Right: 1\,cm image with natural weighting.}
        \label{fig:enter-label}
    \end{figure*}
    
    \subsection{3C\,264}
    The FRI-LEG 3C264 is a well studied object at centimeter radio wavelengths \citep[e.g.,][]{Lister2019}. It is a $\gamma$-ray emitter detected up to TeV energies, with a one-sided VLBI jet characterized by prominent limb-brightening \citep[][and references therein]{Boccardi2019}.  In our images (Fig. A.8), we also detect a one-sided jet oriented in the north-east direction, with hints of limb-brightening at both bands, as well as some emission upstream of the core. As in the case of 3C\,66B, the possible detection of counter-jet emission needs to be confirmed in future studies. 
    \begin{figure*}[!h]
        \centering
         \includegraphics[height=6.8cm]{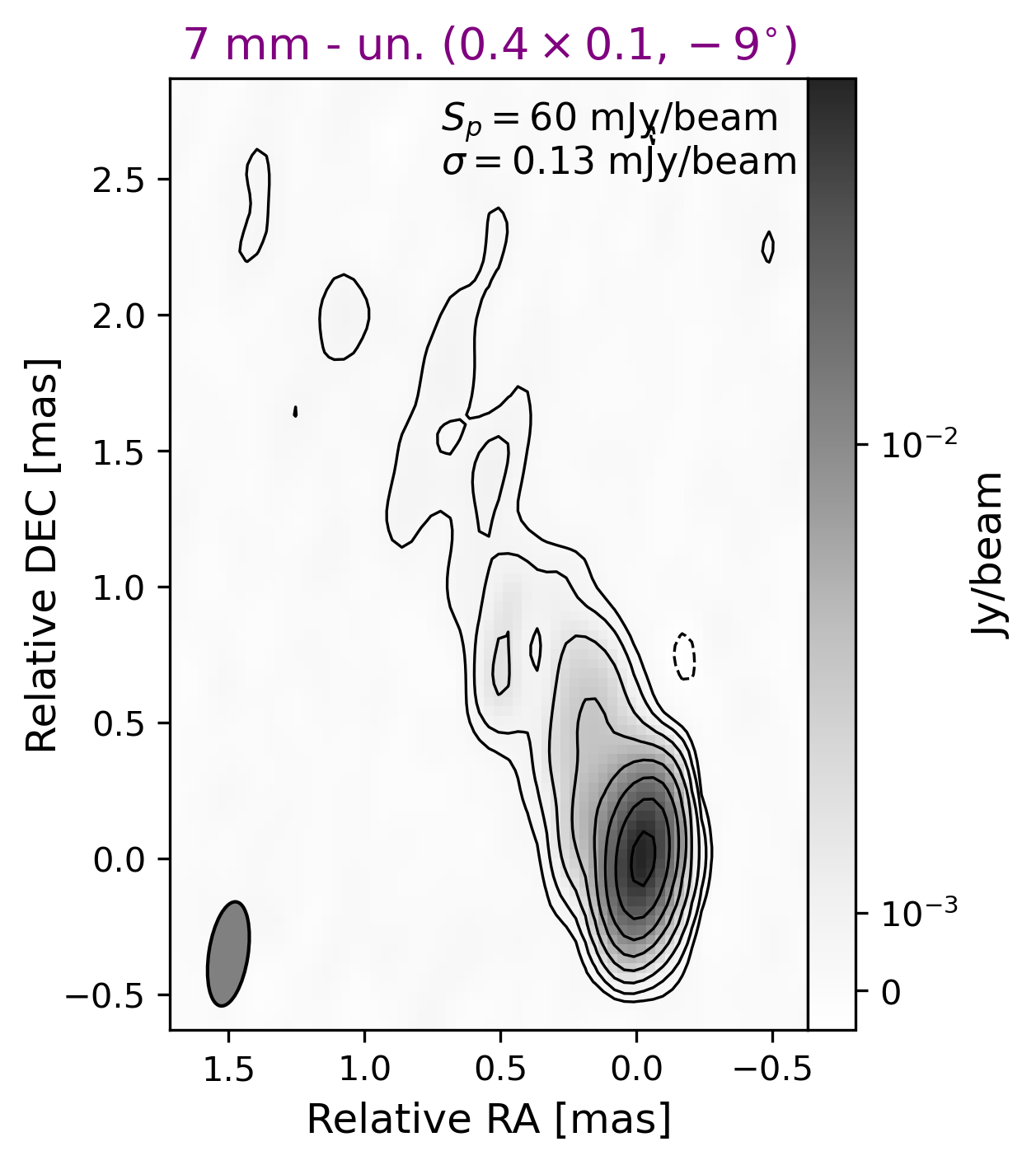}
        \includegraphics[height=6.8cm]{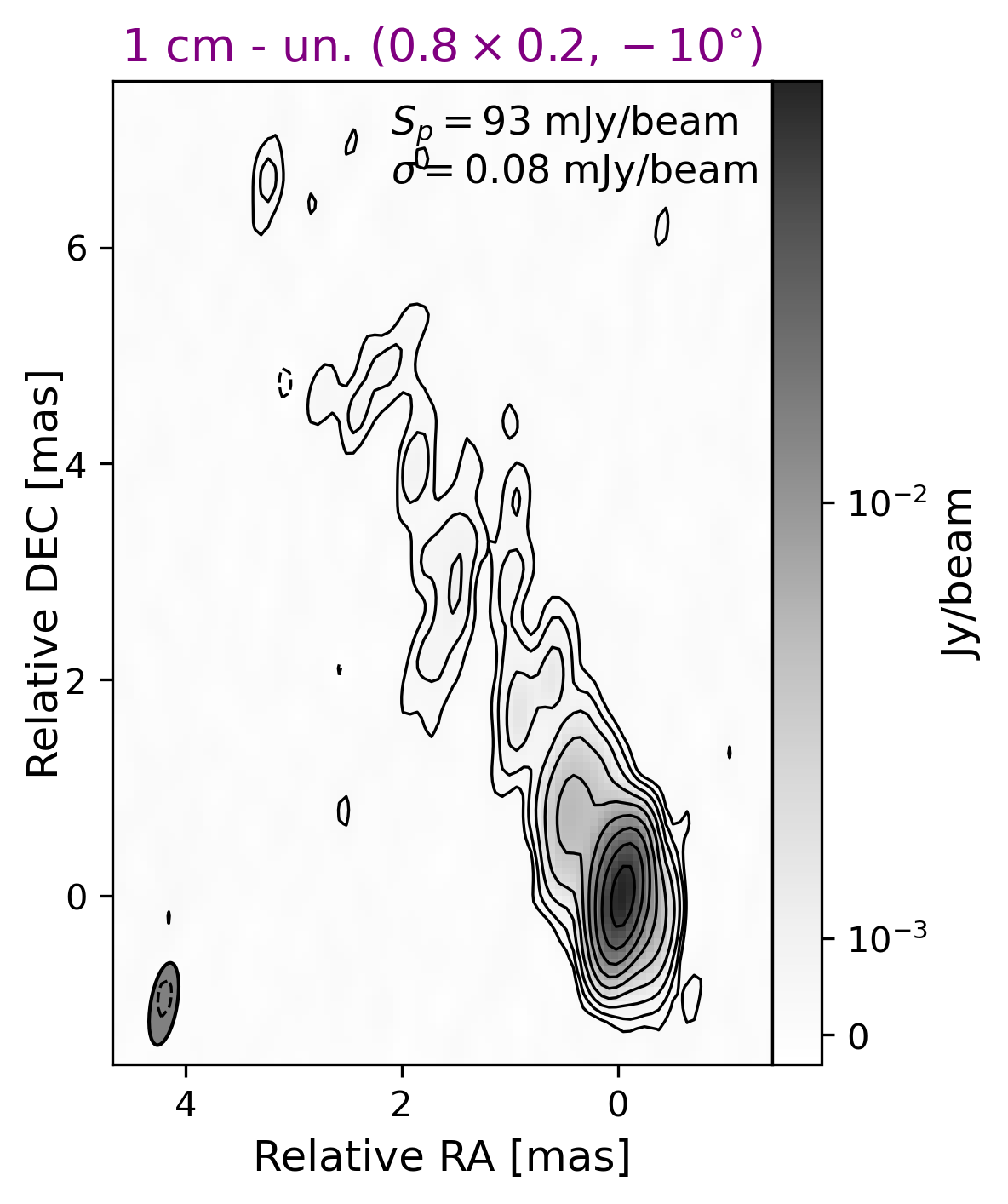}
        \includegraphics[height=6.8cm]{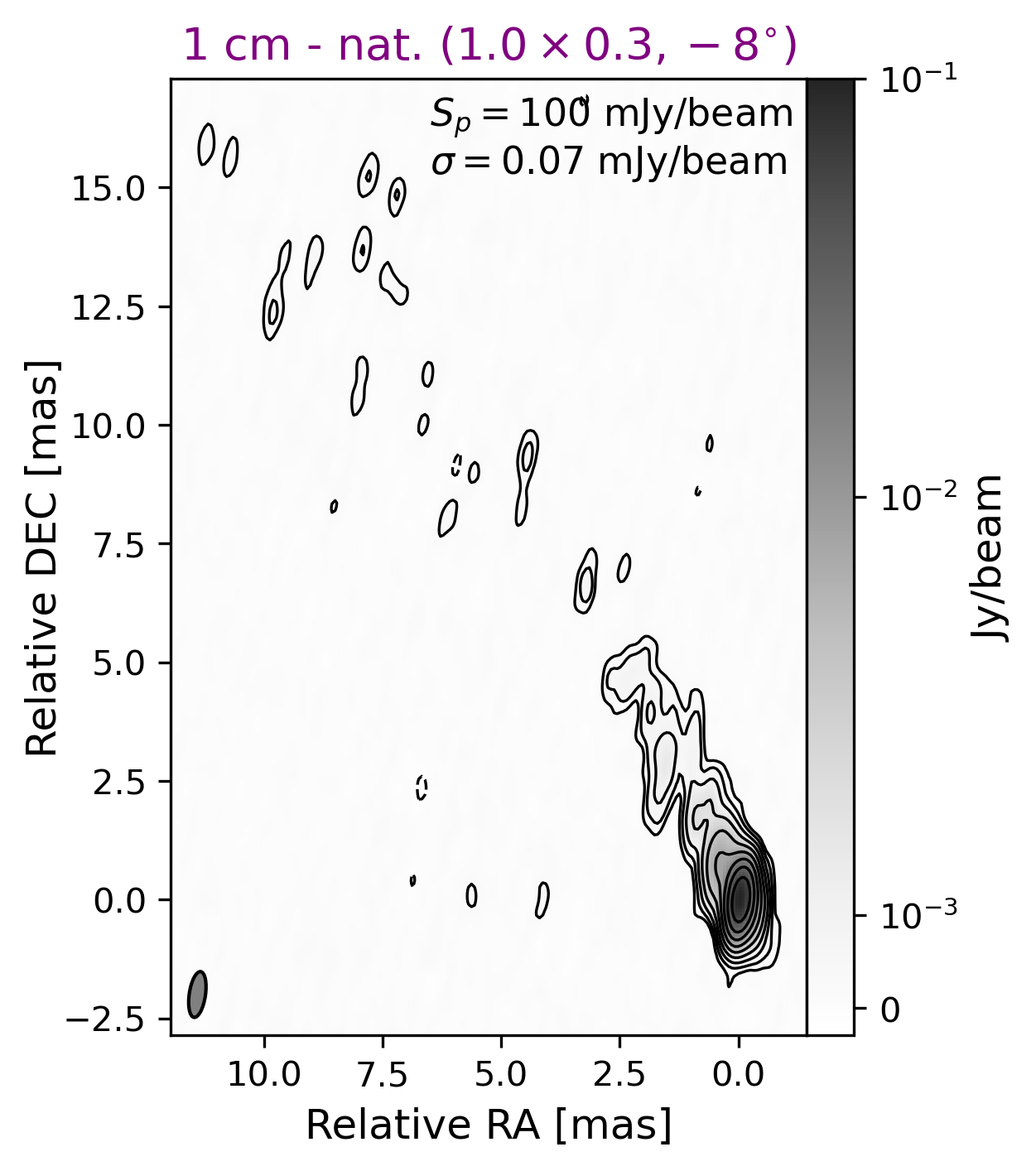}
        \caption{VLBI images of 3C\,264.  Left: 7\,mm image with uniform weighting. Right: 1\,cm image with uniform weighting. Left: 1\,cm image with natural weighting.}
        \label{fig:enter-label}
    \end{figure*}
  \clearpage

        \subsection{NGC4278}
    NGC\,4278 is by far the closest (z=0.0021) and lowest power source ($P_{\rm t}{\sim}6\times 10^{21}\,\rm W/Hz$) in our sample. It is a LEG showing a compact morphology on VLA scales \citep{Giroletti2005}. An early attempt to image this object at 7\,mm was presented by \cite{Ly2004}. Our images (Fig. A.9) reveal a peculiar core structure at both bands and abundant diffuse emission with a two-sided morphology at 1 cm. This is roughly oriented in the north-west/south-east direction, in agreement with cm-VLBI results presented by \cite{Giroletti2005}. The properties of the core component and innermost jet emission at 7\,mm are compatible with the map by \cite{Ly2004}. It is not straightforward to precisely localise,  with the present data, the faint diffuse emission, which seems to also embed the core.  Observations with longer on-source time and better uv-coverage are needed to confirm and improve the imaging of this interesting source. The complex morphology suggested by our data may indicate that the jet is poorly collimated and/or oriented at relatively small angle with respect to the observer, which introduces projection effects. A small viewing angle, which would be consistent with the recent detection of the source at TeV energies \citep{Cao2024}, can be reconciled with the small jet-to-counterjet ratio if the flow has an intrinsically low speed. 
         
          \begin{figure*}[!h]
        \centering
       \includegraphics[height=4.8cm]{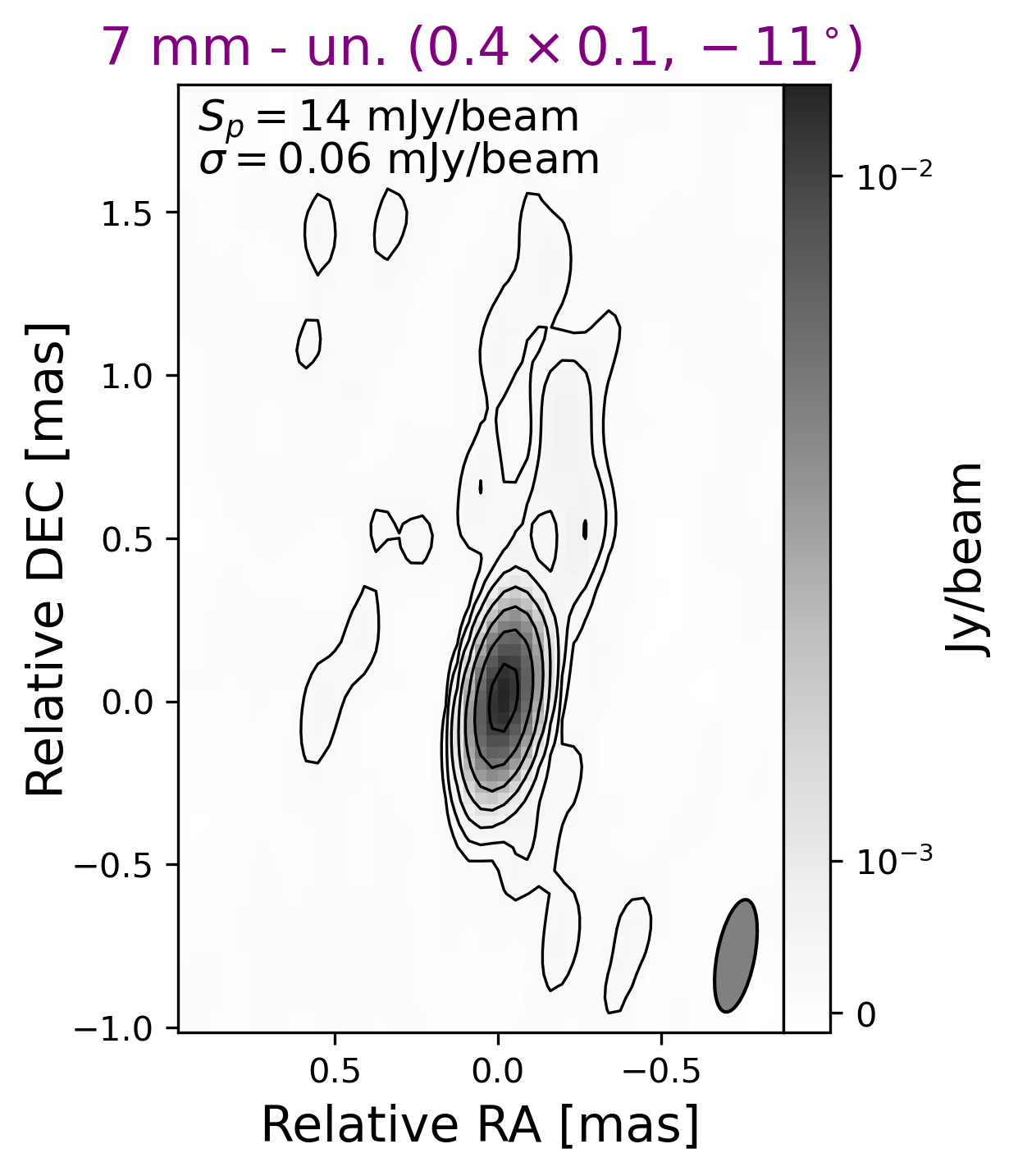}
         \includegraphics[height=4.8cm]{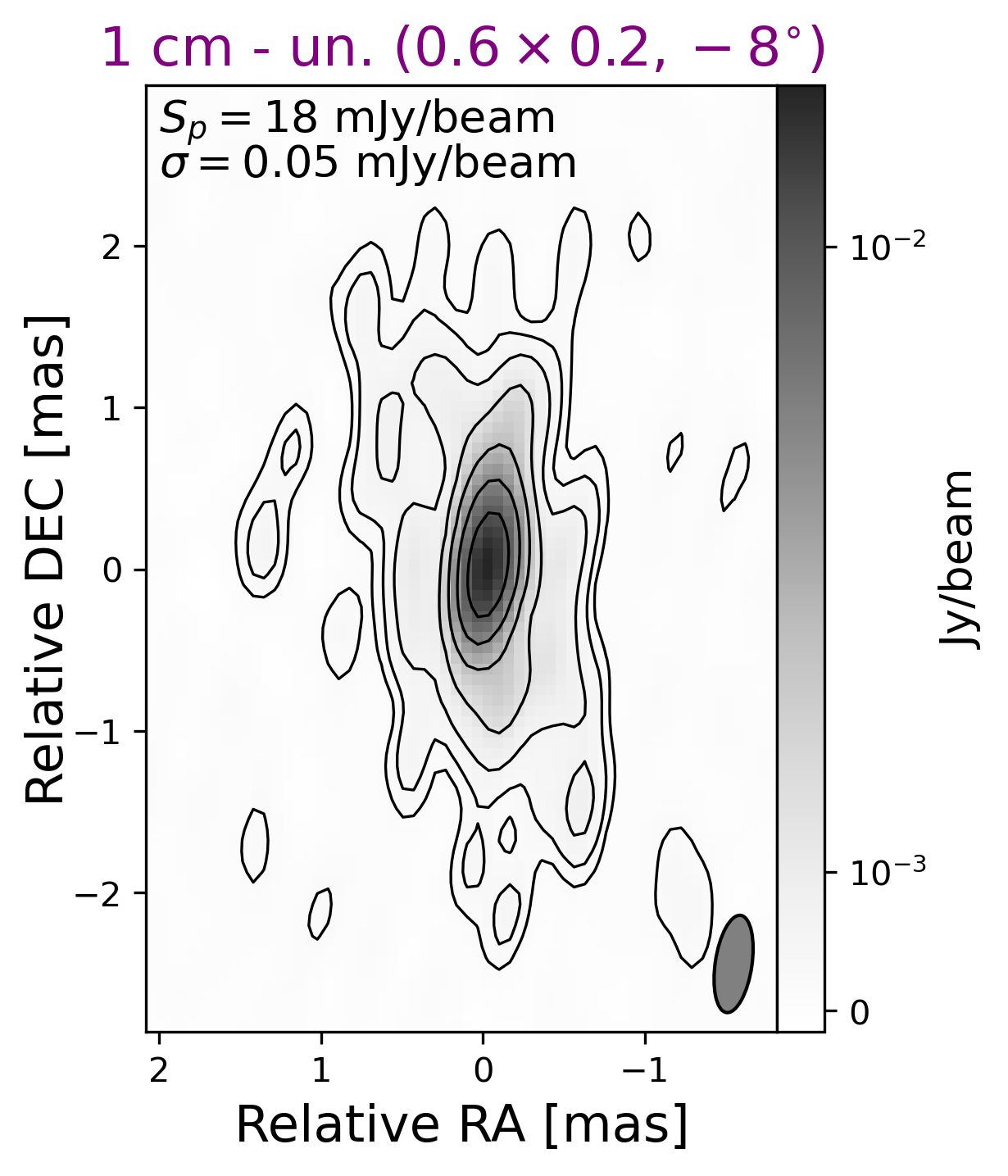}     
        \includegraphics[height=4.4cm]{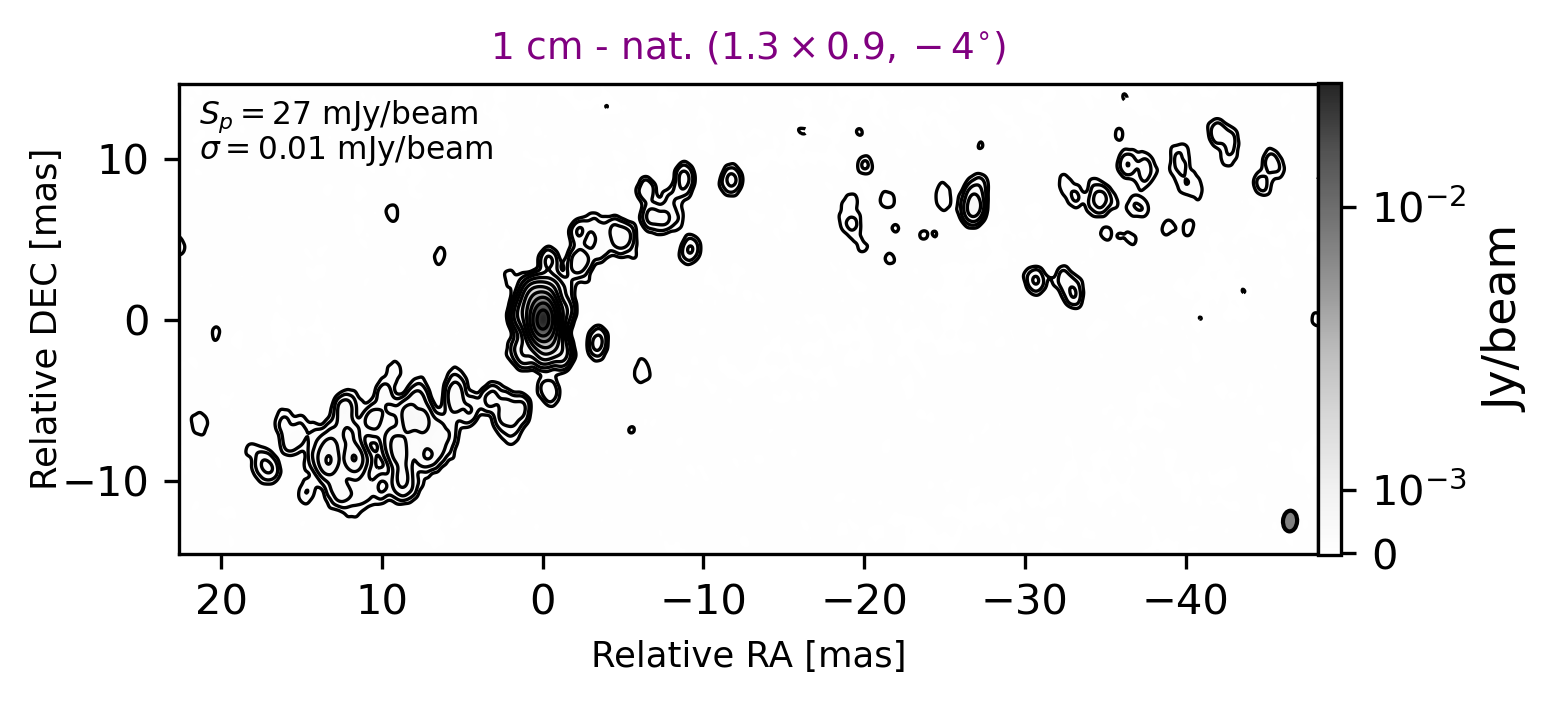}
        \caption{VLBI images of NGC\,4278.  Left: 7\,mm image with uniform weighting. Center: 1\,cm image with uniform weighting. Right: 1\,cm image with natural weighting. A Gaussian taper was also applied in this case to better show the structure of the extended emission.}
        \label{fig:enter-label}
    \end{figure*}
    
         \subsection{3C338}
     The FRI-LEG radio galaxy 3C338 is among the weakest in our sample. Its cm-VLBI properties, as well as its peculiar large scale structure have been discussed in detail by \cite{Giovannini1998}. At 1\,cm, our data shows a quite symmetric two-sided jet oriented in the east-west direction, with the east-side being the brightest. At 7\,mm, the source becomes more core-dominated, but some features in the weak two-sided jet are also detected (Fig. A.10). 
           \begin{figure*}[!h]
        \centering
        \includegraphics[width=9.15cm]{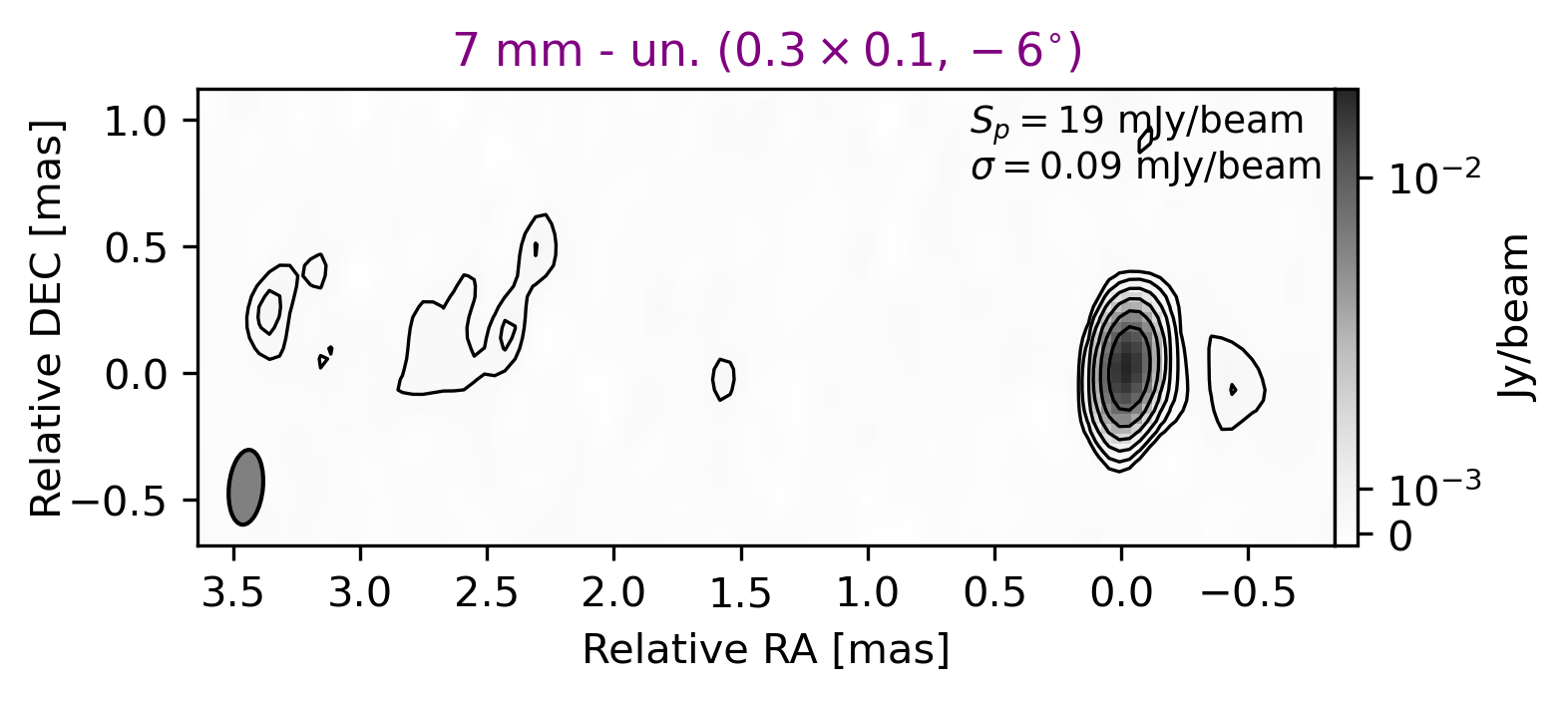}
        \includegraphics[width=9.15cm]{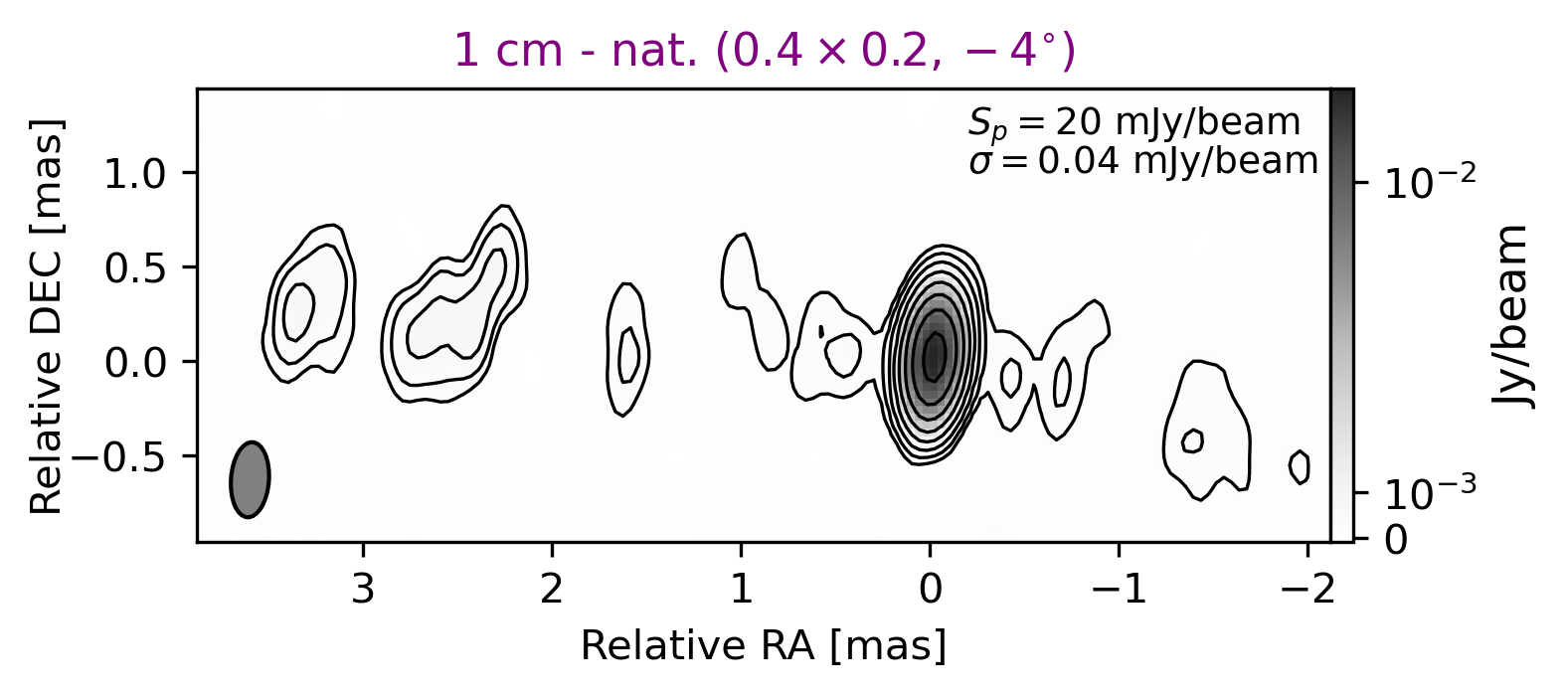}\\
        \includegraphics[width=9.15cm]{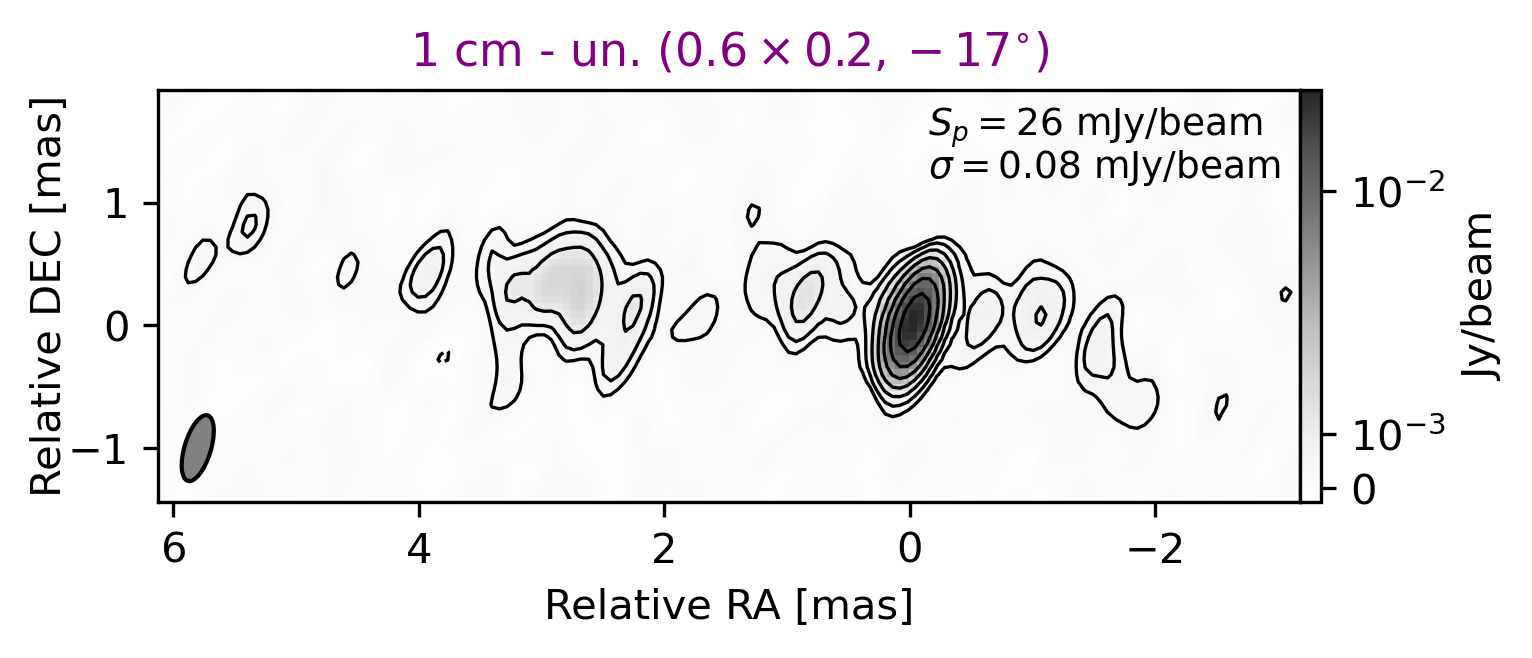}
        \includegraphics[width=9.15cm]{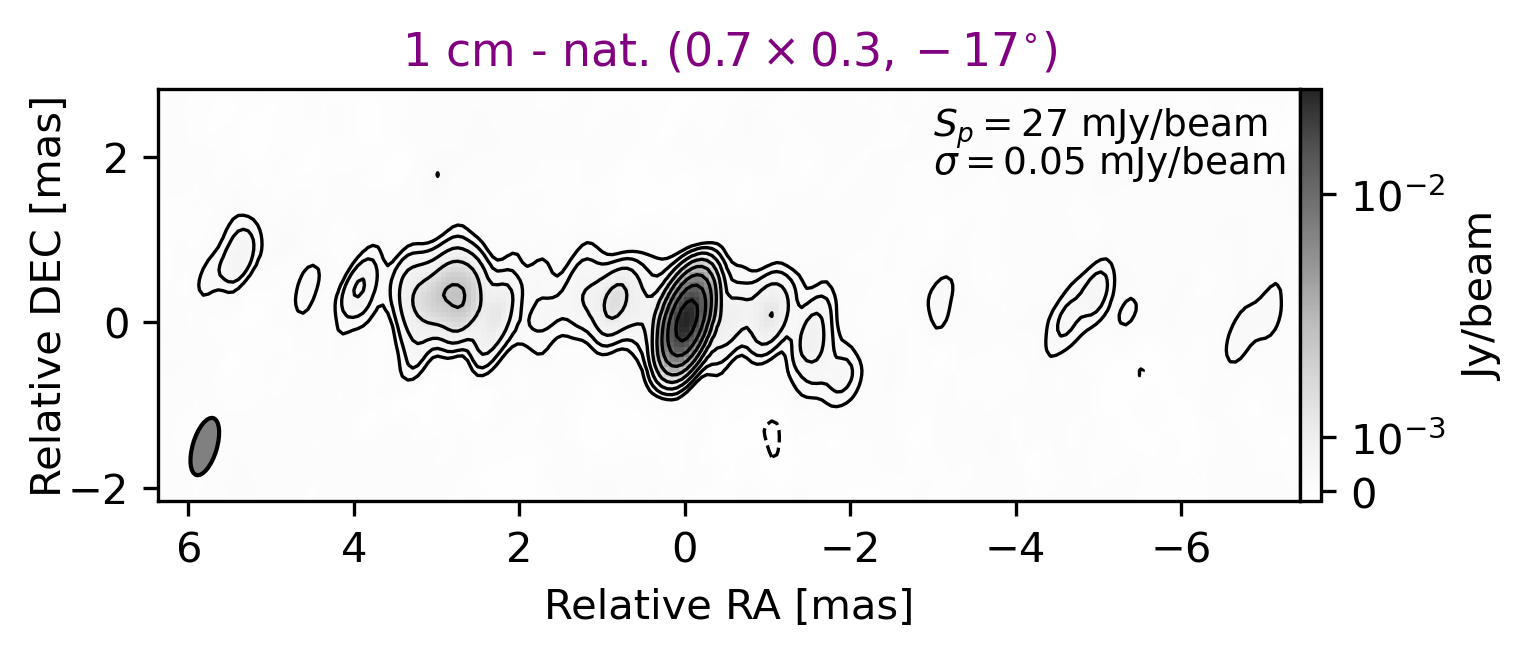}
        \caption{VLBI images of 3C\,338. Top: 7\,mm image with uniform (left) and natural weighting (right). Bottom: 1\,cm image with uniform (left) and natural weighting (right).}
        \label{fig:enter-label}
    \end{figure*}
    
    \clearpage

 \subsection{4C30.31}
 The FRI-LEG 4C30.31 shows at both bands a one-sided jet oriented in the south-west direction (Fig. A.11), in agreement with the lower frequency map presented by \cite{Giovannini2005}. The jet one-sidedness indicates a relatively small viewing angle, consistent with the detection of an optical jet in Hubble-Space-Telescope images of the source \citep{Capetti2000}.
       \begin{figure*}[!h]
        \centering
        \includegraphics[height=5.25cm]{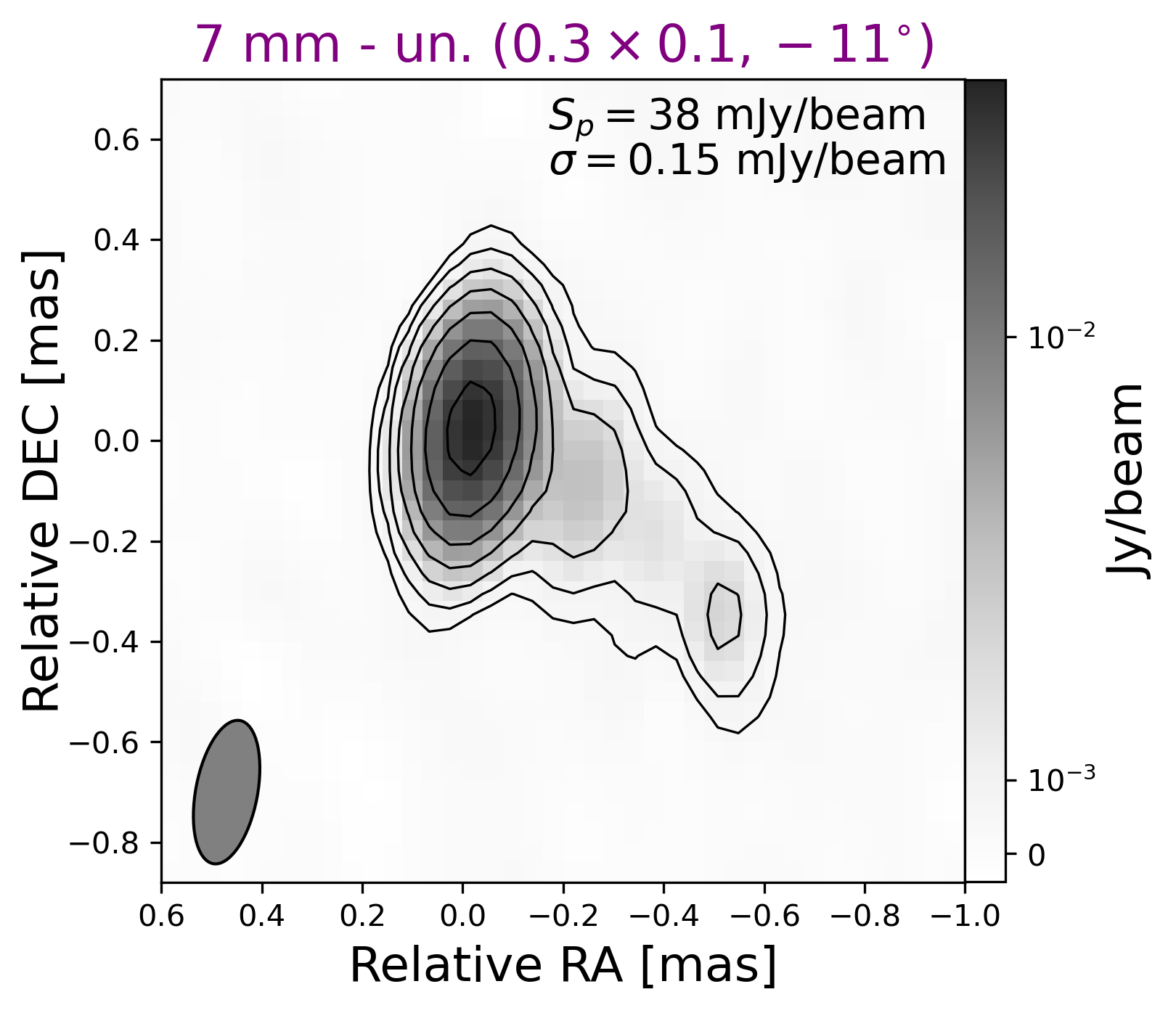}
        \includegraphics[height=5.25cm]{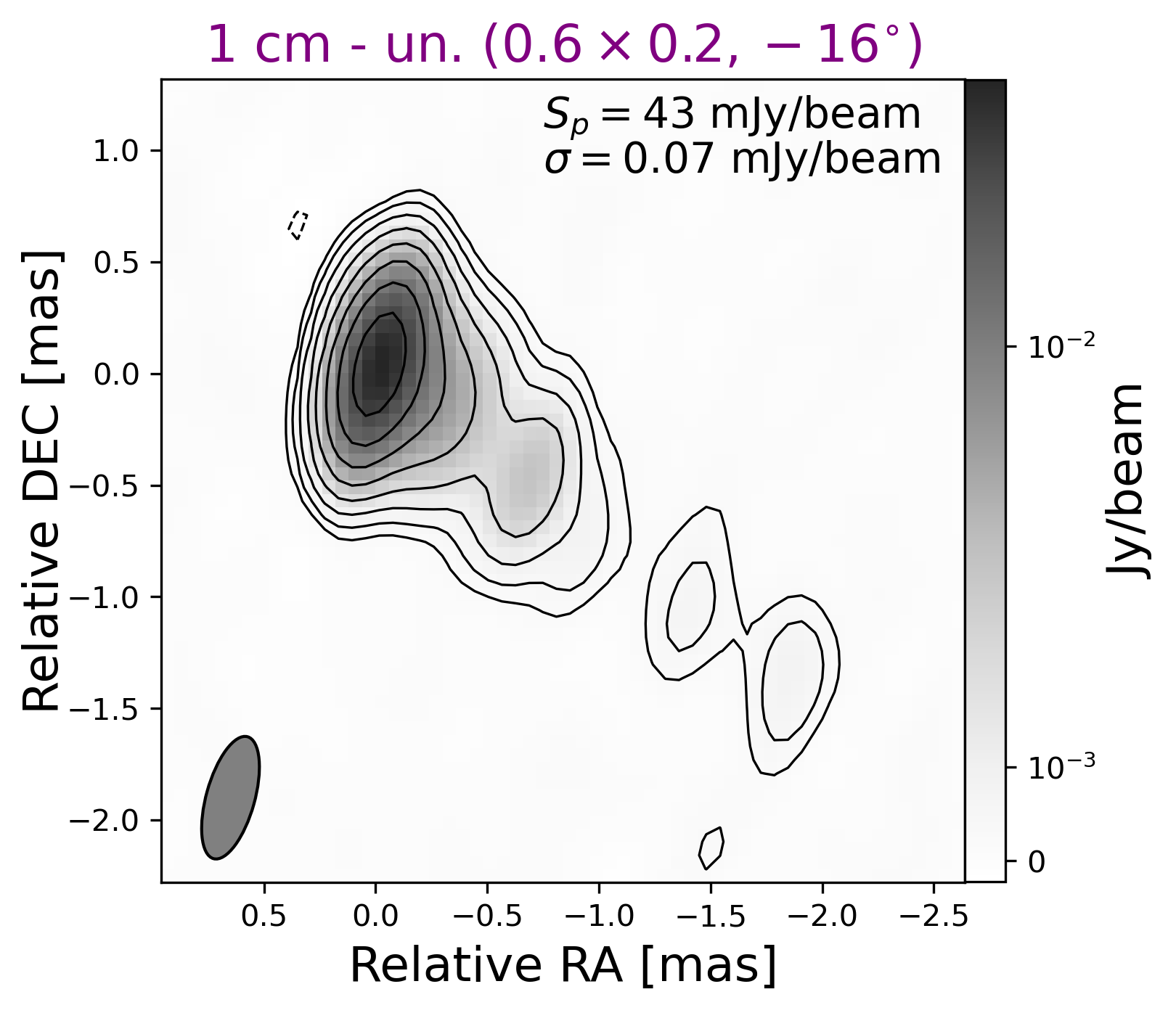}
        \includegraphics[height=5.25cm]{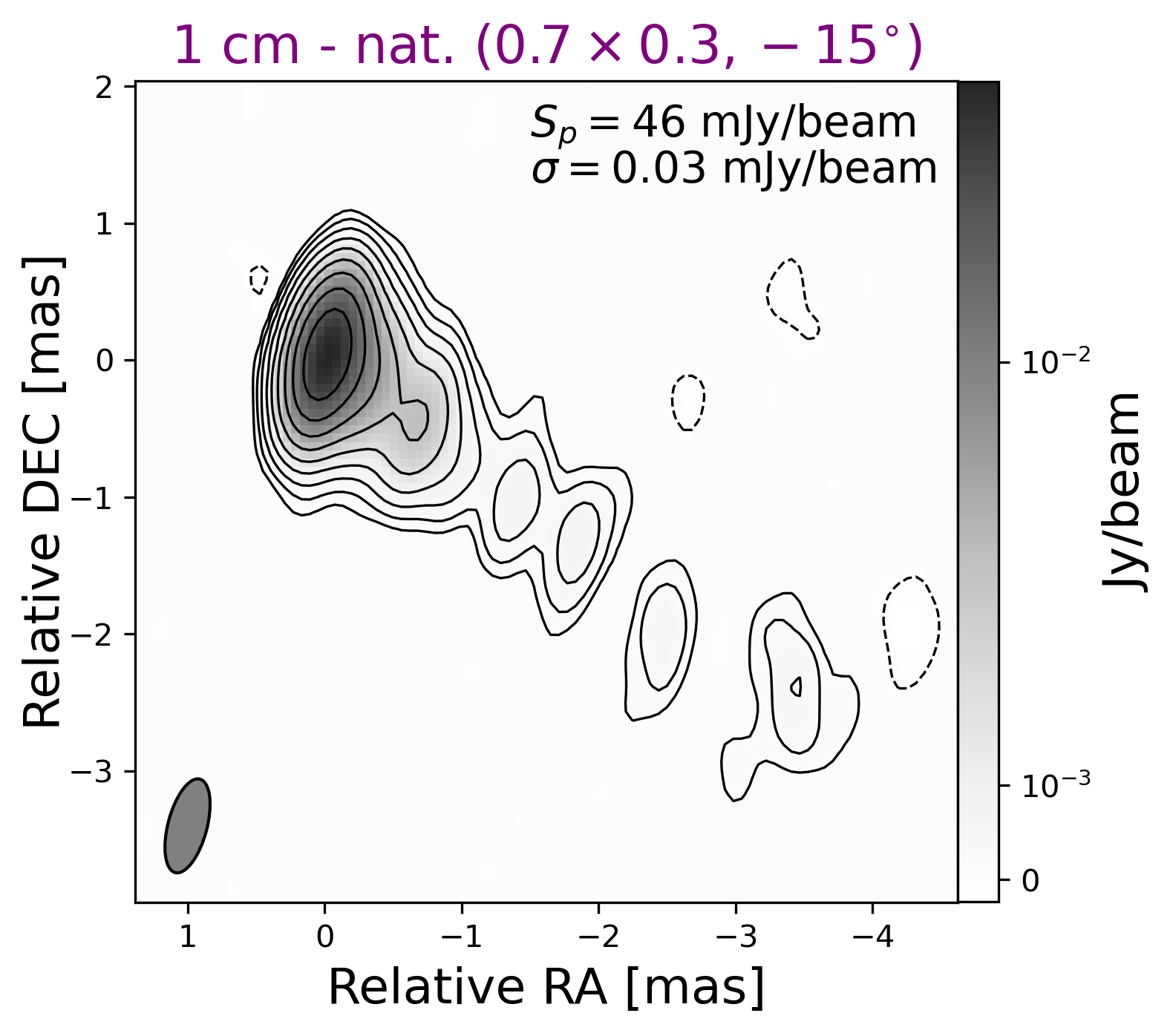}
       \caption{VLBI images of 4C\,30.31. Right: 7\,mm image with uniform weighting. Center: 1\,cm image with uniform weighting. Left: 1\,cm image with natural weighting.}
        \label{fig:enter-label}
    \end{figure*} 

      \subsection{3C\,382}
      The FRII-HEG 3C382 shows at both bands a one-sided jet propagating in the north-east direction (Fig. A.12). At 1\,cm, a weak feature in the receding side is detected at  $\sim$1\,$\rm mas$ from the core. The high jet-to-countrjet intensity ratio, in agreement with results from \cite{Giovannini2005, Lister2019}, indicates a relatively small jet viewing angle, consistent with the classification of the source as a broad-line radio galaxy, i.e., showing an unobscured nucleus \citep[e.g.,][]{Buttiglione2010}.
     \begin{figure*}[!h]
        \centering
        \includegraphics[width=7.4cm]{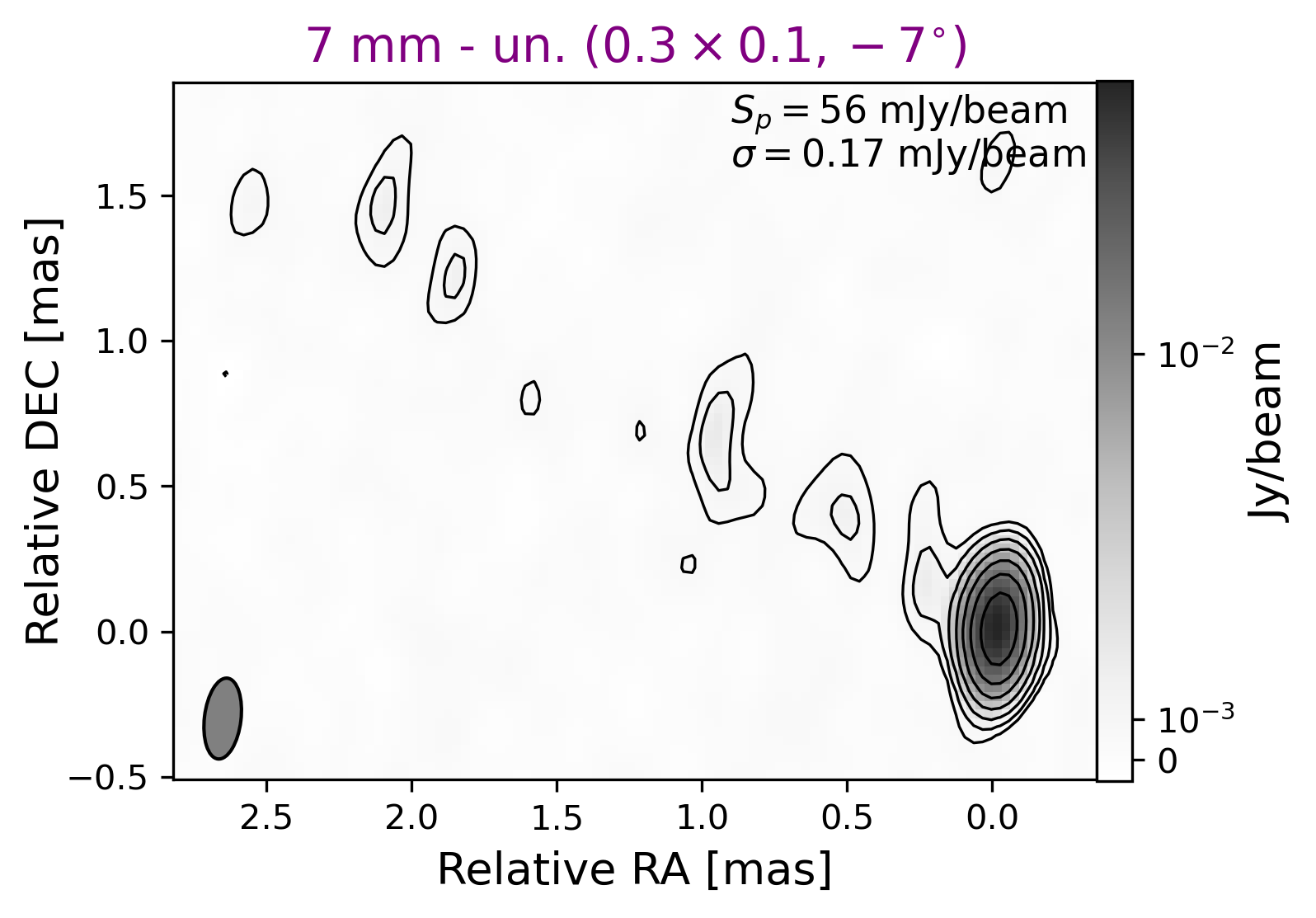}
        \includegraphics[width=7.4cm]{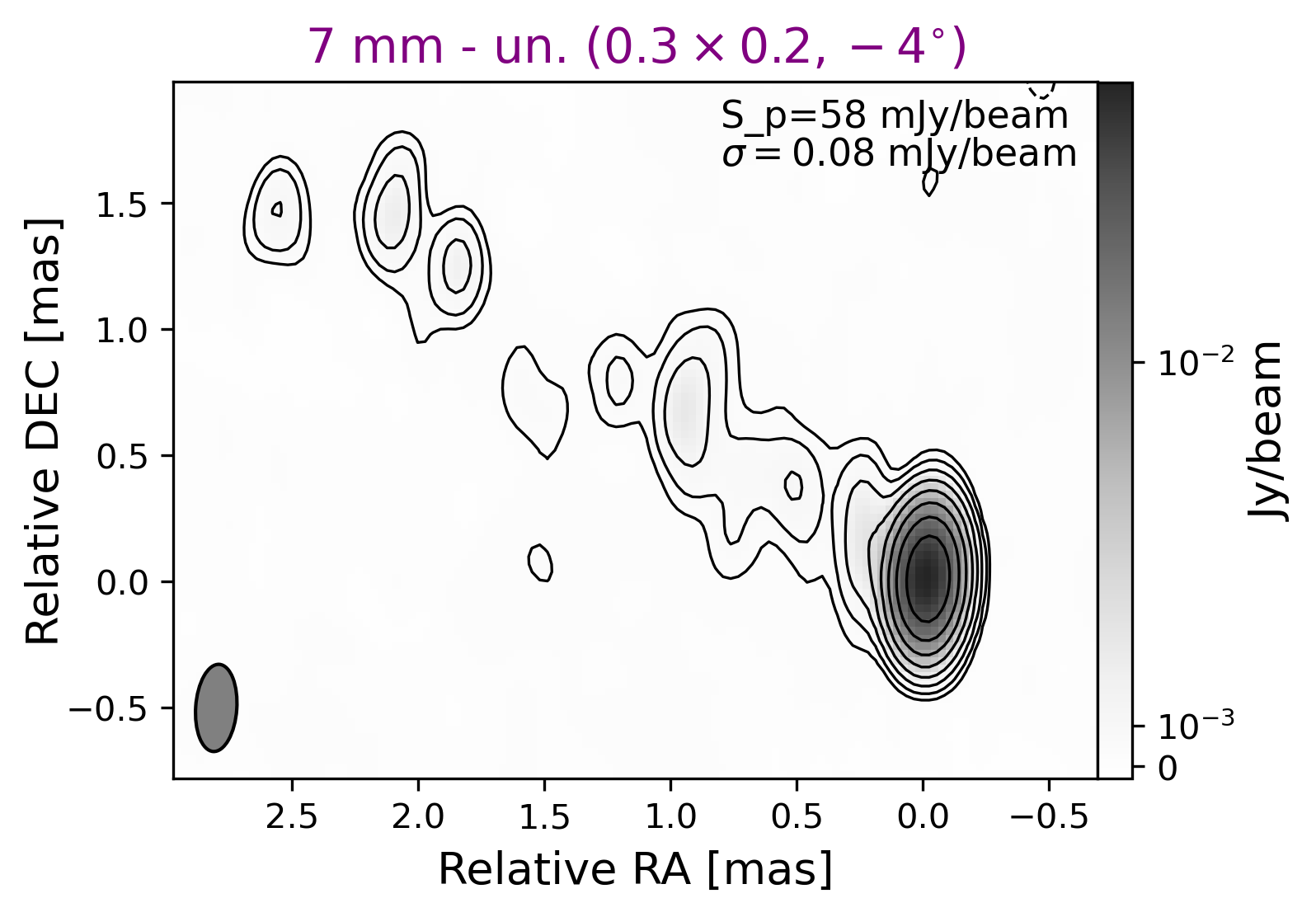}\\
        \includegraphics[width=7.4cm]{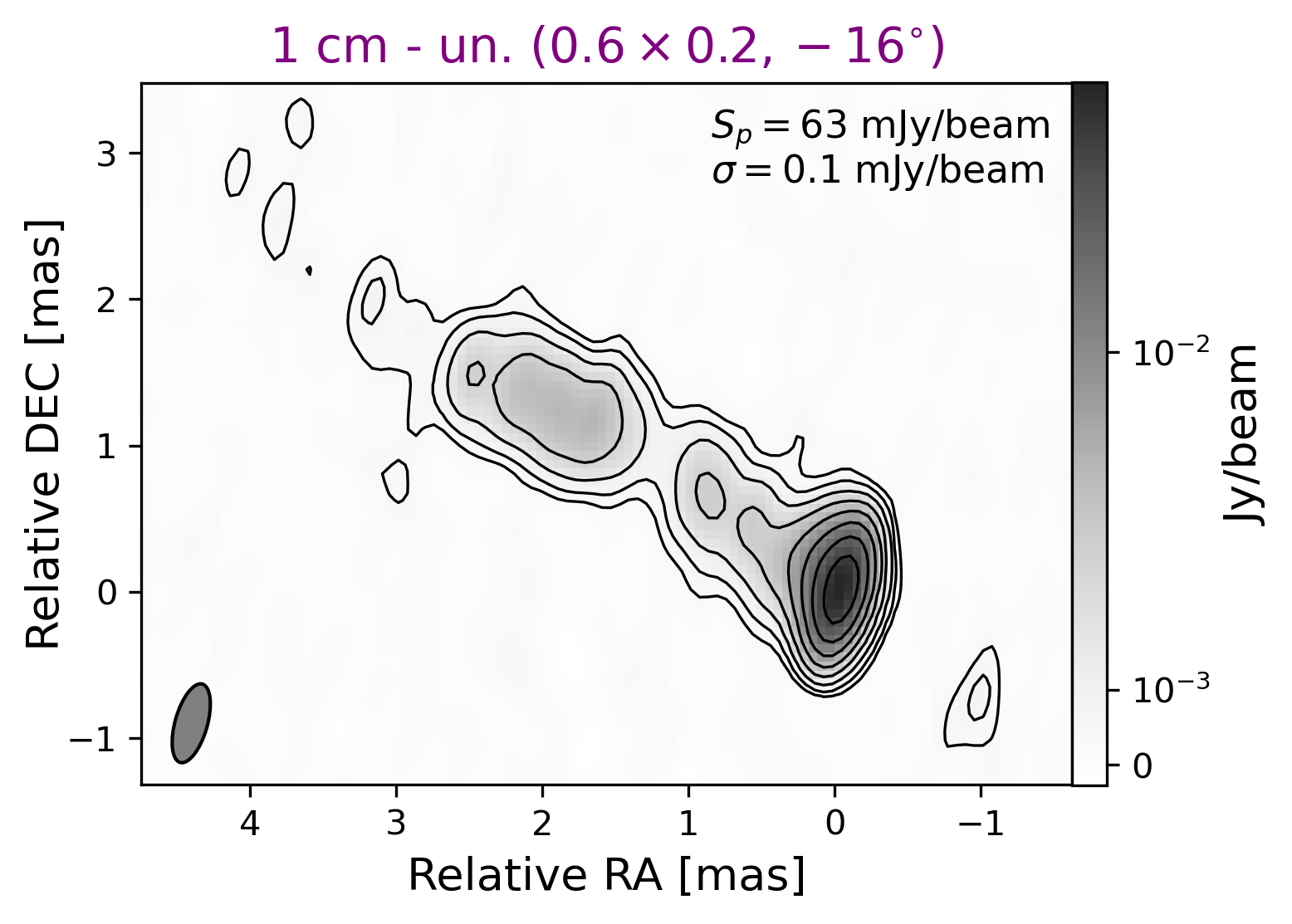}
        \includegraphics[width=7.4cm]{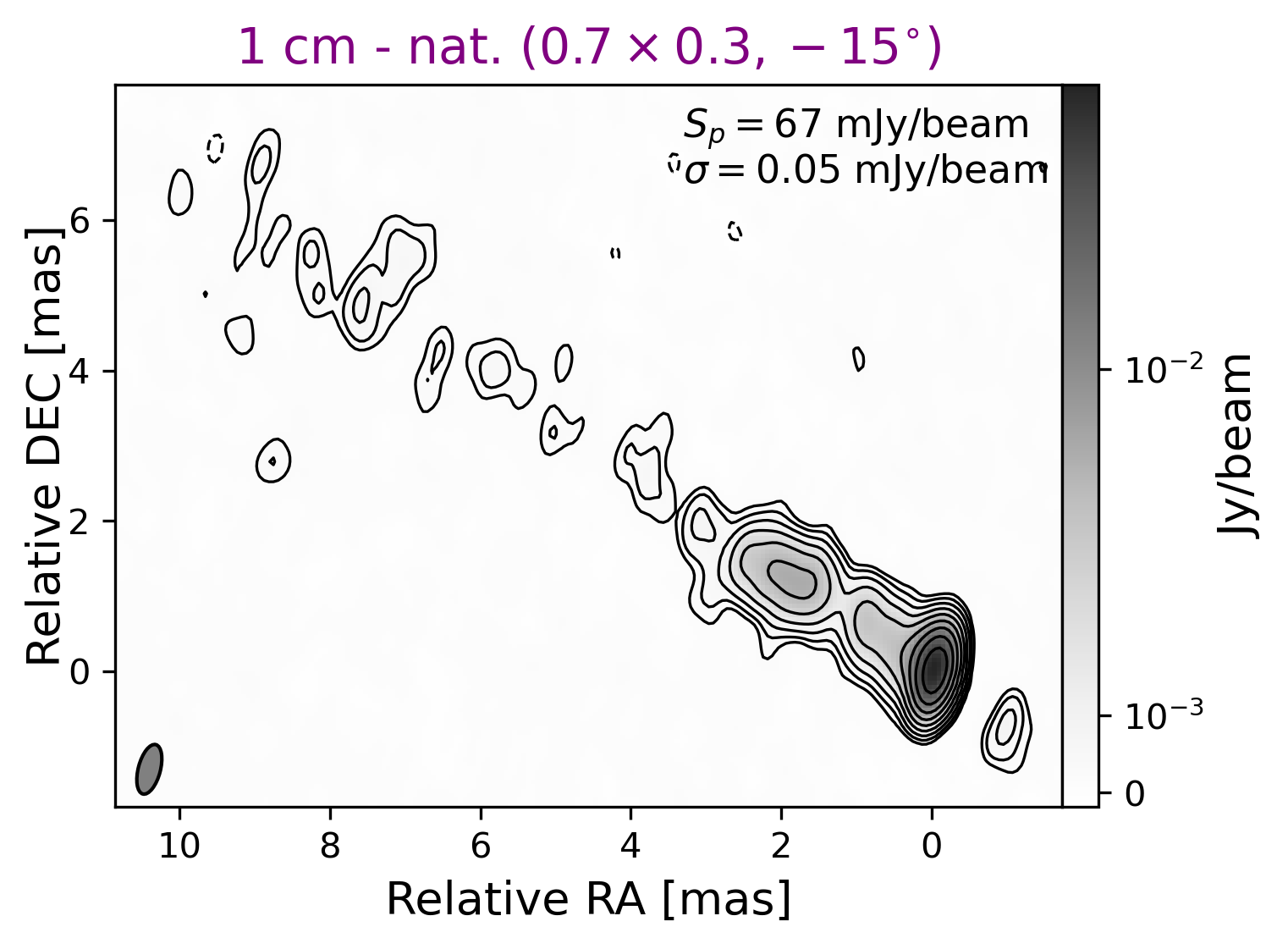}
        \caption{VLBI images of 3C\,382. Top: 7\,mm image with uniform (left) and natural weighting (right). Bottom: 1\,cm image with uniform (left) and natural weighting (right).}
        \label{fig:enter-label}
    \end{figure*}  
    
    \clearpage

        \subsection{3C\,388}
       3C\,388 is a LEG developing an FRII jet, and showing signs of restarted activity based on the detection of extended emission beyond the western hotspot  \citep{Giovannini2005}.  In our VLBI images (Fig. A.13), the source is relatively faint, and appears point-like at 7\,mm. At 1\,cm, we detect a one-sided jet oriented towards south-west, which appears weak but rather extended in the map produced with natural weighting. 

     \begin{figure*}[!h]
        \centering
        \includegraphics[height=4.5cm]{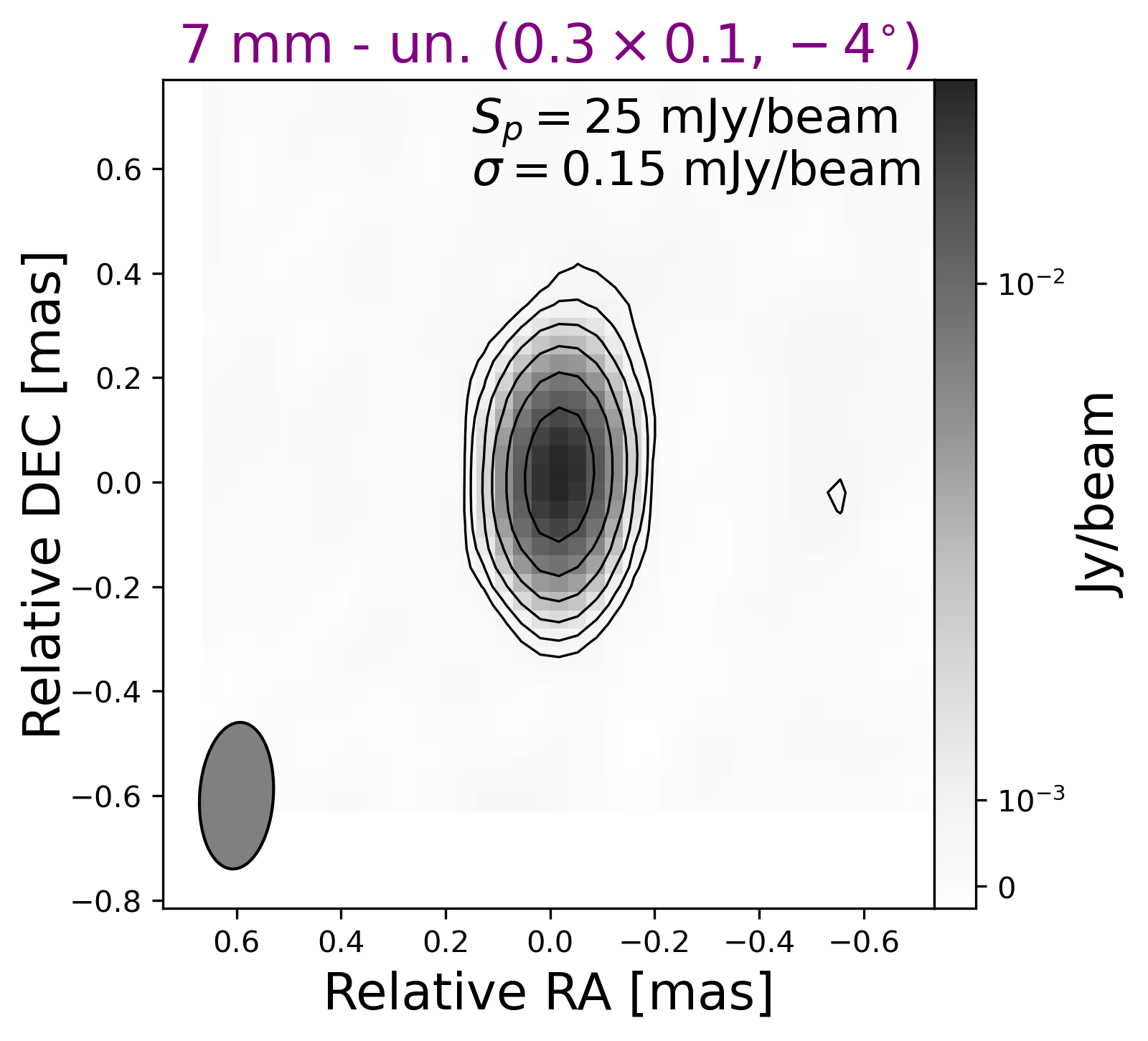}
        \includegraphics[height=4.5cm]{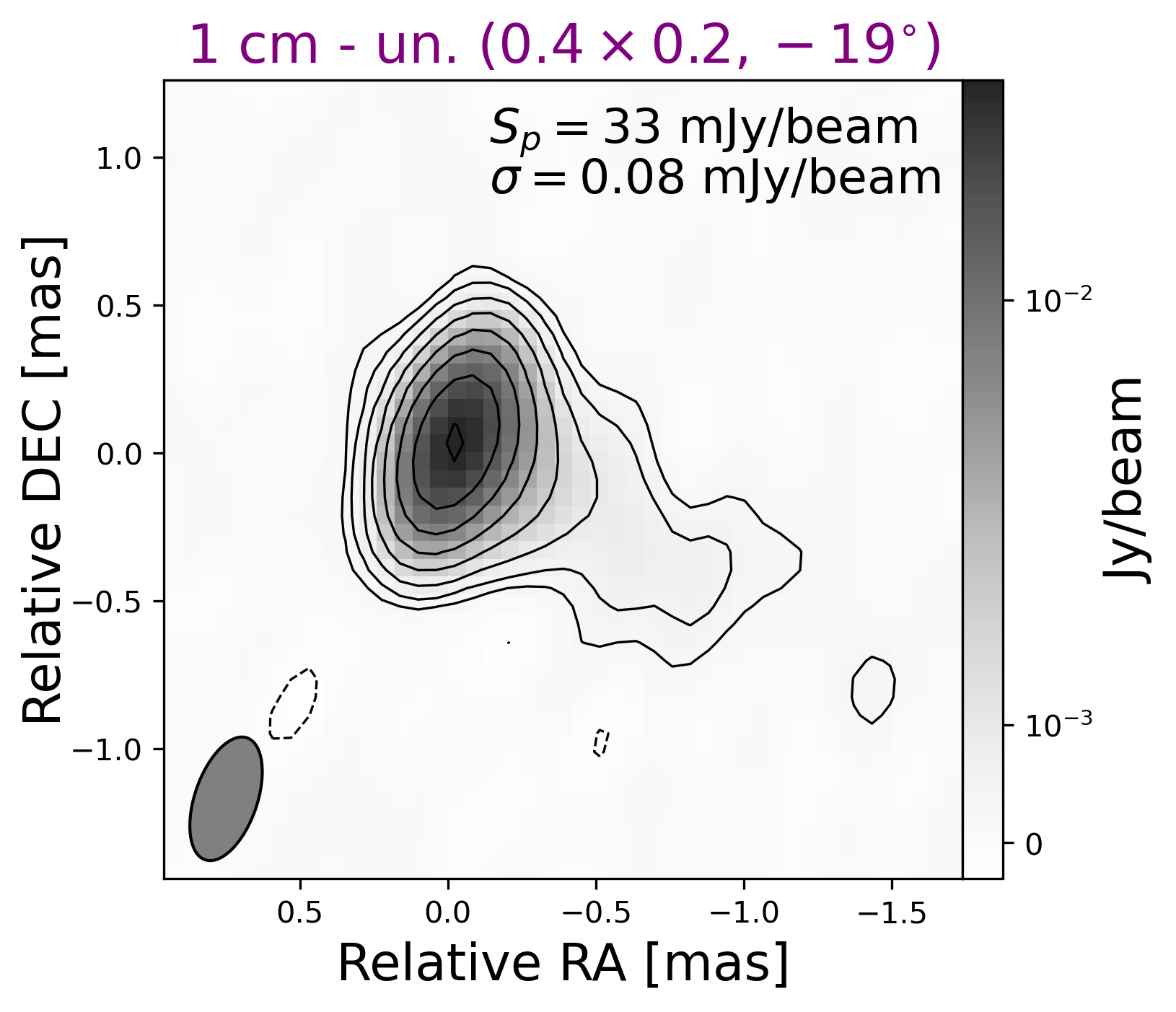}
        \includegraphics[height=4.7cm]{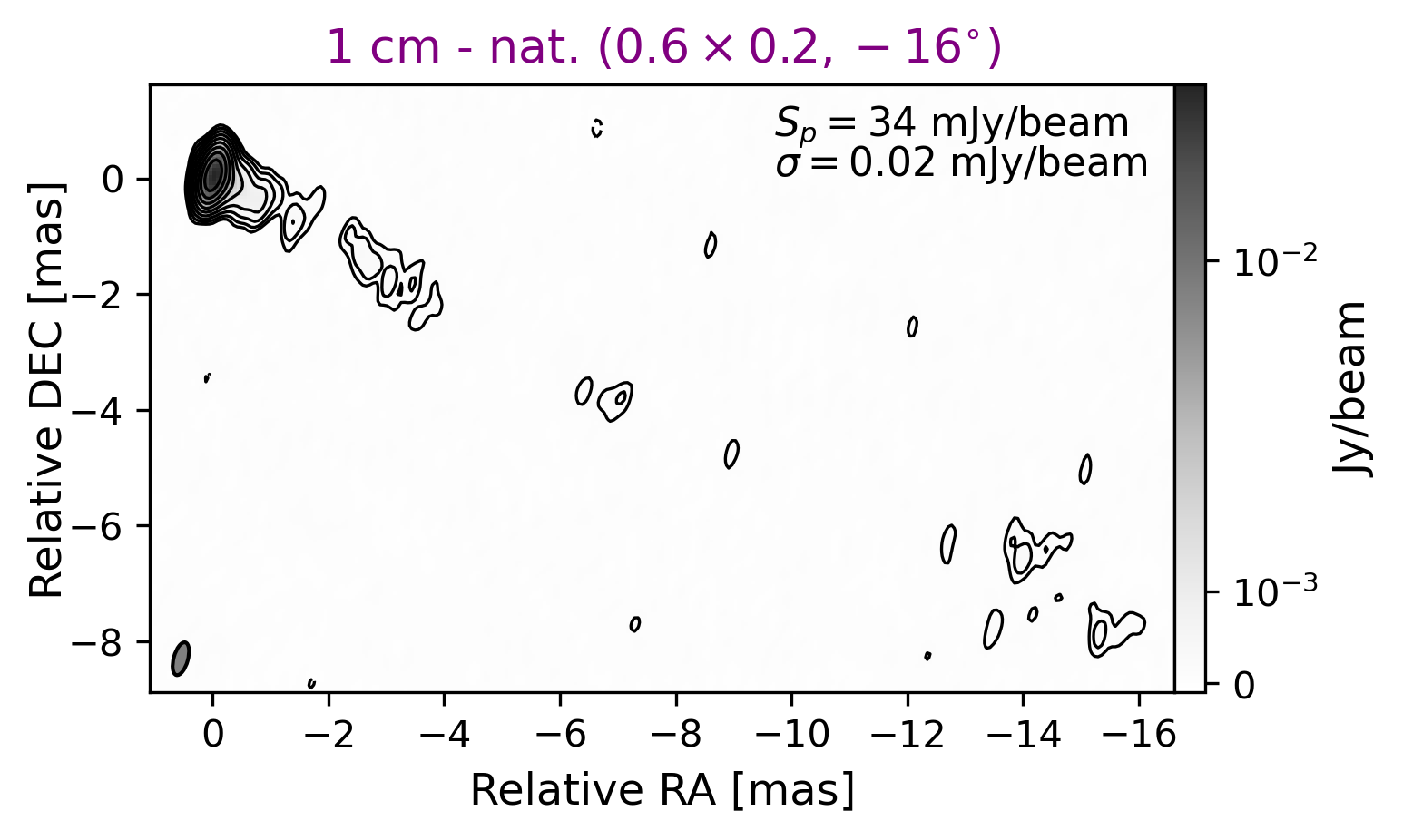}
  \caption{VLBI images of 3C\,388.  Left: 7\,mm image with uniform weighting. Center: 1\,cm image with uniform weighting. Right: 1\,cm image with natural weighting.}
        \label{fig:enter-label}
    \end{figure*}  

     \subsection{3C\,452}
 Among the HEG in our sample, 3C\,452 is the best object to study the properties of nuclear regions in a powerful source. In our images at the two bands (Fig. A.14), the source shows a relatively bright, highly symmetric two-sided jet. The twin-jet structure is oriented in the east-west direction, and appears misaligned with respect to the FRII lobes, which are observed in the north-east/south-west direction \citep{Giovannini2001}. The jets in 3C\,452 are likely oriented close to the plane of the sky, in agreement with the source classification as a narrow-line radio galaxy, i.e., harbouring an obscured nucleus \citep[e.g.,][]{Buttiglione2010}.
  
  \begin{figure*}[!h]
        \centering
        \includegraphics[width=8.7cm]{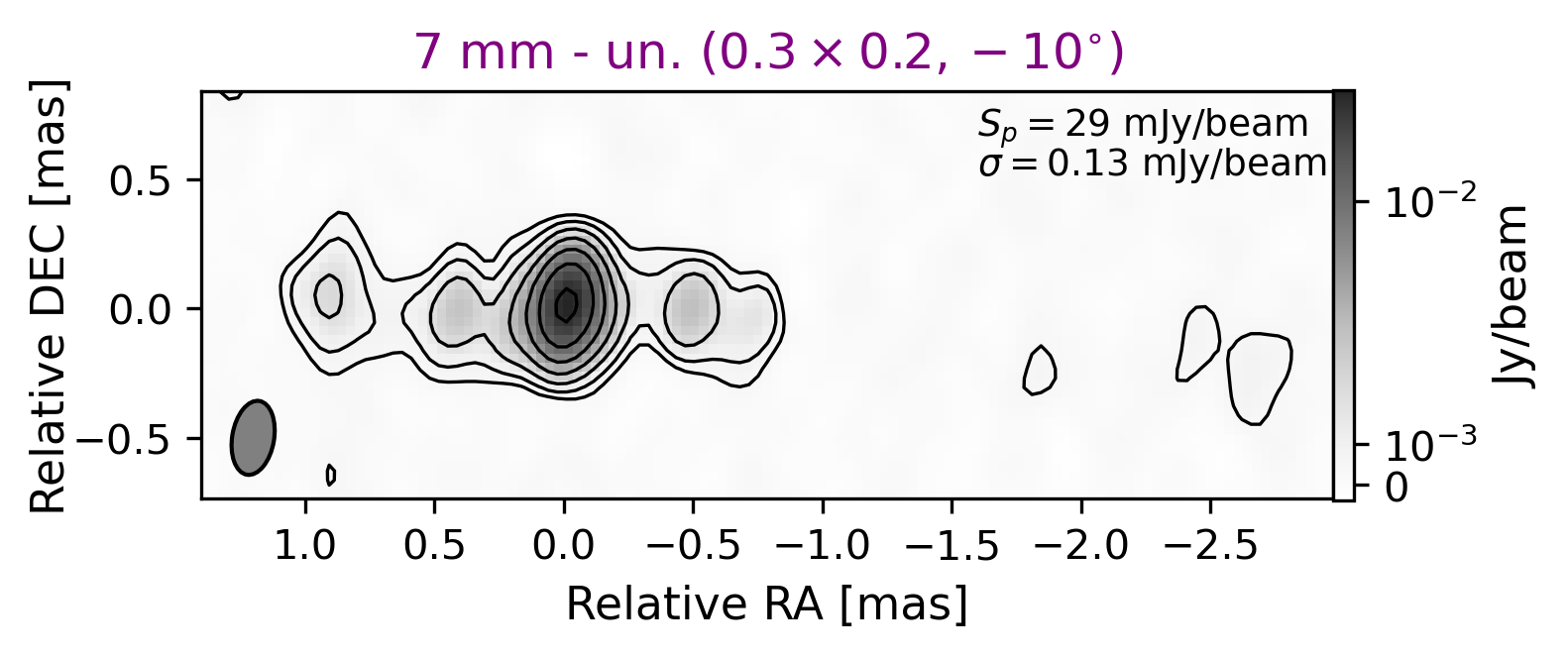}
        \includegraphics[width=8.7cm]{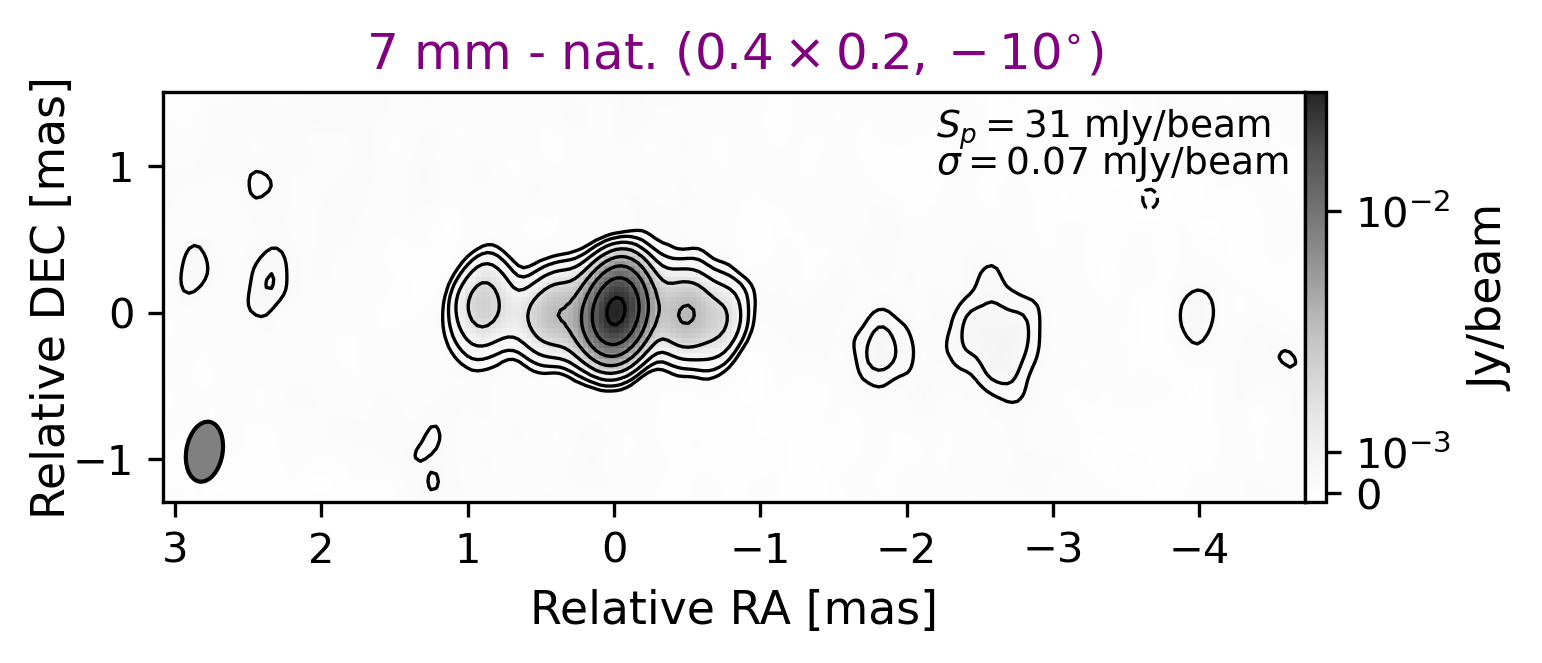}\\
        \includegraphics[width=8.7cm]{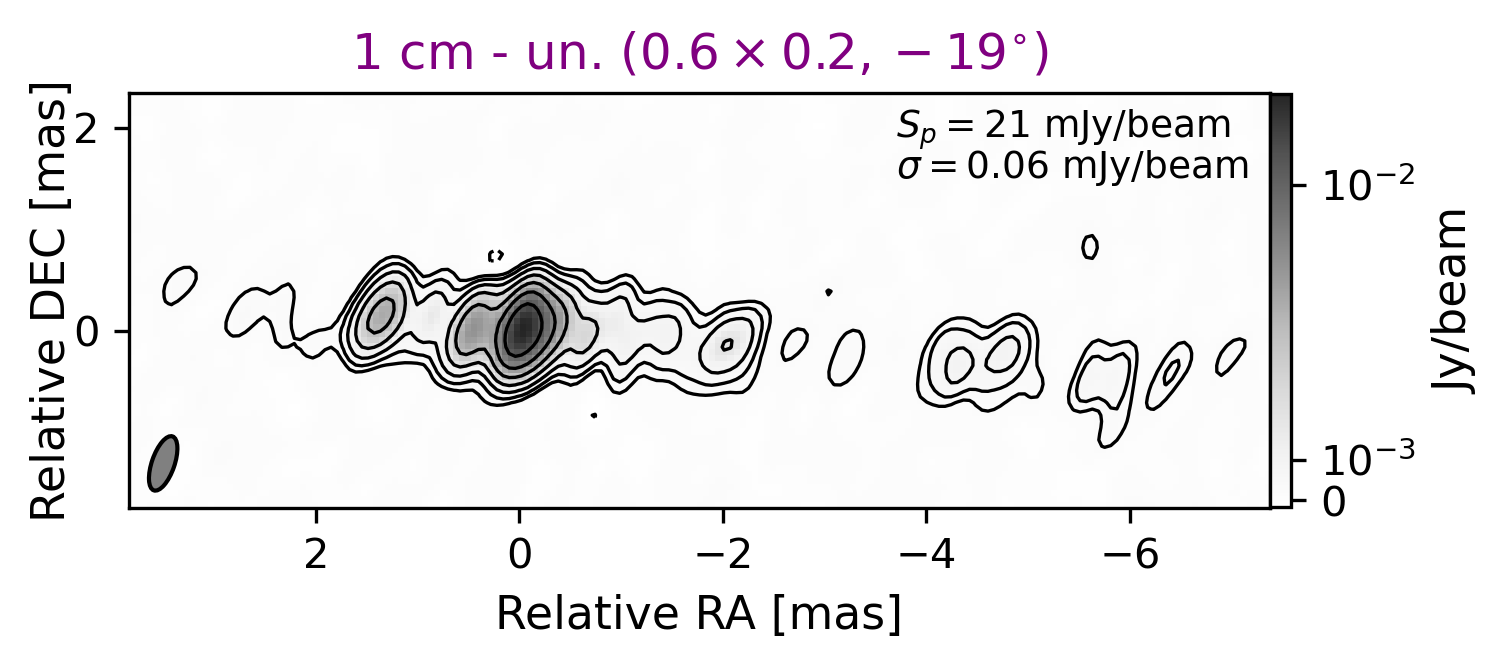}
        \includegraphics[width=8.7cm]{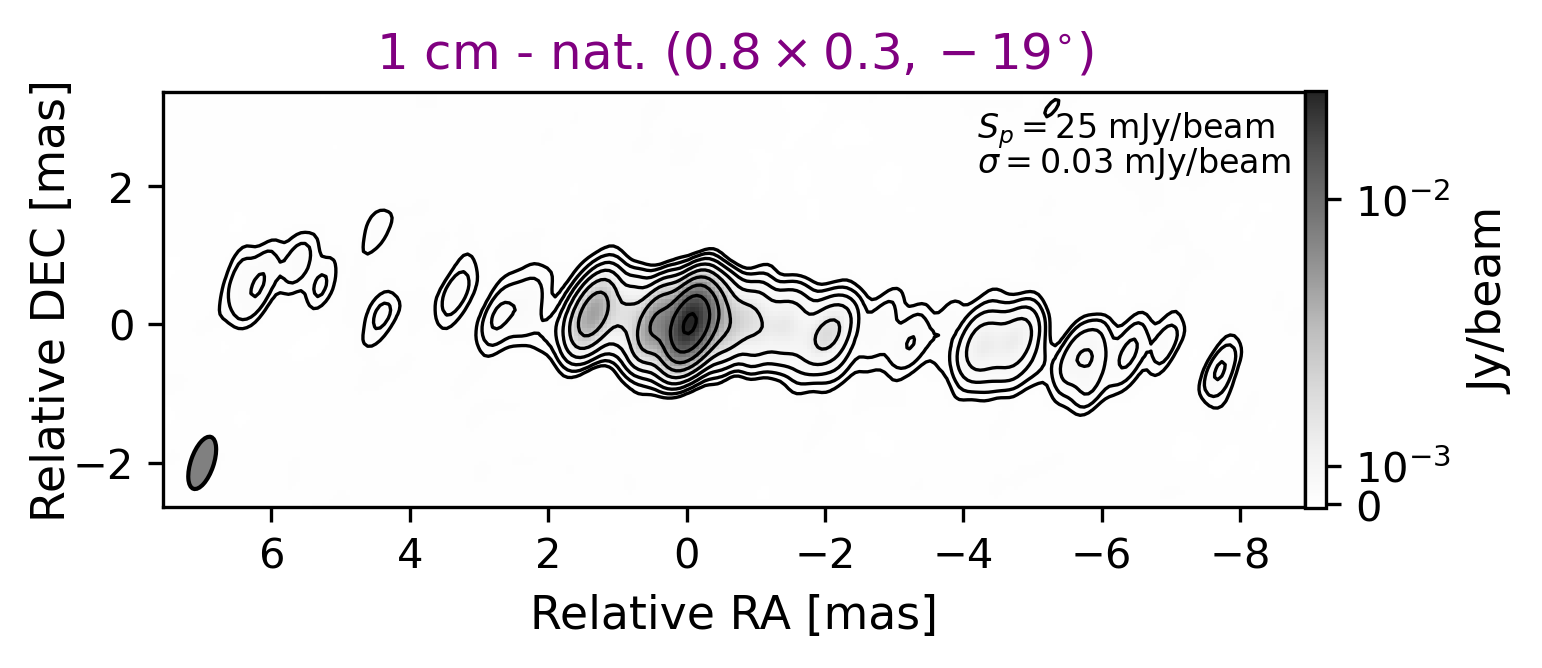}\\
              \caption{VLBI images of 3C\,452. Top: 7\,mm image with uniform (left) and natural weighting (right). Bottom: 1\,cm image with uniform (left) and natural weighting (right).}
        \label{fig:enter-label}
    \end{figure*}
    
    \clearpage
    
   \subsection{3C\,465}
   The FRI-LEG 3C\,465 is a prototypical wide-angle tail radio galaxy \citep{2005MNRAS.359.1007H} . Its jet is one of the best resolved in the sample, since an angular resolution of $0.1$\,$\rm mas$ corresponds to only $451$\,$R_{\rm S}$. The approaching jet is directed towards north-west, and a short counter-jet is detected at both frequencies (Fig. A.15). We observe jet limb-brightening, especially in the innermost jet regions at 7\,mm.
     \begin{figure*}[!h]
        \centering
       \includegraphics[width=6.4cm]{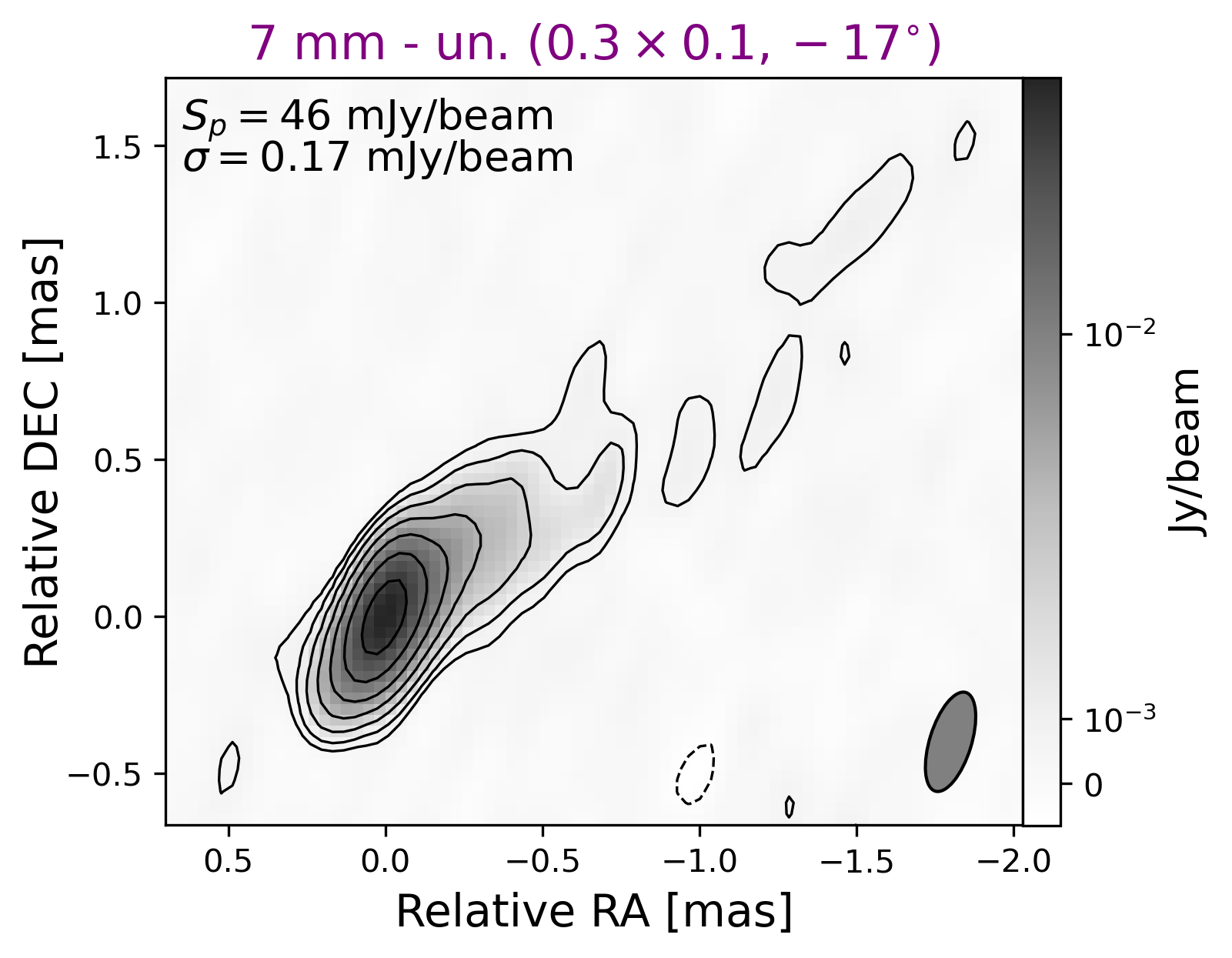}
       \includegraphics[width=6.4cm]{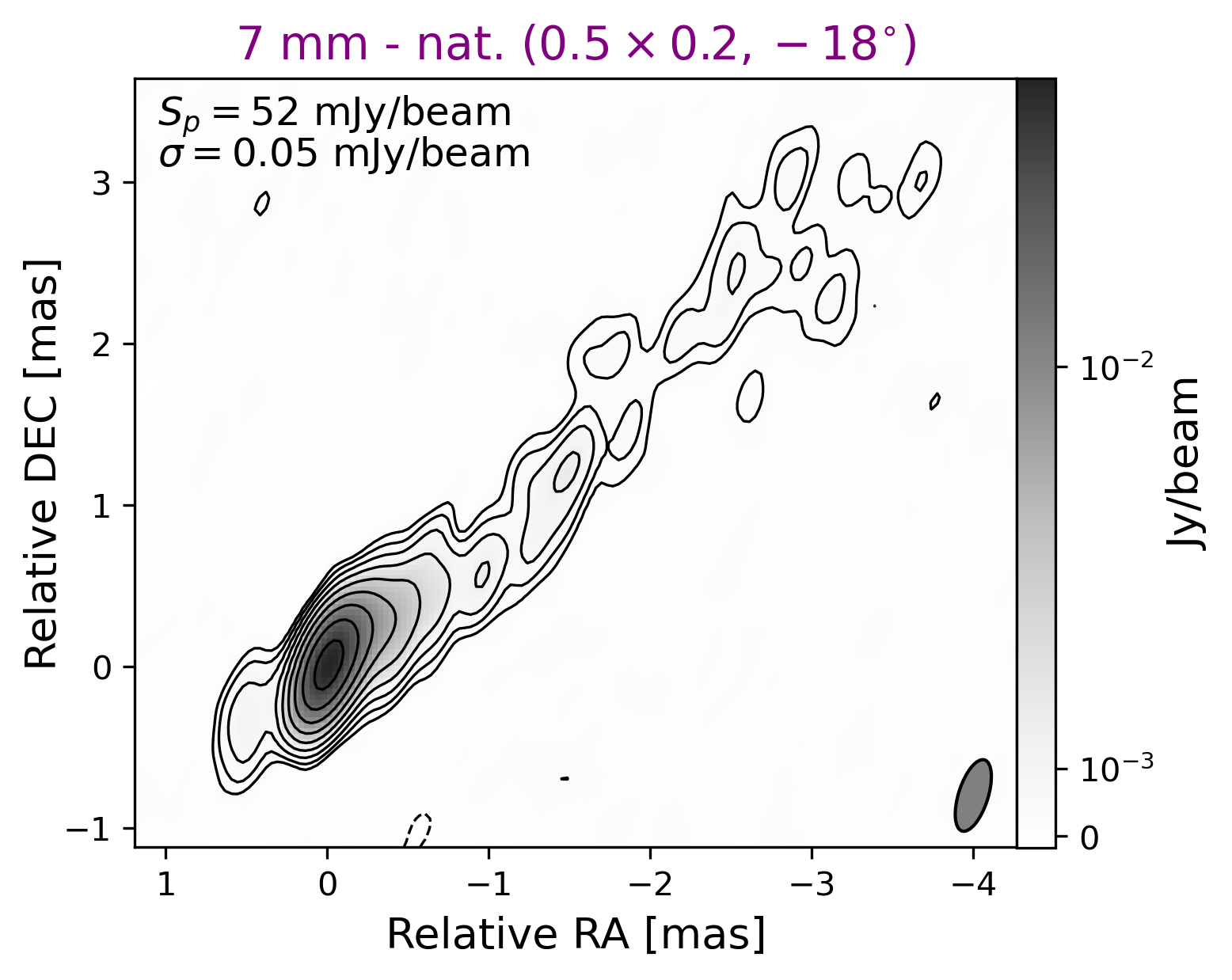}\\
        \includegraphics[width=6.4cm]{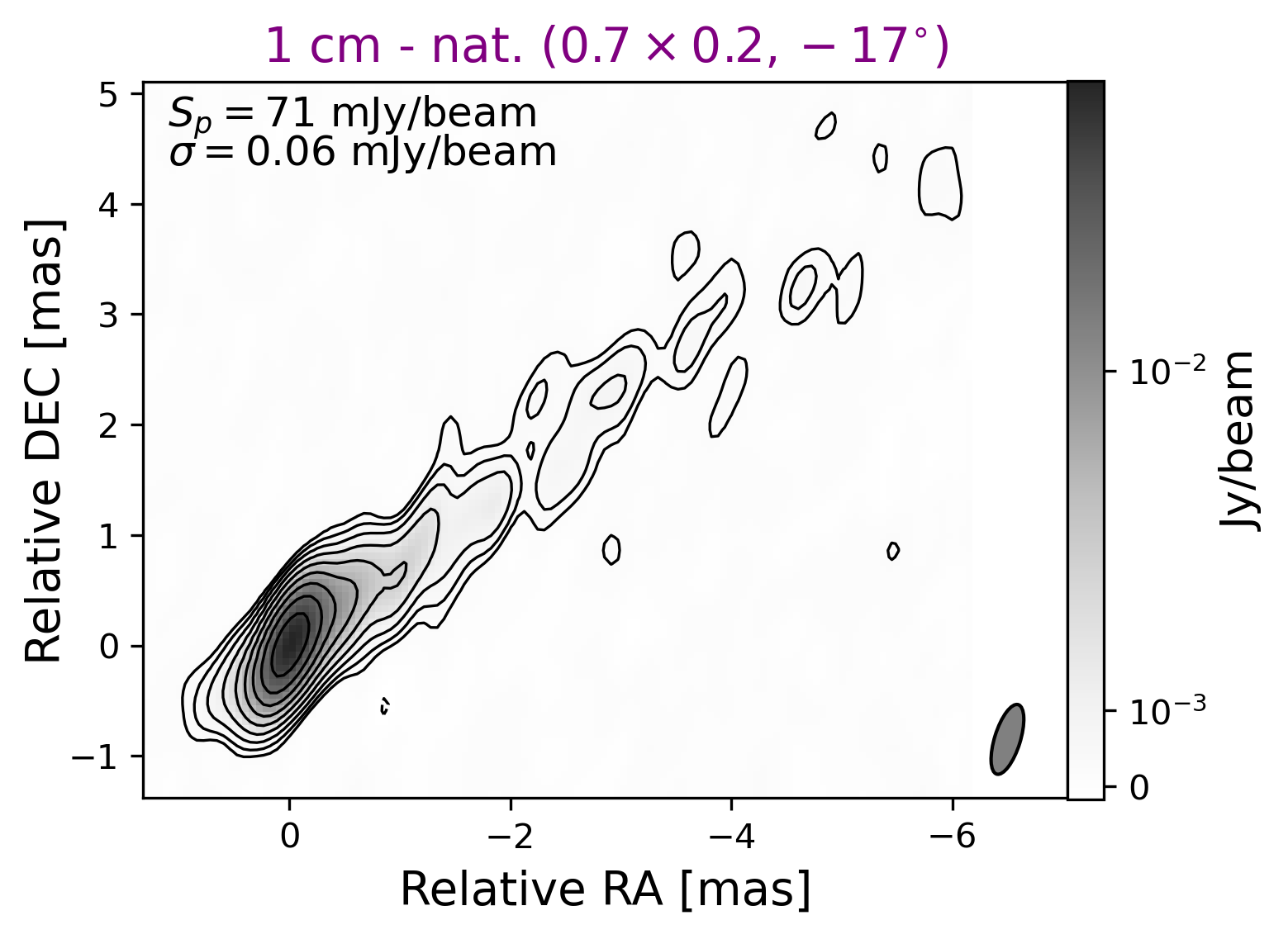}
        \includegraphics[width=6.4cm]{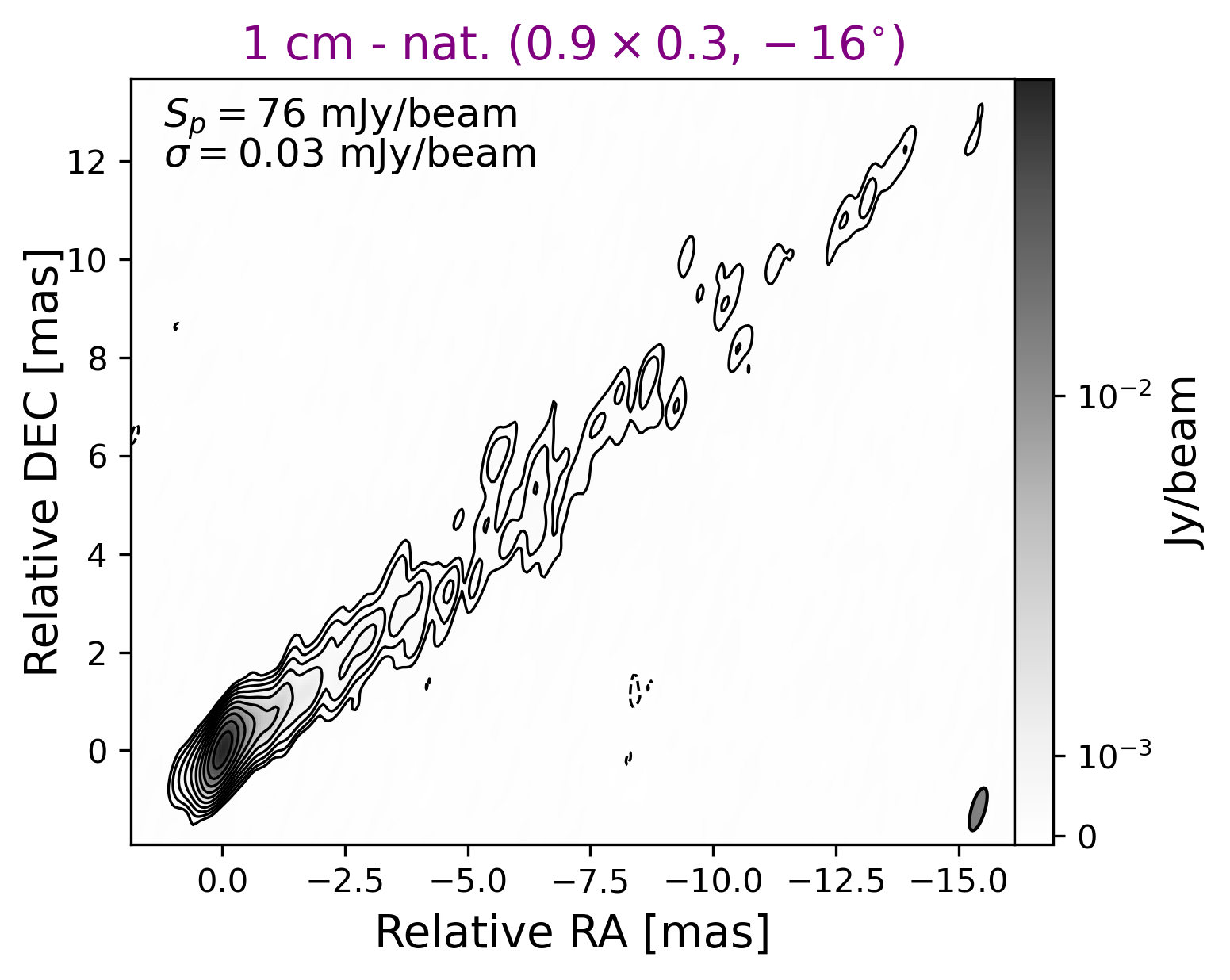}
  \caption{VLBI images of 3C\,465. Top: 7\,mm image with uniform (left) and natural weighting (right). Bottom: 1\,cm image with uniform (left) and natural weighting (right).}
        \label{fig:enter-label}
    \end{figure*}

\section{Modelfit parameters and error analysis}

 In Tables B.1-B.15 we report the parameters of the MODELFIT circular Gaussian components. For uniformity, the MODELFIT was performed in all cases after applying natural weighting to the visibilities, by setting {\it uvweight 0, -2} in DifMap. We computed the statistical errors and the resolution limits based on the S/N of each Gaussian feature, following the approach of \cite{Lee2008}. Specifically, starting from the final MODELFIT map, we have removed the feature of interest and determined its peak intensity $S_{\rm p}^{\rm fit}$, repeating the procedure for all features. We have also determined the post-fit rms $\sigma^{\rm fit}$ of the MODELFIT map, which is typically higher than the CLEAN map rms. The uncertainties on the peak intensity $S_{\rm p}^{\rm fit}$ and on the Gaussian parameters, i.e. integrated flux density $S$, size $d$, radial distance $r$, and position angle $pa$ are derived, respectively, through the following expressions:

\begin{equation}
\Delta S_{\rm p}^{\rm fit}=\sigma^{\rm fit}\left(1+\frac{S_{\rm p}^{\rm fit}}{{\sigma^{\rm fit}}}\right)^{1/2}
\end{equation}

\begin{equation}
\Delta S=\Delta S_{\rm p}^{\rm fit}\left(1+\frac{S^2}{{S_{\rm p}^{\rm fit}}^2}\right)^{1/2}
\end{equation}

\begin{equation}
\Delta d=d\frac{\Delta S_{\rm p}^{\rm fit}}{S_{\rm p}^{\rm fit}}
\end{equation}

\begin{equation}
\Delta r=\frac{\Delta d}{2}
\end{equation}

\begin{equation}
\Delta pa=\arctan\left({\frac{\Delta r}{r}}\right)\cdot 180/\pi
\end{equation}

 The S/N of each Gaussian feature is given by the ratio $S_{\rm p}^{\rm fit}/\sigma^{\rm fit}$, and is used, together with the information on the major and minor beam axes $b_{\rm maj}$ and $b_{\rm min}$, to determine the resolution limit $d_{\rm lim}$. Its expression, valid in the case of natural weighting, is:

\begin{equation}
d_{\rm lim}=\frac{2}{\pi}\left[2\pi b_{\rm maj} b_{\rm min} \ln 2\ln\frac{S/N}{S/N-1}\right]
\end{equation}

 When the size $d$ of a feature was found to be smaller than $d_{lim}$, the latter was assumed in place of $d$ for the calculating the uncertainties through Equations B.1-B.5. Moreover, an upper limit equal to  $d_{lim}$ was reported for the size in Tables B.1-B.15.

In the context of the analysis presented in this paper, the resolution limit is most relevant for determining wether the core component is resolved, and thus its brightness temperature can be calculated. The results of this analysis are presented in Table B.16. The core components were found to be resolved in all cases except one, i.e., in the map of 3C\,33 at 7\,mm. For this we derived a lower limit for the brightness temperature by assuming a core size equal to the resolution limit.

\vspace{1cm}


\begin{table*}[!h]
\centering
\caption{Modelfit parameters of the circular Gaussian features describing the 1\,cm and 7\,mm  VLBI structure of 3C\,31.}
\begin{tabular}{ccccc}
\hline
\hline
1\,cm& $S$   & $r$  &$pa$  & $d$  \\
          & $\mathrm {[mJy]}$& $\mathrm {[mas]}$&  $\rm{[deg]}$ &  $\mathrm {[mas]}$  \\
           \hline
&	0.59	 $\pm$ 	0.26	&	1.093	 $\pm$ 	0.034	&	159.78	 $\pm$ 	1.78	&	0.200	 $\pm$ 	0.068	\\
&	0.77	 $\pm$ 	0.29	&	0.603	 $\pm$ 	0.016	&	165.05	 $\pm$ 	1.50	&	$<0.122$	\\
&     53.55	 $\pm$ 	2.32	&	0.000	 $\pm$ 	0.001	&	         -                              &	0.033	 $\pm$ 	0.001	\\
&     14.59	 $\pm$ 	1.21	&	0.251	 $\pm$ 	0.002	&	-11.63	 $\pm$ 	0.36	&	0.053	 $\pm$ 	0.003	\\
&	3.84	 $\pm$ 	0.64	&	0.735	 $\pm$ 	0.017	&	-7.17	          $\pm$ 	1.31	&	0.252	 $\pm$ 	0.034	\\
&	1.20	 $\pm$ 	0.36	&	1.728	 $\pm$ 	0.020	&	-12.06	 $\pm$ 	0.66	&	0.175	 $\pm$ 	0.040	\\
&	1.22	 $\pm$ 	0.37	&	2.505	 $\pm$ 	0.048	&	-12.70	 $\pm$ 	1.09	&	0.372	 $\pm$ 	0.095	\\
&	0.91	 $\pm$ 	0.36	&	5.171	 $\pm$ 	0.181	&	-12.27	 $\pm$ 	2.01	&	1.033	 $\pm$ 	0.363	\\
&	0.62	 $\pm$ 	0.29	&	9.837	 $\pm$ 	0.189	&	-12.37	 $\pm$ 	1.10	&	0.921	 $\pm$ 	0.378	\\
           \hline
7\,mm&	34.51 $\pm$	2.43	&	0.000	 $\pm$	0.001	&	           -      	        &	0.016	 $\pm$	0.001	\\
&	21.07 $\pm$	1.90	&	0.070	 $\pm$	0.002	&	1.41	 $\pm$	1.33	&	0.050	 $\pm$	0.003	\\
&	4.02	 $\pm$	0.85	&	0.361	 $\pm$	0.013	&	-12.17 $\pm$	2.12	&	0.159	 $\pm$	0.027	\\
&	2.38	 $\pm$	0.71	&	0.896	 $\pm$	0.058	&	-7.51	 $\pm$	3.71	&	0.442	 $\pm$	0.116	\\
&	1.23	 $\pm$	0.55	&	2.597	 $\pm$	0.146	&	-9.35	 $\pm$	3.21	&	0.732	 $\pm$	0.292	\\
     \hline
\end{tabular}
\label{table:3C31}
\end{table*} 

\begin{table*}[!h]
\centering
\caption{Modelfit parameters of the circular Gaussian features describing the 1\,cm and 7\,mm  VLBI structure of 3C\,33.}
\begin{tabular}{ccccc}
\hline
\hline
1\,cm& $ S$   & $ r$  &$ pa$  & $ d$  \\
          & $\mathrm {[mJy]}$& $\mathrm {[mas]}$&  $\rm{[deg]}$ &  $\mathrm {[mas]}$  \\
           \hline
&	0.74	 $\pm$ 	0.27	&	3.751	 $\pm$ 	0.119	&	22.42	 $\pm$ 	1.82	&	0.755	 $\pm$ 	0.238	\\
&	0.27	 $\pm$ 	0.16	&	1.488	 $\pm$ 	0.048	&	19.39	 $\pm$ 	1.85	&	$<0.238$	\\
&	1.53	 $\pm$ 	0.36	&	0.504	 $\pm$ 	0.025	&	20.49	 $\pm$ 	2.82	&	0.275	 $\pm$ 	0.050	\\
&	12.55 $\pm$ 	1.00	&	0.000	 $\pm$ 	0.003	&	          -                               &	0.107	 $\pm$ 	0.006	\\
&	0.63	 $\pm$ 	0.25	&	1.349	 $\pm$ 	0.029	&	201.02	 $\pm$ 	1.22	&	0.293	 $\pm$ 	0.057	\\
&	1.27	 $\pm$ 	0.39	&	0.632	 $\pm$ 	0.035	&	200.33	 $\pm$ 	3.15	&	0.244	 $\pm$ 	0.070	\\
&	0.44	 $\pm$ 	0.20	&	3.951	 $\pm$ 	0.031	&	201.05	 $\pm$ 	0.44	&	$<0.192$	\\
           \hline
7\,mm&	18.58	 $\pm$	2.03	&	0.000	 $\pm$	0.001	&	           -      	&	$<0.033$	\\
     \hline
\end{tabular}
\label{table:3C33}
\end{table*} 

\begin{table*}[]
\centering
\caption{Modelfit parameters of the circular Gaussian features describing the 1\,cm and 7\,mm  VLBI structure of 3C\,66B.}
\begin{tabular}{ccccc}
\hline
\hline
1\,cm& $ S$   & $ r$  &$ pa$  & $ d$  \\
          & $\mathrm {[mJy]}$& $\mathrm {[mas]}$&  $\rm{[deg]}$ &  $\mathrm {[mas]}$  \\
           \hline
&	6.16	 $\pm$ 	0.73	&	0.264	 $\pm$ 	0.012	&	226.71	 $\pm$ 	2.66	&	0.260	 $\pm$ 	0.025	\\
&	69.78 $\pm$ 	2.42	&	0.000	 $\pm$ 	0.001	&	          -      	                 &	0.066	 $\pm$ 	0.002	\\
&	34.98 $\pm$ 	1.72	&	0.157	 $\pm$ 	0.002	&	49.94	 $\pm$ 	0.61	&	0.094	 $\pm$ 	0.003	\\
&	12.20 $\pm$ 	1.02	&	0.464	 $\pm$ 	0.007	&	55.48	 $\pm$ 	0.81	&	0.203	 $\pm$ 	0.013	\\
&	5.45	 $\pm$ 	0.68	&	0.745	 $\pm$ 	0.008	&	53.55	 $\pm$ 	0.59	&	0.163	 $\pm$ 	0.015	\\
&	6.82	 $\pm$ 	0.78	&	1.032	 $\pm$ 	0.013	&	52.01	 $\pm$ 	0.70	&	0.279	 $\pm$ 	0.025	\\
&	4.03	 $\pm$ 	0.63	&	1.625	 $\pm$ 	0.033	&	53.99	 $\pm$ 	1.16	&	0.495	 $\pm$ 	0.066	\\
&	4.52	 $\pm$ 	0.68	&	2.344	 $\pm$ 	0.041	&	56.75	 $\pm$ 	0.99	&	0.617	 $\pm$ 	0.081	\\
&	3.37	 $\pm$ 	0.76	&	5.035	 $\pm$ 	0.213	&	54.71	 $\pm$ 	2.42	&	1.971	 $\pm$ 	0.426	\\
&	0.77	 $\pm$ 	0.27	&	8.952	 $\pm$ 	0.039	&	55.91	 $\pm$ 	0.25	&	0.279	 $\pm$ 	0.078	\\
           \hline
7\,mm&	75.28 $\pm$	6.75	&	0.000	 $\pm$	0.003	&	           -                       	&	0.087	 $\pm$	0.006	\\
&	17.31 $\pm$	3.28	&	0.174	 $\pm$	0.009	&	55.16	 $\pm$	2.81	&	0.118	 $\pm$	0.017	\\
&	12.89 $\pm$	3.10	&	0.619	 $\pm$	0.036	&	53.68	 $\pm$	3.33	&	0.340	 $\pm$	0.072	\\
&	5.33	 $\pm$	2.01	&	1.240	 $\pm$	0.056	&	50.11	 $\pm$	2.57	&	0.343	 $\pm$	0.111	\\
&	3.71	 $\pm$	1.79	&	2.077	 $\pm$	0.124	&	58.95	 $\pm$	3.41	&	0.578	 $\pm$	0.248	\\
     \hline
\end{tabular}
\label{table:3C66B}
\end{table*} 

\begin{table*}[]
\centering
\caption{Modelfit parameters of the circular Gaussian features describing the 1\,cm and 7\,mm  VLBI structure of B2\,0222+36.}
\begin{tabular}{ccccc}
\hline
\hline
1\,cm& $ S$   & $ r$  &$ pa$  & $ d$  \\
          & $\mathrm {[mJy]}$& $\mathrm {[mas]}$&  $\rm{[deg]}$ &  $\mathrm {[mas]}$  \\
           \hline
&	1.86	 $\pm$ 	0.45	&	1.798	 $\pm$ 	0.060	&	-18.05	 $\pm$ 	1.91	&	0.622	 $\pm$ 	0.120	\\
&	6.89	 $\pm$ 	0.83	&	0.560	 $\pm$ 	0.009	&	-21.58	 $\pm$ 	0.95	&	0.209	 $\pm$ 	0.019	\\
&	23.84 $\pm$ 	1.55	&	0.000	 $\pm$ 	0.002	&	           -     	                 &	0.095	 $\pm$ 	0.004	\\
&	7.04	 $\pm$ 	0.84	&	0.696	 $\pm$ 	0.006	&	156.17	 $\pm$ 	0.46	&	0.131	 $\pm$ 	0.011	\\
&	5.46	 $\pm$ 	0.74	&	1.185	 $\pm$ 	0.013	&	157.19	 $\pm$ 	0.64	&	0.261	 $\pm$ 	0.026	\\
&	0.25	 $\pm$ 	0.17	&	3.540	 $\pm$ 	0.067	&	162.34	 $\pm$ 	1.08	&	$<0.297$	\\
           \hline
7\,mm&	0.42	 $\pm$	0.39	&	1.952	 $\pm$	0.246	&	-13.15	 $\pm$	7.18	&	0.801	 $\pm$	0.492	\\
&	4.61	 $\pm$	1.18	&	0.487	 $\pm$	0.027	&	-21.34	 $\pm$	3.14	&	0.260	 $\pm$	0.054	\\
&	17.68 $\pm$	2.24	&	0.000	 $\pm$	0.004	&	           -                       	&	0.080	 $\pm$	0.007	\\
&	3.37	 $\pm$	1.02	&	0.786	 $\pm$	0.036	&	153.68	 $\pm$	2.60	&	0.290	 $\pm$	0.071	\\
&	1.35	 $\pm$	0.64	&	1.254	 $\pm$	0.021	&	151.97	 $\pm$	0.95	&	$<0.128$                            	\\
     \hline
\end{tabular}
\label{table:B20222+36}
\end{table*} 

\begin{table*}[]
\centering
\caption{Modelfit parameters of the circular Gaussian features describing the 1\,cm and 7\,mm  VLBI structure of IC\,310.}
\begin{tabular}{ccccc}
\hline
\hline
1\,cm& $ S$   & $ r$  &$ pa$  & $ d$  \\
          & $\mathrm {[mJy]}$& $\mathrm {[mas]}$&  $\rm{[deg]}$ &  $\mathrm {[mas]}$  \\
           \hline 
&	40.29  $\pm$ 	2.29	&	0.000	 $\pm$ 	0.001	&	           -     	                 &	0.039	 $\pm$ 	0.002	\\
&	16.16 $\pm$ 	1.45	&	0.245	 $\pm$ 	0.004	&	234.21	 $\pm$ 	0.94	&	0.122	 $\pm$ 	0.008	\\
&	5.97	 $\pm$ 	0.90	&	0.626	 $\pm$ 	0.015	&	235.20	 $\pm$ 	1.38	&	0.254	 $\pm$ 	0.030	\\
&	5.35	 $\pm$ 	0.95	&	1.262	 $\pm$ 	0.045	&	230.35	 $\pm$ 	2.03	&	0.559	 $\pm$ 	0.089	\\
&	2.75	 $\pm$ 	0.68	&	2.446	 $\pm$ 	0.060	&	226.67	 $\pm$ 	1.41	&	0.544	 $\pm$ 	0.121	\\
&	1.16	 $\pm$ 	0.42	&	3.640	 $\pm$ 	0.049	&	220.62	 $\pm$ 	0.78	&	0.337	 $\pm$ 	0.099	\\
           \hline
7\,mm&	36.04 $\pm$	5.33	&	0.000	 $\pm$	0.003	&	           -                      	&	0.065	 $\pm$	0.007	\\
&	4.10	 $\pm$	1.87	&	0.279	 $\pm$	0.018	&	234.97	 $\pm$	3.66	&     $<0.111$	\\
&	3.32	 $\pm$	1.70	&	0.561	 $\pm$	0.024	&	232.77	 $\pm$	2.42	&	$<0.125$	\\
&	7.23	 $\pm$	3.44	&	1.432	 $\pm$	0.138	&	224.52	 $\pm$	5.52	&	0.609	 $\pm$	0.277	\\
     \hline
\end{tabular}
\label{table:IC310}
\end{table*}

\begin{table*}[]
\centering
\caption{Modelfit parameters of the circular Gaussian features describing the 1\,cm and 7\,mm  VLBI structure of 4C\,39.12.}
\begin{tabular}{ccccc}
\hline
\hline
1\,cm& $ S$   & $ r$  &$ pa$  & $ d$  \\
          & $\mathrm {[mJy]}$& $\mathrm {[mas]}$&  $\rm{[deg]}$ &  $\mathrm {[mas]}$  \\
           \hline
&	38.41$\pm$ 	2.00	&	0.000	 $\pm$ 	0.001	&	          -      	                 &	0.050	 $\pm$ 	0.002	\\
&	18.65 $\pm$ 	1.39	&	0.174	 $\pm$ 	0.001	&	169.31	 $\pm$ 	0.31	&	0.035	 $\pm$ 	0.002	\\
&	9.29	 $\pm$ 	0.99	&	0.496	 $\pm$ 	0.004	&	165.87	 $\pm$ 	0.52	&	0.114	 $\pm$ 	0.009	\\
&	3.10	 $\pm$ 	0.58	&	1.063	 $\pm$ 	0.017	&	168.42	 $\pm$ 	0.90	&	0.224	 $\pm$ 	0.033	\\
&	1.53	 $\pm$ 	0.43	&	1.681	 $\pm$ 	0.040	&	167.20	 $\pm$ 	1.36	&	0.341	 $\pm$ 	0.080	\\
&	0.95	 $\pm$ 	0.38	&	3.772	 $\pm$ 	0.178	&	171.79	 $\pm$ 	2.71	&	0.981	 $\pm$ 	0.357	\\
&	1.40	 $\pm$ 	0.55	&	9.621	 $\pm$ 	0.387	&	165.19	 $\pm$ 	2.30	&	2.047	 $\pm$ 	0.774	\\
&	1.45	 $\pm$ 	0.55	&	6.044	 $\pm$ 	0.418	&	169.24	 $\pm$ 	3.96	&	2.302	 $\pm$ 	0.837	\\
           \hline
7\,mm&	27.12 $\pm$	2.45	&	0.000	 $\pm$	0.001	&	           -     	                 &	0.026	 $\pm$	0.002	\\
&	27.34 $\pm$	2.46	&	0.067	 $\pm$	0.001	&	17.08	 $\pm$	0.94	&	0.034	 $\pm$	0.002	\\
&	7.50	 $\pm$	1.31	&	0.252	 $\pm$	0.008	&	164.03	 $\pm$	1.80	&	0.116	 $\pm$	0.016	\\
&	4.81	 $\pm$	1.09	&	0.524	 $\pm$	0.018	&	165.99	 $\pm$	1.98	&	0.193	 $\pm$	0.036	\\
&	2.40	 $\pm$	0.88	&	1.386	 $\pm$	0.096	&	170.15	 $\pm$	3.96	&	0.575	 $\pm$	0.192	\\
&	1.71	 $\pm$	0.79	&	2.242	 $\pm$	0.142	&	169.01	 $\pm$	3.62	&	0.660	 $\pm$	0.284	\\
     \hline
\end{tabular}
\label{table:4C39.12}
\end{table*} 

\begin{table*}[]
\centering
\caption{Modelfit parameters of the circular Gaussian features describing the 1\,cm and 7\,mm  VLBI structure of IC\,2402.}
\begin{tabular}{ccccc}
\hline
\hline
1\,cm& $ S$   & $ r$  &$ pa$  & $ d$  \\
          & $\mathrm {[mJy]}$& $\mathrm {[mas]}$&  $\rm{[deg]}$ &  $\mathrm {[mas]}$  \\
           \hline
&	0.22	 $\pm$ 	0.17	&	1.649	 $\pm$ 	0.079	&	163.16	 $\pm$ 	2.74	&	$<0.290$	\\
&	1.03	 $\pm$ 	0.35	&	0.618	 $\pm$ 	0.015	&	164.70	 $\pm$ 	1.43	&	$<0.130$	\\
&	33.59 $\pm$ 	1.92	&	0.000	 $\pm$ 	0.001	&	           -     	                 &	0.029	 $\pm$ 	0.001	\\
&	3.54	 $\pm$ 	0.63	&	0.341	 $\pm$ 	0.008	&	-13.54	 $\pm$ 	1.30	&	0.119	 $\pm$ 	0.015	\\
&	1.68	 $\pm$ 	0.44	&	1.065	 $\pm$ 	0.028	&	-18.06	 $\pm$ 	1.48	&	0.268	 $\pm$ 	0.055	\\
&	0.86	 $\pm$ 	0.36	&	6.379	 $\pm$ 	0.214	&	-15.21	 $\pm$ 	1.92	&	1.157	 $\pm$ 	0.428	\\
&	0.52	 $\pm$ 	0.27	&	11.176	 $\pm$ 	0.121	&	-9.27	         $\pm$ 	0.62	&	0.556	 $\pm$ 	0.241	\\
           \hline
7\,mm&	0.76	 $\pm$	0.36	&	0.571	 $\pm$	0.017	&	159.36	 $\pm$	1.69	&	$<0.097$	\\
&	0.87	 $\pm$	0.39	&	0.256	 $\pm$	0.014	&	168.42	 $\pm$	3.06	&	$<0.089$	\\
&	29.47 $\pm$	2.16	&	0.000	 $\pm$	0.002	&	           -                      	&	0.073	 $\pm$	0.004	\\
&	1.89	 $\pm$	0.58	&	0.424	 $\pm$	0.033	&	-20.34	 $\pm$	4.50	&	0.266	 $\pm$	0.067	\\
&	0.56	 $\pm$	0.32	&	1.044	 $\pm$	0.026	&	-21.41	 $\pm$	1.42	&	$<0.124$	\\
     \hline
\end{tabular}
\label{table:IC2402}
\end{table*} 

\begin{table*}[]
\centering
\caption{Modelfit parameters of the circular Gaussian features describing the 1\,cm and 7\,mm  VLBI structure of 3C\,264. }
\begin{tabular}{ccccc}
\hline
\hline
1\,cm& $ S$   & $ r$  &$ pa$  & $ d$  \\
          & $\mathrm {[mJy]}$& $\mathrm {[mas]}$&  $\rm{[deg]}$ &  $\mathrm {[mas]}$  \\
           \hline
&	9.26	 $\pm$ 	1.58	&	0.363	 $\pm$ 	0.015	&	225.06	 $\pm$ 	2.33	&	$<0.105$	\\
&	94.37 $\pm$ 	4.96	&	0.000	 $\pm$ 	0.002	&	           -     	                 &	0.098	 $\pm$ 	0.004	\\
&	15.88 $\pm$ 	2.05	&	0.293	 $\pm$ 	0.008	&	28.21	 $\pm$ 	1.65	&	0.175	 $\pm$ 	0.017	\\
&	13.63 $\pm$ 	2.00	&	0.955	 $\pm$ 	0.032	&	27.02	 $\pm$ 	1.92	&	0.515	 $\pm$ 	0.064	\\
&	6.35	 $\pm$ 	1.45	&	2.187	 $\pm$ 	0.102	&	25.81	 $\pm$ 	2.68	&	1.002	 $\pm$ 	0.204	\\
&	1.34	 $\pm$ 	0.65	&	3.706	 $\pm$ 	0.130	&	22.48	 $\pm$ 	2.01	&	0.656	 $\pm$ 	0.260	\\
&	2.60	 $\pm$ 	0.92	&	5.155	 $\pm$ 	0.138	&	27.53	 $\pm$ 	1.53	&	0.903	 $\pm$ 	0.276	\\
&	2.46	 $\pm$ 	1.17	&	10.658	 $\pm$ 	0.889	&	32.48	 $\pm$ 	4.77	&	3.914	 $\pm$ 	1.779	\\
&	6.75	 $\pm$ 	2.63	&	16.964	 $\pm$ 	1.510	&	30.92	 $\pm$ 	5.09	&	7.833	 $\pm$ 	3.019	\\
           \hline
7\,mm&	2.25	 $\pm$	0.94	&	0.213	 $\pm$	0.015	&	213.16	 $\pm$	4.14	&	0.094	 $\pm$	0.031	\\
&	62.72 $\pm$	4.76	&	0.000	 $\pm$	0.002	&	           -     	                &	0.071	 $\pm$	0.004	\\
&	15.38 $\pm$	2.42	&	0.198	 $\pm$	0.015	&	40.08	 $\pm$	4.27	&	0.235	 $\pm$	0.030	\\
&	8.37	 $\pm$	1.85	&	0.738	 $\pm$	0.036	&	24.60	 $\pm$	2.78	&	0.386	 $\pm$	0.072	\\
&	7.02	 $\pm$	1.85	&	1.632	 $\pm$	0.079	&	25.16	 $\pm$	2.78	&	0.667	 $\pm$	0.159	\\
&	2.12	 $\pm$	1.09	&	3.495	 $\pm$	0.232	&	22.66	 $\pm$	3.79	&	0.988	 $\pm$	0.463	\\
     \hline
\end{tabular}
\label{table:3C264}
\end{table*}

\begin{table*}[]
\centering
\caption{Modelfit parameters of the circular Gaussian features describing the 1\,cm and 7\,mm  VLBI structure of NGC\,4278.}
\begin{tabular}{ccccc}
\hline
\hline
1\,cm& $ S$   & $ r$  &$ pa$  & $ d$  \\
          & $\mathrm {[mJy]}$& $\mathrm {[mas]}$&  $\rm{[deg]}$ &  $\mathrm {[mas]}$  \\
           \hline
&	6.23	 $\pm$ 	1.49	&	13.748	 $\pm$ 	0.608	&	126.53	 $\pm$ 	2.53	&	5.305	 $\pm$ 	1.216	\\
&	2.67	 $\pm$ 	0.89	&	5.119	 $\pm$ 	0.590	&	146.56	 $\pm$ 	6.57	&	3.807	 $\pm$ 	1.180	\\
&	12.13 $\pm$ 	1.53	&	0.503	 $\pm$ 	0.052	&	189.69	 $\pm$ 	5.94	&	1.013	 $\pm$ 	0.105	\\
&	14.39 $\pm$ 	1.61	&	0.000	 $\pm$ 	0.003	&	           -     	                 &	0.072	 $\pm$ 	0.006	\\
&	6.85	 $\pm$ 	1.12	&	0.391	 $\pm$ 	0.011	&	-5.14	          $\pm$ 	1.59	&	0.185	 $\pm$ 	0.022	\\
&	9.20	 $\pm$ 	1.34	&	0.808	 $\pm$ 	0.065	&	5.50	          $\pm$ 	4.60	&	1.079	 $\pm$ 	0.130	\\
&	2.44	 $\pm$ 	0.82	&	5.373	 $\pm$ 	0.499	&	-36.38	 $\pm$ 	5.30	&	3.236	 $\pm$ 	0.997	\\
&	4.57	 $\pm$ 	1.28	&	12.145	 $\pm$ 	0.704	&	-49.86	 $\pm$ 	3.32	&	5.215	 $\pm$ 	1.407	\\
&	0.60	 $\pm$ 	0.36	&	27.656	 $\pm$ 	0.069	&	-75.41	 $\pm$ 	0.14	&	0.316	 $\pm$ 	0.138	\\
&	0.76	 $\pm$ 	0.42	&	36.044	 $\pm$ 	0.361	&	-78.47	 $\pm$ 	0.57	&	1.548	 $\pm$ 	0.722	\\
           \hline
7\,mm&	3.91	 $\pm$	1.68	&	0.693	 $\pm$	0.123	&	-2.14	 $\pm$	10.08	&	0.593	 $\pm$	0.246	\\
&	16.51 $\pm$	2.27	&	0.000	 $\pm$	0.005	&	           -     	                 &	0.085	 $\pm$	0.009	\\
&	1.69	 $\pm$	0.94	&	0.524	 $\pm$	0.099	&	190.43	 $\pm$	10.70&	0.385	 $\pm$	0.198	\\
\end{tabular}
\label{table:NGC4278}
\end{table*}

\begin{table*}[]
\centering
\caption{Modelfit parameters of the circular Gaussian features describing the 1\,cm and 7\,mm  VLBI structure of 3C\,338. }
\begin{tabular}{ccccc}
\hline
\hline
1\,cm& $ S$   & $ r$  &$ pa$  & $ d$  \\
          & $\mathrm {[mJy]}$& $\mathrm {[mas]}$&  $\rm{[deg]}$ &  $\mathrm {[mas]}$  \\
           \hline
&	5.09	 $\pm$ 	1.89	&	5.430	 $\pm$ 	0.463	&	82.94	 $\pm$ 	4.87	&	2.569	 $\pm$ 	0.926	\\
&	1.55	 $\pm$ 	0.73	&	3.364	 $\pm$ 	0.036	&	83.80	 $\pm$ 	0.62	&	0.194	 $\pm$ 	0.072	\\
&	8.08	 $\pm$ 	1.79	&	2.706	 $\pm$ 	0.062	&	85.20	 $\pm$ 	1.31	&	0.618	 $\pm$ 	0.123	\\
&	3.99	 $\pm$ 	1.22	&	0.937	 $\pm$ 	0.056	&	76.40	 $\pm$ 	3.44	&	0.424	 $\pm$ 	0.113	\\
&	28.71 $\pm$ 	2.95	&	0.000	 $\pm$ 	0.004	&	            -    	                 &	0.117	 $\pm$ 	0.009	\\
&	2.98	 $\pm$ 	1.06	&	0.460	 $\pm$ 	0.067	&	-80.58	 $\pm$ 	8.23	&	0.431	 $\pm$ 	0.133	\\
&	2.33	 $\pm$ 	0.91	&	1.556	 $\pm$ 	0.051	&	263.42	 $\pm$ 	1.87	&	0.310	 $\pm$ 	0.102	\\
&	2.09	 $\pm$ 	0.98	&	4.817	 $\pm$ 	0.203	&	-89.99	 $\pm$ 	2.41	&	0.955	 $\pm$ 	0.406	\\
           \hline
7\,mm&	5.26	 $\pm$	1.76	&	2.814	 $\pm$	0.147	&	85.24	 $\pm$	3.00	&	0.924	 $\pm$	0.295	\\
&	1.57	 $\pm$	0.80	&	0.503	 $\pm$	0.085	&	101.89	 $\pm$	9.55	&	0.384	 $\pm$	0.169	\\
&	21.47 $\pm$	2.55	&	0.000	 $\pm$	0.003	&	           -                       	&	0.066	 $\pm$	0.006	\\
&	1.14	 $\pm$	0.65	&	0.653	 $\pm$	0.103	&	258.03	 $\pm$	9.00	&	0.448	 $\pm$	0.207	\\
     \hline
\end{tabular}
\label{table:3C338}
\end{table*}

\begin{table*}[]
\centering
\caption{Modelfit parameters of the circular Gaussian features describing the 1\,cm and 7\,mm  VLBI structure of 4C\,30.31. }
\begin{tabular}{ccccc}
\hline
\hline
1\,cm& $ S$   & $ r$  &$ pa$  & $ d$  \\
          & $\mathrm {[mJy]}$& $\mathrm {[mas]}$&  $\rm{[deg]}$ &  $\mathrm {[mas]}$  \\
           \hline
&	49.28 $\pm$ 	3.99	&	0.000	 $\pm$ 	0.004	&	          -      	                 &	0.129	 $\pm$ 	0.008	\\
&	10.26 $\pm$ 	1.89	&	0.287	 $\pm$ 	0.018	&	239.53	 $\pm$ 	3.63	&	0.242	 $\pm$ 	0.036	\\
&	5.24	 $\pm$ 	1.37	&	0.847	 $\pm$ 	0.028	&	232.55	 $\pm$ 	1.91	&	0.262	 $\pm$ 	0.056	\\
&	1.48	 $\pm$ 	0.77	&	1.758	 $\pm$ 	0.056	&	229.92	 $\pm$ 	1.81	&	0.261	 $\pm$ 	0.111	\\
&	1.12	 $\pm$ 	0.66	&	4.242	 $\pm$ 	0.043	&	233.97	 $\pm$ 	0.58	&	$<0.187$	\\
           \hline
7\,mm&	46.96 $\pm$	4.36	&	0.000	 $\pm$	0.003	&	           -                      	&	0.077	 $\pm$	0.005	\\
&	5.47	 $\pm$	1.52	&	0.296	 $\pm$	0.012	&	236.30	 $\pm$	2.23	&	0.108	 $\pm$	0.023	\\
&	2.48	 $\pm$	1.04	&	0.653	 $\pm$	0.012	&	231.20	 $\pm$	1.08	&	$<0.075$	\\
&	1.39	 $\pm$	0.84	&	1.352	 $\pm$	0.110	&	229.91	 $\pm$	4.66	&	0.450	 $\pm$	0.220	\\
     \hline
\end{tabular}
\label{table:4C30.31}
\end{table*} 

\begin{table*}[]
\centering
\caption{Modelfit parameters of the circular Gaussian features describing the 1\,cm and 7\,mm  VLBI structure of 3C\,382. }
\begin{tabular}{ccccc}
\hline
\hline
1\,cm& $ S$   & $ r$  &$ pa$  & $ d$  \\
          & $\mathrm {[mJy]}$& $\mathrm {[mas]}$&  $\rm{[deg]}$ &  $\mathrm {[mas]}$  \\
           \hline
&	1.70	 $\pm$ 	0.81	&	1.322	 $\pm$ 	0.025	&	228.55	 $\pm$ 	1.10	&	$<0.145$	\\
&	64.85 $\pm$ 	4.71	&	0.000	 $\pm$ 	0.002	&	           -                       	&	0.093	 $\pm$ 	0.005	\\
&	16.67 $\pm$ 	2.46	&	0.230	 $\pm$ 	0.013	&	48.31	 $\pm$ 	3.36	&	0.226	 $\pm$ 	0.027	\\
&	10.33 $\pm$ 	2.11	&	0.992	 $\pm$ 	0.043	&	55.10	 $\pm$ 	2.47	&	0.469	 $\pm$ 	0.085	\\
&	8.55   $\pm$ 	1.83	&	2.013	 $\pm$ 	0.028	&	55.57	 $\pm$ 	0.80	&	0.309	 $\pm$ 	0.056	\\
&	10.49 $\pm$ 	2.08	&	2.460	 $\pm$ 	0.033	&	57.74	 $\pm$ 	0.76	&	0.376	 $\pm$ 	0.065	\\
&	1.89	 $\pm$ 	0.84	&	2.940	 $\pm$ 	0.019	&	59.64	 $\pm$ 	0.38	&	$<0.125$	\\
&	5.72	 $\pm$ 	2.11	&	8.885	 $\pm$ 	0.437	&	55.34	 $\pm$ 	2.81	&	2.449	 $\pm$ 	0.873	\\
&	3.48	 $\pm$ 	1.52	&	3.390	 $\pm$ 	0.164	&	58.40	 $\pm$ 	2.77	&	0.788	 $\pm$ 	0.328	\\
&	2.08	 $\pm$ 	0.97	&	4.980	 $\pm$ 	0.138	&	55.51	 $\pm$ 	1.59	&	0.685	 $\pm$ 	0.277	\\
           \hline
7\,mm&	63.65 $\pm$	5.31	&	0.000	 $\pm$	0.002	&	          -                       	&	0.059	 $\pm$	0.004	\\
&	2.50	 $\pm$	1.09	&	0.310	 $\pm$	0.011	&	53.65	 $\pm$	1.96	&	$<0.068$	\\
&	2.00	 $\pm$	0.99	&	0.628	 $\pm$	0.014	&	54.50	 $\pm$	1.31	&	$<0.079$	\\
&	3.09	 $\pm$	1.23	&	1.189	 $\pm$	0.015	&	55.99	 $\pm$	0.73	&	0.099	 $\pm$	0.030	\\
&	6.74	 $\pm$	2.18	&	2.467	 $\pm$	0.083	&	57.18	 $\pm$	1.92	&	0.548	 $\pm$	0.165	\\
     \hline
\end{tabular}
\label{table:3C382}
\end{table*} 

\begin{table*}[]
\centering
\caption{Modelfit parameters of the circular Gaussian features describing the 1\,cm and 7\,mm  VLBI structure of 3C\,388. }
\begin{tabular}{ccccc}
\hline
\hline
1\,cm& $ S$   & $ r$  &$ pa$  & $ d$  \\
          & $\mathrm {[mJy]}$& $\mathrm {[mas]}$&  $\rm{[deg]}$ &  $\mathrm {[mas]}$  \\
           \hline
&	36.31 $\pm$ 	2.71	&	0.000	 $\pm$ 	0.003	&	           -     	                 &	0.104	 $\pm$ 	0.006	\\
&	4.93	 $\pm$ 	1.03	&	0.220	 $\pm$ 	0.018	&	244.01	 $\pm$ 	4.60	&	0.209	 $\pm$ 	0.035	\\
&	2.70	 $\pm$ 	0.85	&	0.806	 $\pm$ 	0.065	&	242.49	 $\pm$ 	4.62	&	0.467	 $\pm$ 	0.130	\\
&	1.05	 $\pm$ 	0.55	&	3.273	 $\pm$ 	0.098	&	240.11	 $\pm$ 	1.72	&	0.428	 $\pm$ 	0.196	\\
&	0.65	 $\pm$ 	0.43	&	7.965	 $\pm$ 	0.122	&	241.28	 $\pm$ 	0.88	&	0.438	 $\pm$ 	0.245	\\
&	1.42	 $\pm$ 	0.75	&	15.913	 $\pm$ 	0.382	&	244.00	 $\pm$ 	1.37	&	1.521	 $\pm$ 	0.763	\\
           \hline
7\,mm&	25.96 $\pm$	3.24	&	0.000	 $\pm$	0.002	&	          -                       	&	0.039	 $\pm$	0.004	\\
     \hline
\end{tabular}
\label{table:3C388}
\end{table*}

\begin{table*}[]
\centering
\caption{Modelfit parameters of the circular Gaussian features describing the 1\,cm and 7\,mm  VLBI structure of 3C\,452.}
\begin{tabular}{ccccc}
\hline
\hline
1\,cm& $ S$   & $ r$  &$ pa$  & $ d$  \\
          & $\mathrm {[mJy]}$& $\mathrm {[mas]}$&  $\rm{[deg]}$ &  $\mathrm {[mas]}$  \\
           \hline
&	2.18	 $\pm$ 	0.62	&	5.645	 $\pm$ 	0.248	&	82.22	 $\pm$ 	2.51	&	1.890	 $\pm$ 	0.495	\\
&	0.79	 $\pm$ 	0.32	&	2.699	 $\pm$ 	0.019	&	85.60	 $\pm$ 	0.40	&	$<0.133$	\\
&	6.33	 $\pm$ 	0.89	&	1.395	 $\pm$ 	0.014	&	84.31	 $\pm$ 	0.58	&	0.254	 $\pm$ 	0.028	\\
&	7.89	 $\pm$ 	0.98	&	0.511	 $\pm$ 	0.008	&	91.98	 $\pm$ 	0.92	&	0.173	 $\pm$ 	0.016	\\
&	26.98 $\pm$ 	1.82	&	0.000	 $\pm$ 	0.005	&	           -                       	&	0.192	 $\pm$ 	0.010	\\
&	9.16	 $\pm$ 	1.07	&	0.260	 $\pm$ 	0.011	&	268.40	 $\pm$ 	2.36	&	0.233	 $\pm$ 	0.021	\\
&	5.03	 $\pm$ 	0.82	&	1.072	 $\pm$ 	0.029	&	266.69	 $\pm$ 	1.53	&	0.423	 $\pm$ 	0.057	\\
&	2.74	 $\pm$ 	0.60	&	2.042	 $\pm$ 	0.025	&	265.56	 $\pm$ 	0.69	&	0.280	 $\pm$ 	0.049	\\
&	4.48	 $\pm$ 	0.80	&	4.486	 $\pm$ 	0.057	&	266.05	 $\pm$ 	0.72	&	0.733	 $\pm$ 	0.114	\\
&	1.82	 $\pm$ 	0.48	&	5.805	 $\pm$ 	0.024	&	265.47	 $\pm$ 	0.23	&	0.226	 $\pm$ 	0.047	\\
           \hline
7\,mm&	4.81	 $\pm$	0.96	&	0.900	 $\pm$	0.017	&	87.74	 $\pm$	1.11	&	0.220	 $\pm$	0.035	\\
&	9.47	 $\pm$	1.36	&	0.207	 $\pm$	0.016	&	97.14	 $\pm$	4.42	&	0.273	 $\pm$	0.032	\\
&	28.50 $\pm$	2.27	&	0.000	 $\pm$	0.002	&	           -                       	&	0.056	 $\pm$	0.003	\\
&	8.83	 $\pm$	1.31	&	0.512	 $\pm$	0.018	&	265.74	 $\pm$	1.97	&	0.288	 $\pm$	0.035	\\
&	0.68	 $\pm$	0.37	&	2.650	 $\pm$	0.022	&	263.90	 $\pm$	0.47	&	$<0.115$	\\
     \hline
\end{tabular}
\label{table:3C452}
\end{table*}

\begin{table*}[]
\centering
\caption{Modelfit parameters of the circular Gaussian features describing the 1\,cm and 7\,mm  VLBI structure of 3C\,465.}
\begin{tabular}{ccccc}
\hline
\hline
1\,cm& $ S$   & $ r$  &$ pa$  & $ d$  \\
          & $\mathrm {[mJy]}$& $\mathrm {[mas]}$&  $\rm{[deg]}$ &  $\mathrm {[mas]}$  \\
           \hline
&	2.47	 $\pm$ 	0.66	&	0.624	 $\pm$ 	0.026	&	123.62	 $\pm$ 	2.36	&	0.242	 $\pm$ 	0.051	\\
&	15.24 $\pm$ 	1.57	&	0.219	 $\pm$ 	0.003	&	126.21	 $\pm$ 	0.78	&	0.080	 $\pm$ 	0.006	\\
&	58.35 $\pm$ 	3.06	&	0.000	 $\pm$ 	0.001	&	           -     	                 &	0.030	 $\pm$ 	0.001	\\
&	23.53 $\pm$ 	1.95	&	0.230	 $\pm$ 	0.004	&	-54.27	 $\pm$ 	0.92	&	0.120	 $\pm$ 	0.007	\\
&	8.59	 $\pm$ 	1.20	&	0.627	 $\pm$ 	0.012	&	-52.78	 $\pm$ 	1.08	&	0.216	 $\pm$ 	0.024	\\
&	5.32	 $\pm$ 	0.97	&	1.158	 $\pm$ 	0.023	&	-57.09	 $\pm$ 	1.14	&	0.306	 $\pm$ 	0.046	\\
&	3.72	 $\pm$ 	0.87	&	1.961	 $\pm$ 	0.057	&	-53.28	 $\pm$ 	1.67	&	0.556	 $\pm$ 	0.115	\\
&	3.16	 $\pm$ 	0.86	&	3.360	 $\pm$ 	0.123	&	-52.53	 $\pm$ 	2.09	&	0.985	 $\pm$ 	0.246	\\
&	2.66	 $\pm$ 	0.82	&	5.016	 $\pm$ 	0.199	&	-53.07	 $\pm$ 	2.27	&	1.382	 $\pm$ 	0.398	\\
&	2.43	 $\pm$ 	0.82	&	7.850	 $\pm$ 	0.272	&	-49.54	 $\pm$ 	1.98	&	1.706	 $\pm$ 	0.544	\\
&	2.53	 $\pm$ 	0.96	&	12.361	 $\pm$ 	0.547	&	-50.99	 $\pm$ 	2.54	&	2.995	 $\pm$ 	1.095	\\
&	0.79	 $\pm$ 	0.42	&	16.656	 $\pm$ 	0.131	&	-49.91	 $\pm$ 	0.45	&	0.565	 $\pm$ 	0.262	\\
&	0.58	 $\pm$ 	0.35	&	21.658	 $\pm$ 	0.263	&	-52.23	 $\pm$ 	0.70	&	1.043	 $\pm$ 	0.527	\\
           \hline
7\,mm&	2.08	 $\pm$	0.94	&	0.524	 $\pm$	0.086	&	119.67	 $\pm$	9.27	&	0.438	 $\pm$	0.171	\\
&	5.72	 $\pm$	1.37	&	0.146	 $\pm$	0.004	&	127.95	 $\pm$	1.54	&	$<0.046$	\\
&	47.66 $\pm$	3.91	&	0.000	 $\pm$	0.002	&	           -     	                 &	0.057	 $\pm$	0.003	\\
&	10.15 $\pm$	1.82	&	0.160	 $\pm$	0.005	&	-58.31	 $\pm$	1.86	&	0.078	 $\pm$	0.010	\\
&	10.22 $\pm$	1.91	&	0.394	 $\pm$	0.020	&	-51.57	 $\pm$	2.86	&	0.251	 $\pm$	0.039	\\
&	4.32	 $\pm$	1.30	&	0.877	 $\pm$	0.049	&	-56.88	 $\pm$	3.17	&	0.376	 $\pm$	0.097	\\
&	3.43	 $\pm$	1.18	&	1.874	 $\pm$	0.074	&	-51.50	 $\pm$	2.27	&	0.493	 $\pm$	0.149	\\
&	5.53	 $\pm$	1.88	&	3.492	 $\pm$	0.219	&	-48.21	 $\pm$	3.59	&	1.348	 $\pm$	0.438	\\
     \hline
\end{tabular}
\label{table:3C465}
\end{table*}

\begin{table*}[]
\centering
\caption{MODELFIT core component and map parameters used to derive the core resolution limit and brightness temperature at the two wavelengths. }
\begin{tabular}{cc|ccccccc}
\hline
\hline
	&		& $S_{\rm p}^{\rm fit}$    & $\sigma^{\rm fit}$	&     $b_{\rm maj}$	&     $b_{\rm min}$	&	$d_{\rm lim}$	&	$d/d_{\rm lim}$&	$T_{\rm B}^{\rm c}$		\\	
	&		& $\rm [mJy/beam]$&$\rm [mJy/beam]$     &     $\rm [mas]$	&      $\rm [mas]$	&	$\rm [mas]$	&		       &	$[10^{10} \rm K]$	\\
\hline	
NGC\,315	&	1\,cm	&	155.56	 $\pm$	5.85	&0.22	&	0.93	&	0.27	&	0.018	&	3.803	&	8.83	 $\pm$	0.91	\\
	&	7\,mm	&	78.22	 $\pm$	5.61	&0.40	&	0.55	&	0.37	&	0.030	&	1.504	&	2.65	 $\pm$	0.44	\\ \hline
3C\,31	&	1\,cm	&	52.76	 $\pm$	1.62	&0.05	&	0.92	&	0.28	&	0.015	&	2.255	&	12.12	 $\pm$	0.91	\\
	&	7\,mm	&	34.32	 $\pm$	1.71	&0.09	&	0.48	&	0.18	&	0.014	&	1.152	&	9.31	 $\pm$	1.13	\\ \hline
3C\,33	&	1\,cm	&	12.08	 $\pm$	0.70	&0.04	&	1.05	&	0.40	&	0.035	&	3.063	&	0.28	 $\pm$	0.04	\\
	&	7\,mm	&	18.46	 $\pm$	1.43	&0.11	&	0.65	&	0.32	&	0.033	&	0.694	&	$\geq1.50$	\\ \hline
3C\,66B	&	1\,cm	&	66.87	 $\pm$	1.68	&0.04	&	0.68	&	0.33	&	0.011	&	5.904	&	4.07	 $\pm$	0.25	\\
	&	7\,mm	&	68.34	 $\pm$	4.54	&0.30	&	0.37	&	0.22	&	0.018	&	4.934	&	0.69	 $\pm$	0.11	\\ \hline
B2\,0222+36	&	1\,cm	&	23.41	 $\pm$	1.08	&0.05	&	0.92	&	0.58	&	0.032	&	3.012	&	0.67	 $\pm$	0.08	\\
	&	7\,mm	&	16.97	 $\pm$	1.55	&0.14	&	0.53	&	0.34	&	0.036	&	2.203	&	0.19	 $\pm$	0.04	\\ \hline
IC\,310	&	1\,cm	&	39.30	 $\pm$	1.60	&0.07	&	0.63	&	0.40	&	0.019	&	2.023	&	6.68	 $\pm$	0.66	\\
	&	7\,mm	&	34.95	 $\pm$	3.71	&0.39	&	0.41	&	0.34	&	0.037	&	1.740	&	0.59	 $\pm$	0.15	\\ \hline
4C\,39.12	&	1\,cm	&	37.59	 $\pm$	1.40	&0.05	&	0.67	&	0.29	&	0.015	&	3.247	&	3.84	 $\pm$	0.35	\\
	&	7\,mm	&	26.47	 $\pm$	1.71	&0.11	&	0.38	&	0.17	&	0.015	&	1.710	&	2.76	 $\pm$	0.43	\\ \hline
IC\,2402	&	1\,cm	&	33.42	 $\pm$	1.36	&0.06	&	1.02	&	0.34	&	0.023	&	1.286	&	10.49	 $\pm$	1.04	\\
	&	7\,mm	&	27.48	 $\pm$	1.48	&0.08	&	0.50	&	0.20	&	0.016	&	4.607	&	0.40	 $\pm$	0.05	\\ \hline
3C\,264	&	1\,cm	&	90.43	 $\pm$	3.43	&0.13	&	0.97	&	0.34	&	0.021	&	4.768	&	2.47	 $\pm$	0.23	\\
	&	7\,mm	&	59.73	 $\pm$	3.28	&0.18	&	0.59	&	0.23	&	0.019	&	3.730	&	0.88	 $\pm$	0.12	\\ \hline
NGC\,4278	&	1\,cm	&	14.34	 $\pm$	1.14	&0.09	&	0.94	&	0.30	&	0.040	&	1.809	&	0.68	 $\pm$	0.13	\\
	&	7\,mm	&	13.18	 $\pm$	1.41	&0.15	&	0.35	&	0.12	&	0.020	&	4.215	&	0.16	 $\pm$	0.04	\\ \hline
3C\,338	&	1\,cm	&	25.88	 $\pm$	1.98	&0.15	&	0.76	&	0.26	&	0.032	&	3.672	&	0.53	 $\pm$	0.10	\\
	&	7\,mm	&	20.55	 $\pm$	1.76	&0.15	&	0.40	&	0.20	&	0.023	&	2.902	&	0.35	 $\pm$	0.07	\\ \hline
4C\,30.31	&	1\,cm	&	43.35	 $\pm$	2.64	&0.16	&	0.82	&	0.25	&	0.026	&	4.981	&	0.76	 $\pm$	0.11	\\
	&	7\,mm	&	42.32	 $\pm$	2.92	&0.20	&	0.42	&	0.17	&	0.017	&	4.510	&	0.56	 $\pm$	0.09	\\ \hline
3C\,382	&	1\,cm	&	60.33	 $\pm$	3.21	&0.17	&	0.75	&	0.25	&	0.022	&	4.275	&	1.95	 $\pm$	0.25	\\
	&	7\,mm	&	58.66	 $\pm$	3.60	&0.22	&	0.38	&	0.15	&	0.014	&	4.291	&	1.33	 $\pm$	0.20	\\ \hline
3C\,388	&	1\,cm	&	32.93	 $\pm$	1.82	&0.10	&	0.59	&	0.25	&	0.020	&	5.292	&	0.90	 $\pm$	0.12	\\
	&	7\,mm	&	24.80	 $\pm$	2.24	&0.20	&	0.28	&	0.14	&	0.017	&	2.318	&	1.25	 $\pm$	0.27	\\ \hline
3C\,452	&	1\,cm	&	22.30	 $\pm$	1.16	&0.06	&	0.89	&	0.28	&	0.024	&	7.884	&	0.19	 $\pm$	0.02	\\
	&	7\,mm	&	27.57	 $\pm$	1.58	&0.09	&	0.45	&	0.24	&	0.018	&	3.169	&	0.68	 $\pm$	0.10	\\ \hline
3C\,465	&	1\,cm	&	57.79	 $\pm$	2.15	&0.08	&	0.90	&	0.28	&	0.017	&	1.725	&	16.17	 $\pm$	1.46	\\
	&	7\,mm	&	45.23	 $\pm$	2.69	&0.16	&	0.46	&	0.18	&	0.016	&	3.505	&	1.04	 $\pm$	0.15	\\
 \hline
\end{tabular}
\tablefoot{Col. 1: Source name. Col. 2: Wavelength. Col. 3: Peak intensity of the modelfit core component and associated uncertainty. Col. 4: Post-fit rms of the modelfit map. Col. 5: Beam major axis. Col. 6: Beam minor axis.
Col. 7: Resolution limit. Col. 8: Ratio between the core size and the resolution limit. Col. 9: Core brightness temperature and associated uncertainty. A lower limit is given when the core size is smaller than the resolution limit.}
\label{table:TB}
\end{table*}

\end{appendix}

\end{document}